\documentclass[twocolumn, floatfix, prb, aps, showpacs, longbibliography]{revtex4-2}
\usepackage{graphicx, amssymb, color, amsmath}
\usepackage{nicefrac}
\usepackage{multirow, array, booktabs}
\usepackage{longtable}
\usepackage{float}
\usepackage{romannum}
\usepackage{mathtools}
\usepackage{bbm}
\usepackage{mathrsfs}
\usepackage{IEEEtrantools}
\usepackage{relsize}
\usepackage{mathrsfs}
\usepackage{mathtools}
\usepackage{xparse}
\usepackage[titletoc, title]{appendix}
\usepackage[colorlinks, bookmarks=true, citecolor=blue, linkcolor=red, urlcolor=blue]{hyperref}
\usepackage{orcidlink}
\AtBeginDocument{\pagenumbering{arabic}}

\begin{document}
\newcommand{\sump}{\sideset{}{'}\sum}
\newcommand{\addPRD}[1]{{\bf \color{blue}{#1}}}
\newcommand{\addACB}[1]{{\bf \color{red}{#1}}}

\title{Dispersion of collective modes in spinful fractional quantum Hall states on the sphere}
 
\author{Rakesh K. Dora\orcidlink{0009-0009-0043-2982}}
\email{prakeshdora@imsc.res.in}
\affiliation{Institute of Mathematical Sciences, CIT Campus, Chennai, 600113, India}
\affiliation{Homi Bhabha National Institute, Training School Complex, Anushaktinagar, Mumbai 400094, India}
\affiliation{Department of Physics, Indian Institute of Technology Bombay, Mumbai, MH 400076, India}

\author{Ajit C. Balram\orcidlink{0000-0002-8087-6015}}
\email{cb.ajit@gmail.com}
\affiliation{Institute of Mathematical Sciences, CIT Campus, Chennai 600113, India}
\affiliation{Homi Bhabha National Institute, Training School Complex, Anushaktinagar, Mumbai 400094, India} 

\date{\today}

\begin{abstract}
Collective modes capture the dynamical aspects of fractional quantum Hall (FQH) fluids. Depending on the active degrees of freedom, different types of collective modes can arise in a FQH state. In this work, we consider spinful FQH states in the lowest Landau level (LLL) along the Jain sequence of fillings $\nu{=}n/(2n{\pm}1)$ and compute the Coulomb dispersion of their spin-flip and spin-conserving collective modes in the spherical geometry. We use the LLL-projected density-wave and composite fermion (CF) exciton states as trial wave functions for these modes. To evaluate the dispersion of density-wave states, we derive the commutation algebra of spinful LLL-projected density operators on the sphere, which enables us to extract the gap of the density-wave excitations from the numerically computed density-density correlation function, i.e., the static structure factor, of the FQH ground state. We find that the CF excitons provide an accurate description of the collective modes at all wavelengths, while the density-wave states fail to do so. Specifically, the spin-flip density wave reliably captures the spin-flip collective mode only for the Laughlin and Halperin states, and that too only in the long-wavelength limit. Interestingly, for spin-singlet primary Jain states, the spin-conserving density mode is inaccurate even in the long-wavelength regime. We show that this discrepancy stems from the presence of an additional high-energy spin-conserving parton mode, similar to that found in fully polarized secondary Jain states at $\nu{=}n/(4n{\pm}1)$. We propose an ansatz for this parton mode and compute its Coulomb dispersion in the singlet state at $\nu{=}2/5$. The predicted parton mode can be observed in circularly polarized inelastic light scattering experiments.
\end{abstract}

\maketitle
\section{Introduction}
A fractional quantum Hall (FQH) state is a strongly interacting phase of two-dimensional electrons, first discovered at cryogenic temperatures under a strong perpendicular magnetic field~\cite{Tsui82}. At high magnetic fields, the electronic motion is restricted to a single Landau level (LL)---a macroscopically degenerate manifold of single-particle states with identical kinetic energy---and, in particular, to the lowest LL (LLL). Moreover, at high magnetic fields, the Zeeman energy is large, resulting in the complete spin polarization of the electrons, thereby rendering their dynamics to be entirely governed by Coulomb interactions within the LLL. With improvements in sample quality and electron mobility, FQH states were later observed at low magnetic fields~\cite{Mendez84, Stormer99, Pan02}. This development enabled the observation of spinful FQH states with various spin polarizations, arising from the delicate competition between the Coulomb and Zeeman energies~\cite{Du95, Kukushkin99}.  

Interestingly, at certain LL fillings, even for vanishing Zeeman energy, the exchange part of the interaction energetically favors aligning the spins of all electrons, a phenomenon dubbed quantum Hall ferromagnetism~\cite{Sondhi93}, leading to a fully polarized state. In such cases, even if the ground state is fully spin polarized, the lowest-lying excitation may involve a spin-flip~\cite{Kallin84, Mandal01, Wurstbauer11, Amet15, An25}. Specifically, at LL filling $\nu{=}1/3,$ and in its vicinity, a rich variety of low-energy spinful excitations, such as spin-waves~\cite{Pinczuk93}, a far-separated pair of quasihole and spin-reversed quasiparticle~\cite{Boebinger85, Chakraborty86}, and topological spin textures called skyrmions~\cite{Sondhi93, Groshaus08, Balram15d}, have been observed. Here, we consider the spin-conserving and spin-flip collective excitations, which arise from bound states of a quasihole and spin-conserving and spin-flip quasiparticle pair, respectively. This bound state is charge-neutral; thus, the collective excitations can propagate with a well-defined momentum, giving rise to a dispersion characteristic of the underlying interactions and the parent FQH ground state. 

The dispersion of collective modes in FQH states was first computed with density-wave ansatzes that describe density-modulated states built atop the uniform ground state~\cite{Girvin85, Girvin86, Rasolt86}. These states are generated by acting with the LLL-projected density operators on the ground state, and the gap relative to the ground state determines the mode dispersion. For fully spin-polarized Laughlin states~\cite{Laughlin83}, Girvin, MacDonald, and Platzman (GMP) computed the spin-conserving collective-mode dispersion for the Coulomb interaction and found a pronounced minimum in its dispersion---referred to as the magnetoroton minimum~\cite{Girvin85, Girvin86}. Subsequent studies extended the density-wave construction to a few spinful FQH states, leading to variational estimates for their spin-conserving and spin-flip gaps~\cite{Rasolt86, Moon95}. All of these computations were performed in planar geometry.

The composite fermion (CF) theory provides an alternative framework to compute the collective modes in spinful FQH states at Jain fractions $\nu{=}n/(2np{\pm}1)$ in the LLL, where $n$ and $p$ are positive integers~\cite{Jain89}. CFs are weakly interacting topological quasiparticles, which can be envisioned as bound states of electrons and an even number, $2p$, of vortices. The various spin polarizations of the strongly interacting electronic FQH ground states can be understood qualitatively and quantitatively as integer quantum Hall (IQH) states of CFs, with certain numbers of filled spin-up and spin-down ``Landau-like” levels of CFs, called $\Lambda$ levels ($\Lambda$Ls). Likewise, the collective neutral excitations can be understood as arising from CF excitons (CFEs), where the constituent CF hole and CF particle reside in different $\Lambda$Ls, either within the same spin sector or across different spin sectors. Previous works have shown that in the fully polarized $p{=}1$ primary Jain states at $\nu{=}n/(2n{+}1)$, the spin-conserving CFE and spin-flip CFE modes provide an accurate description of the actual low-lying collective excitations~\cite{Dev92, Jain07, Balram24a}. In contrast, the spin-conserving density-wave mode, referred to as the GMP mode, is accurate and reliable only up to the roton minimum for Laughlin states, and only in the long-wavelength limit for $n{>}1$ primary Jain states~\cite{Girvin85, Girvin86, Scarola00, Dora24}. 

In this paper, we compute the dispersion of various density-wave excitations in spinful Jain states in the spherical geometry~\cite{Haldane83}, extending the earlier results obtained on the plane. The spherical geometry is ideally suited to capture the bulk properties, such as the dispersion of the collective modes in the bulk FQH state, as it eliminates boundary effects that arise on the plane due to confinement. Moreover, the topological triviality of the zero-genus sphere ensures that the ground state of FQH systems is unique, in contrast to other compact topologically nontrivial manifolds such as the torus~\cite{Haldane85, Haldane85b}. We extend the spherical algebra of spinless density operators that was previously worked out by us~\cite{Dora24} to spinful density operators, which enables a computation of the spin-conserving and spin-flip density modes for spinful FQH states on the sphere. 

We compare these density-wave dispersions with the CFE modes that are also computed on the sphere. For the spinful primary Jain states, we find that the spin-flip density-wave gaps are reliable only in the long-wavelength limit, where they lie close to the CF spin-flip gaps, which remain accurate at all wavelengths. Surprisingly, for the GMP mode in the spin-singlet primary Jain states, we find that it is inaccurate even in the long-wavelength limit. This failure can be attributed to the presence of multiple density modes, which a single GMP mode, obtained from a single-mode approximation~\cite{Girvin85, Girvin86}, cannot capture, very analogous to the case in fully polarized secondary Jain states at $\nu{=}n/(4n{\pm}1)$ with $n{>}1$~\cite{Balram21d, Nguyen22, Wang22, Balram24}. Motivated by this observation, we show that the long-wavelength GMP mode in the Jain spin-singlet states at $\nu{=}2/(2m{-}1)$ [equivalent to the Halperin-$(m,m,m{-}1)$ states~\cite{Halperin83}] with $m{\geq}3$ decomposes into a low-energy CFE mode and a high-energy parton mode. For the singlet state at $\nu{=}2/5$, we compute the dispersion of both the CFE and parton modes, and predict that they can be experimentally detected via circularly polarized inelastic light-scattering~\cite{Liang24}. For the bosonic singlet state at  $\nu{=}2/3$, the long-wavelength GMP mode closely matches the CFE mode and therefore remains accurate. Analogous to the singlet states, partially polarized states may also support a parton mode, as their long-wavelength GMP mode is found to be inaccurate. 
 
The rest of the article is organized as follows. In Sec.~\ref{sec: model_and_formulation}, we define the LLL-projected density operators on a sphere, the Hamiltonian under consideration, and the different spin-polarized FQH states considered in this work. Next, in Sec.~\ref{sec: dispersion_spin_density_mode_polarized_QH} we compute the spin-wave dispersion in fully polarized FQH states using the density-wave ansatz and the CF theory. The results on the comparison between these two approaches are presented in Sec.~\ref{ssec: spin_wave_dispersion_parallel_vortex_attached_Jain_states} and Sec.~\ref{ssec: spin_wave_dispersion_reverse_vortex_attached_Jain_states} for the $n/(2n{+}1)$ [parallel-vortex attached] and $n/(2n{-}1)$ [reverse-vortex attached] Jain states, respectively. The spin-flip and spin-conserving charge-neutral collective mode's dispersion in unpolarized states based on density-wave ansatz and CF theory is computed in Sec.~\ref{ssec: Spin_gap_from_density_wave_ansatz_wave_functions} and Sec.~\ref{ssec: CFE_gaps}, respectively. The results on the spin-flip gaps and a comparison of these two approaches are discussed in Sec.~\ref{ssec: Spin_flip_gaps_in_singlet_partially_polarized_states}. In Sec.~\ref{ssec: spin_conserving_gaps_unpolarized_states} we present a similar comparison for the spin-conserving gaps and demonstrate the presence of a parton mode in the $\nu{=}2/(2m{-}1)$ singlet states for $m{\geq}3$. The paper is concluded in Sec.~\ref{sec: conclusion} with a summary of our main results and an outlook for the future. The appendices (App.~\ref{app: Antisymmetrized_symmetrized_Halperin_states}-\ref{app: gaps_Haldane_Rezayi}) include many technical details and some additional results.

\section{Model and Formulation}
\label{sec: model_and_formulation}
\subsection{Model}
We consider $N$ spin $s{=}1/2$ particles on a sphere moving under a radially outward magnetic field $B$, emanating from a magnetic monopole of strength $Q{>}0$ at the center of the sphere~\cite{Haldane83}. The radius $R$ of the sphere is tied to the monopole strength as $R{=}\sqrt{Q}\ell$, which follows from the fact that total magnetic flux through the surface of the sphere is $2Q\phi_{0}$. Here, the magnetic length $\ell{=}\sqrt{\hbar c/(eB)}$ and the flux quantum $\phi_{0}{=}hc/e$, where $e$ is the magnitude of the charge of the particles. The Dirac quantization condition, or equivalently gauge invariance, demands that $2Q$ be an integer; as a result, $Q$ could be an integer or a half-integer. The quantum mechanics of a particle on a sphere under a perpendicular magnetic field results in a discrete set of energy levels, called LLs, indexed by an angular momentum quantum number $l$. The eigenstates associated with each LL are given by monopole spherical harmonics $Y^{Q}_{l, m}$~\cite{Wu77}, where $m$ is the azimuthal quantum number, ranging from ${-}l$ to $l$ in steps of one. Due to the presence of a monopole, $l$ is lower bounded and can take values $l{=}Q,Q{+}1,Q{+}2,\cdots$. Thus, $l{=}Q$ corresponds to the lowest LL (LLL), and in general, $l{=}Q{+}n_{\rm LL}$ corresponds to the LL indexed by $n_{\rm LL}$. Throughout this paper, we work in the $s^z$ basis for spins, and label each LL by an additional $s^z$ quantum number: $s^{z}{=}1/2~(\uparrow)$ or $s^{z}{=}{-}1/2~(\downarrow)$. Next, we define the spinful density operators on a sphere, which will be useful in computing the spin-conserving and spin-flip, collectively referred to as spin excitation gaps. 

\subsection{Spinful density operators on the sphere}
On the sphere, the $\alpha$-component of a spinful density operator $\rho$ is defined as~\cite{Rasolt86}
\begin{align}
\label{eq: real_space_density_operator}
    \rho^{\alpha}\left(\boldsymbol{\Omega}\right)&=\sum_{j{=}1}^{N} \sigma^{\alpha}_{j}\otimes \delta\left(\boldsymbol{\Omega}-\boldsymbol{\Omega}_{j}\right).
\end{align}
Here, $\sigma_{j}^{\alpha}$ denotes the Pauli matrices acting on the $j^{\rm th}$ particle, with $\alpha{=}I,x,y,z$, and $\sigma^{I}{\equiv} I$ is the identity matrix in spin space. The coordinate on the sphere is denoted by $\boldsymbol{\Omega}{\equiv}(\theta, \phi){\equiv}(u, v)$, where $u{=}\cos(\theta/2)e^{\iota\phi/2}$ and $v{=}\sin(\theta/2)e^{-\iota\phi/2}$ are the spinor coordinates corresponding to $\theta$ and $\phi$ which are the polar and azimuthal angles~\cite{Haldane83}, respectively, and $\iota{=}\sqrt{{-}1}$ is the imaginary unit. The spin operator $s^{\alpha}{=}(1/2)\sigma^{\alpha}$ for $\alpha{=}x, y, z$ [throughout this work we set $\hbar{=}1$]. Next, since we are interested in collective states carrying definite total orbital angular momentum [referred to as angular momentum from here on in], we decompose Eq.~\eqref{eq: real_space_density_operator} into operators with well-defined angular momentum, i.e.,   
\begin{align}
    \label{eq: angular_momentum_decomposition}
     \rho^{\alpha}\left(\boldsymbol{\Omega}\right)&= \sum_{j{=}1}^{N} \sigma_{j}^{\alpha} ~\otimes~\sum_{L{=}0}^{\infty}~\sum_{M{=}-L}^{L} \frac{\left[Y_{L, M} \left(\boldsymbol{\Omega} \right)\right]^{*}~Y_{L, M} \left(\boldsymbol{\Omega}_{i} \right)}{Q}\nonumber\\
     &=\sum_{L{=}0}^{\infty}~\sum_{M{=}-L}^{L} \frac{\left[Y_{L, M} \left(\boldsymbol{\Omega} \right)\right]^{*}}{Q}~\sum_{j{=}1}^{N}\sigma_{j}^{\alpha} ~\otimes Y_{L, M} \left(\boldsymbol{\Omega}_{i} \right)\nonumber\\
     &\equiv\sum_{L{=}0}^{\infty}~\sum_{M{=}-L}^{L} \frac{\left[Y_{L, M} \left(\boldsymbol{\Omega} \right)\right]^{*}}{Q} \rho^{\alpha}_{L, M}.
\end{align}
To obtain the first line of the above equation, we have inserted into Eq.~\eqref{eq: real_space_density_operator} a representation of the delta function arising from the completeness of the spherical harmonics, which is
\begin{align}
    \delta\left(\boldsymbol{\Omega}-\boldsymbol{\Omega}_{j}\right)&=\sum_{L{=}0}^{\infty}~\sum_{M{=}-L}^{L} \frac{\left[Y_{L, M} \left(\boldsymbol{\Omega} \right)\right]^{*}~Y_{L, M} \left(\boldsymbol{\Omega}_{j} \right)}{Q}.
\end{align}
In the last line of Eq.~\eqref{eq: angular_momentum_decomposition}, we have defined $\rho_{L, M}^{\alpha}$ as the spin-$\alpha$ component of the angular-momentum-space density operator, i.e.,
\begin{align}
\label{eq: first_quantized_angular_momentum_space_density_operator}
   \rho^{\alpha}_{L, M}&= \sum_{j{=}1}^{N}\sigma_{j}^{\alpha} ~\otimes Y_{L, M} \left(\boldsymbol{\Omega}_{i} \right),
\end{align}
where $Y_{L, M}$ denotes the usual spherical harmonic function. The operator $\rho^{\alpha}_{0,0}$ is proportional to the $\alpha$-component of the total spin operator $\vec{\mathbb{S}}$, since $Y_{0,0}$ is just a constant equal to $1/\sqrt{4\pi}$. Specifically,
\begin{align}
\label{eq: spin_angular_momentum_components}
\frac{\sqrt{4\pi}}{2}~\rho^{\alpha}_{0,0}&=\mathbb{S}^{\alpha}\equiv\sum_{j{=}1}^{N}s_{j}^{\alpha}.
\end{align}
For later convenience, we define the total spin raising and lowering operators, $\mathbb{S}^{\pm}$, as
\begin{align}
\label{eq: spin_angular_momentum_raising_lowering_operator}
    \mathbb{S}^{\pm}&= \sum_{j{=}1}^{N}s_{j}^{\pm}\equiv\sum_{j{=}1}^{N}\left(s_{j}^{x}\pm \iota s_{j}^{y}\right).
\end{align}

As we are interested in the spin excitation gaps within the LLL, we project the operator $\rho^{\alpha}_{L, M}$ in Eq.~\eqref{eq: first_quantized_angular_momentum_space_density_operator} onto the LLL. For convenience, we perform the LLL projection in the Fock space. To begin with, let us first write the unprojected operator $\rho^{\alpha}_{L, M}$ in the second-quantized form, in terms of $\chi^{\dagger}_{L, M,\uparrow}$ and $\chi^{\dagger}_{L, M,\downarrow}$, which are LL creation operators for spin-$\uparrow$ and $\downarrow$ particles in the state $|L, M;\uparrow\rangle$ and $|L, M;\downarrow\rangle$, respectively. The formalism presented in this section applies equally to bosons and fermions. Accordingly, LL operators $\chi$ satisfy the usual canonical commutation or anticommutation algebra. To this end, $\rho^{\alpha}_{L, M}$ can be written in the Fock-space as
\begin{align}
\label{eq: Fock_space_unprojected_angular_momentum_density_operator }
    \rho^{\alpha}_{L, M}&=\sum_{\substack{l_1, m_1,\\l_2, m_2,\\\lambda_1,\lambda_2}} \mathcal{W}_{l_1, m_1,\lambda_1}^{\alpha,L, M;l_2, m_2,\lambda_2} ~\chi^{\dagger}_{l_1,m_1,\lambda_1}~\chi_{l_2,m_2,\lambda_2}.
\end{align}
Here, indices $\lambda_{1},\lambda_2{=}\uparrow, \downarrow$ and the coefficient $\mathcal{W}_{l_1, m_1,\lambda_1}^{\alpha;l_2, m_2,\lambda_2}$ is given by
\begin{align}
    \mathcal{W}_{l_1, m_1,\lambda_1}^{\alpha,L, M ;l_2, m_2,\lambda_2}&=\big\langle l_1, m_1;\lambda_1 \big|\sigma^{\alpha} \otimes Y_{L, M}\big |l_2,m_2;\lambda_2\big\rangle\nonumber\\
    &=\big\langle\lambda_1\big|\sigma^{\alpha}\big|\lambda_2\big\rangle ~ \big\langle l_1,m_1\big|Y_{L, M}\big|l_2,m_2\big\rangle.
\end{align}
The LLL projection of $\rho^{\alpha}_{L, M}$ is carried out by setting $l_1{=}l_2{=}Q$ in Eq.~\eqref{eq: Fock_space_unprojected_angular_momentum_density_operator }, which yields
\begin{align}
\label{eq: Fock_space_projected_angular_momentum_density_operator}
    \bar{\rho}^{~\alpha}_{L, M}&\equiv\mathcal{P}_{\rm LLL}~ \rho^{\alpha}_{L, M}~ \mathcal{P}_{\rm LLL}\nonumber\\[1ex]
    &=\sum_{\substack{m_1,m_2\\\lambda_1,\lambda_2}} W_{ m_1,\lambda_1}^{\alpha,L, M;m_2,\lambda_2} ~\chi^{\dagger}_{m_1,\lambda_1}~\chi_{m_2,\lambda_2}.
\end{align}
Here, $\mathcal{P}_{\rm LLL}$ is the LLL projection operator, and we define $W_{ m_1,\lambda_1}^{\alpha,L, M;m_2,\lambda_2}{\equiv} \mathcal{W}_{Q,m_1,\lambda_1}^{\alpha,L, M;Q,m_2,\lambda_2}$. Furthermore, the LL index $l$ has been omitted from $\chi$ and $\chi^{\dagger}$, as we will consider particles confined to the LLL Hilbert space in what follows. The allowed values of $L$ in $\bar{\rho}^{~\alpha}_{L, M}$ range from $0$ to $2Q$ in steps of one, which follows from the addition of angular momenta of two particles in the LLL~\cite{He94, Dora24}. Therefore, the coefficients $W_{ m_1,\lambda_1}^{\alpha,L, M;m_2,\lambda_2}$ vanish for $L{>}2Q$, rendering $\bar{\rho}^{~\alpha}_{L, M}$ a null operator in that range.

Upon projection to the LLL, density operators at different angular momenta do not commute with each other; instead, they form a closed commutation algebra~\cite{Dora24}. In particular, the commutator of $\bar{\rho}_{L_1,M_1}^{~\alpha}$ and $\bar{\rho}_{L_2,M_2}^{~\beta}$, where $\alpha,\beta{=}I,x,y,z$, is~\cite{Dora24}
\begin{align}
\label{eq: spinful_operator_commutation_algebra}
\left[\bar{\rho}_{L_1,M_1}^{~\alpha},~\bar{\rho}_{L_2,M_2}^{~\beta}\right]&=\sum_{L{=}0}^{2Q}\bigg[\iota \varepsilon^{I\alpha\beta\gamma} A^{L_1,L_2,M_1,M_2}_{L, M}~\bar{\rho}_{L, M}^{~\gamma} \nonumber \\
&+\delta_{\alpha,\beta}\left(1-\delta_{\alpha,I}\delta_{\beta,I}\right) C^{L_1,L_2,M_1,M_2}_{L, M}~\bar{\rho}_{L, M}^{~I} \nonumber \\
&+\delta_{\alpha,I}~C^{L_1,L_2,M_1,M_2}_{L, M}~\bar{\rho}_{L, M}^{~\beta}\nonumber \\
&+\delta_{\beta,I}~C^{L_1,L_2,M_1,M_2}_{L, M}~\bar{\rho}_{L, M}^{~\alpha}\nonumber \\
&+\delta_{\alpha,I}~\delta_{\beta,I} ~C^{L_1,L_2,M_1,M_2}_{L, M}~\bar{\rho}_{L, M}^{~I}\bigg].
\end{align}
Here, $\varepsilon^{\tau\alpha\beta\gamma}$ denotes the four-dimensional Levi-Civita symbol with the convention that $\varepsilon^{Ixyz}{=}1$, the indices $\tau,\gamma{=}I,x,y,z$, and $M{=}M_1 {+}M_2$. Furthermore, in the above commutation algebra, the coefficients $ A^{L_1,L_2,M_1,M_2}_{L, M}{=}D_{+}$ and $C^{L_1,L_2,M_1,M_2}_{L, M}{=}D_{-}$, where $D_{\pm}$ are expressible in terms of Wigner$-3j$ and Wigner$-6j$ symbols as
\begin{align}
   D_{\pm}&= (-1)^{M}\bigg[(-1)^{L_1+L_2+L}\pm1\bigg]\frac{\mathcal{F}(L_1)\mathcal{F}(L_2)}{\mathcal{F}(L)}\nonumber\\
  &\times(2L+1)\left(\begin{array}{ccc}
L_1 & L_2 & L \\
M_1 & M_2 & -M
\end{array}\right)\left\{\begin{array}{lll}
L_1 & L_2 & L \\
Q & Q & Q
\end{array}\right\},
\end{align}
with the LLL form factor $\mathcal{F}(K)$ given as
\begin{align}
\label{eq: LL_form_factor}
   \mathcal{F}(K)=(2Q+1)\sqrt{\frac{2K+1}{4\pi}} \left(\begin{array}{ccc}
Q & Q & K \\
-Q & Q & 0
\end{array}\right).
\end{align}

In the following, we define specific density operators used to compute different types of spin excitation gaps, thereby establishing the terminology used in the remainder of the paper. We define the operators $\bar{\rho}_{L, M}^{~\pm}$ as
\begin{align}
\label{eq: spin_density_operartor_first_quantized}
    \bar{\rho}_{L, M}^{~\pm}&= \mathcal{P}_{\rm LLL}\left[\sum_{j{=}1}^{N}\sigma^{\pm}_{j} ~\otimes Y_{L, M} \left(\boldsymbol{\Omega}_{j} \right)\right]\mathcal{P}_{\rm LLL}\nonumber\\[1ex]
    &= \bar{\rho}_{L, M}^{~x}~ \pm~\iota~ \bar{\rho}_{L, M}^{~y},
\end{align}
which flips a spin-$\downarrow$ particle into a spin-$\uparrow$ particle or vice-versa. The corresponding Fock space expressions of $\bar{\rho}_{L, M}^{~\pm}$ in Eq.~\eqref{eq: spin_density_operartor_first_quantized} are
\begin{align}
\label{eq: down_up_fock_space_SFD}
\bar{\rho}_{L, M}^{~+}&=\sum_{m_1,m_2} B_{ m_1,\uparrow}^{+;m_2,\downarrow} ~\chi^{\dagger}_{m_1,\uparrow}~\chi_{m_2,\downarrow}, ~~\text{and},\\[1.5ex]
\bar{\rho}_{L, M}^{~-}&= (-1)^{M}\left[\bar{\rho}_{L,{-}M}^{~+}\right]^{\dagger},
\label{eq: up_down_fock_space_SFD}
\end{align}
where
\begin{align}
B_{ m_1,\uparrow}^{+;m_2,\downarrow}&= \big\langle\uparrow\big|\sigma^{+}\big|\downarrow\big\rangle ~ \big\langle Q,m_1\big|Y_{L, M}\big|Q,m_2\big\rangle\nonumber\\
&=\langle Q,m_1\big|Y_{L, M}\big|Q,m_2\big\rangle.
\end{align}
In obtaining Eq.~\eqref{eq: up_down_fock_space_SFD}, we have used the identity $\left[Y_{L, M}\right]^{*}{=}(-1)^{M}~Y_{L,-M}$. The operators $\bar{\rho}_{L, M}^{~\pm}$ act as spin-wave (SW) generators, and we therefore call them the SW operators. 

Next, we consider the operator $\bar{\rho}_{L, M}^{~z}$ [defined in Eq.~\eqref{eq: Fock_space_projected_angular_momentum_density_operator} for $\alpha{=}z$], which generates a different type of spin excitation compared to the SW density operators. The operator $\bar{\rho}_{L, M}^{~z}$ can be expressed in terms of spin-up and spin-down density operators, $\bar{\rho}_{L, M}^{~\uparrow}$ and  $\bar{\rho}_{L, M}^{~\downarrow}$, as
\begin{align}
\label{eq: projected_anti_symmetric_density_operator}
   \bar{\rho}_{L, M}^{~z}&= \bar{\rho}_{L, M}^{~\uparrow}~-~\bar{\rho}_{L, M}^{~\downarrow},
\end{align}
where
\begin{align}
    \bar{\rho}_{L, M}^{\lambda}&=\sum_{m_1,m_2}B_{ m_1,\lambda}^{z;m_2,\lambda} ~\chi^{\dagger}_{m_1,\lambda}~\chi_{m_2,\lambda},
\end{align}
with $\lambda{=}\uparrow,\downarrow$. In what follows, we refer to $\bar{\rho}_{L, M}^{~z}$ as the antisymmetric density wave (ADW) operator.

Finally, the operator $\bar{\rho}_{L, M}^{~I}$  [defined in Eq.~\eqref{eq: Fock_space_projected_angular_momentum_density_operator} for $\alpha{=}I$ and noting $s^{I}{=}\mathbb{I}$] acts as identity in spin space and therefore, when applied to the ground state, generates a density wave excitation with the same spin as the ground state. Thus, $\bar{\rho}_{L, M}^{~I}$ is the total spin density operator. In other words, $\bar{\rho}_{L, M}^{~I}$ can be expressed as
\begin{align}
\label{eq: projected_symmetric_density_operator}
  \bar{\rho}_{L, M}^{~I}& \equiv \bar{\rho}_{L, M} = \bar{\rho}_{L, M}^{~\uparrow}~+~\bar{\rho}_{L, M}^{~\downarrow}.
\end{align}  
Thus, we refer to $\bar{\rho}_{L, M}^{~I}$ as the symmetric density wave (SDW) operator. Moreover, $\bar{\rho}_{L, M}^{~I}$ is equivalent to the usual GMP density operator~\cite{Girvin85, Girvin86} generalized to spinful particles. Therefore, we will refer to $\bar{\rho}_{L, M}^{~I}$ as the GMP operator in certain contexts.

The operators $\bar{\rho}_{L, M}^{~-}$, $\bar{\rho}_{L, M}^{~z}$, and $\bar{\rho}_{L, M}^{~+}$ constitute the triplet of operators in the spin $\mathbb{S}{=}1$ sector, creating excitations with $\mathbb{S}^{z}{=}{-}1,0,1$, respectively. On the other hand, $\bar{\rho}_{L, M}^{~I}$ belongs to the $\mathbb{S}{=}0$ singlet sector. This follows from the fact that adding the spins of the constituent electron and hole states, each carrying spin-$1/2$, in the density operators results in either a singlet ($\mathbb{S}{=}0$) or a triplet ($\mathbb{S}{=}1$) state. Acting with the density operators in the triplet sector on a state with spin $\mathbb{S}$ generally produces a linear superposition of states with spin $\left|\mathbb{S}{-}1\right|$, $\mathbb{S}$, and $\mathbb{S}{+}1$. Thus, $\bar{\rho}_{L, M}^{~-}$, $\bar{\rho}_{L, M}^{~z}$, and $\bar{\rho}_{L, M}^{~+}$ when acted upon a singlet, i.e., $\mathbb{S}{=}0$, state, yield a definite $\mathbb{S}{=}1$ spin state. This is analogous to the action of $\bar{\rho}_{L, M}^{~\alpha}$ on a uniform state with $L{=}0$, which generates a definite angular momentum $L$ state. As we discuss in the subsequent sections, for non-singlet ground states with $\mathbb{S}{>}0$, density operators in the triplet sector generally do not produce states with a definite spin. In this paper, $\bar{\rho}_{L, M}^{~-}$, $\bar{\rho}_{L, M}^{~z}$, and $\bar{\rho}_{L, M}^{~+}$ are referred to as spin-flip density operators, since, generically, they alter the spin of the ground state.

Before concluding this section, we note some useful commutation algebra of density operators, which follows directly from Eq.~\eqref{eq: spinful_operator_commutation_algebra}:
\begin{align}
\label{eq: sigma_plus_sigm_minus_density_commutation}
\left[\bar{\rho}_{L_1,M_1}^{~-},\bar{\rho}_{L_2,M_2}^{~+}\right]&=2\sum_{L{=}0}^{2Q}\left[C_{L, M}^{L_1,L_2,M_1,M_2}\bar{\rho}^{~I}_{L, M}\right.\nonumber\\
 &\qquad\left. -A_{L, M}^{L_1,L_2,M_1,M_2}\bar{\rho}_{L, M}^{~z} \right],\\
\label{eq: sigma_minus_sigm_plus_density_commutation}
\left[\bar{\rho}_{L_1,M_1}^{~+},\bar{\rho}_{L_2,M_2}^{~-}\right]&=2\sum_{L{=}0}^{2Q}\left[C_{L, M}^{L_1,L_2,M_1,M_2}\bar{\rho}^{~I}_{L, M} \right.\nonumber\\
 &\qquad\left. +A_{L, M}^{L_1,L_2,M_1,M_2}\bar{\rho}_{L, M}^{~z} \right],\\
\label{eq: identity_sigm_plus_density_commutation}
\left[\bar{\rho}_{L_1,M_1}^{~I},\bar{\rho}_{L_2,M_2}^{~+}\right]&=\sum_{L{=}0}^{2Q}C_{L, M}^{L_1,L_2,M_1,M_2} \bar{\rho}_{L, M}^{~+},\\
\label{eq: identity_sigm_minus_density_commutation}
\left[\bar{\rho}_{L_1,M_1}^{~I},\bar{\rho}_{L_2,M_2}^{~-}\right]&=\sum_{L{=}0}^{2Q}C_{L, M}^{L_1,L_2,M_1,M_2} \bar{\rho}_{L, M}^{~-}.
\end{align}


\subsection{Lowest Landau level projected Hamiltonian}
\label{ssec: interaction_Hamiltonian}
The Hamiltonian of the system consists of two parts: the two-body radially symmetric inter-particle interactions and the one-body Zeeman coupling of spins to the external magnetic field $B$. We consider an $SU(2)$ symmetric interaction between particles, wherein the intra- and inter-spin species interactions are all identical. Since we are interested only in the dynamics of electrons within the LLL, we project the Hamiltonian to the LLL and write it in the angular momentum space as~\cite{He94, Dora24}:
\begin{align}
\label{eq: normal_ordered_projected_interaction_Hamiltonian}
\bar{H}=\frac{4\pi}{2}\sum_{L{=}0}^{2Q}v_{L}\sum_{M{=}-L}^{L}{\colon}~\left[\bar{\rho}^{~I}_{L, M}\right]^{\dagger}\bar{\rho}^{~I}_{L, M}~{\colon} - \mu B \mathbb{S}^{z}.
\end{align}
Here, $\mu{=}g\mu_{B}$, where $g$ is the Land\'e $g$ factor and $\mu_{B}$ is the Bohr magneton that represents the magnetic dipole moment of the particle due to its spin. Furthermore, $v_{L}$ is the weight of the angular momentum space decomposition of the real space interaction $v\left(\left|\boldsymbol{\Omega}{-}\boldsymbol{\Omega}^{\prime}\right|\right)$ of two particles positioned at $\boldsymbol{\Omega}$ and $\boldsymbol{\Omega}^{\prime}$ on the sphere, where the decomposition is carried out as~\cite{Wooten14}
\begin{align}
\label{eq: interaction_angular_momentum_space}
    v\big(|\boldsymbol{\Omega} {-}\boldsymbol{\Omega^{'}}|\big)=4\pi\sum_{L{=}0}^{\infty}~v_{L}\sum_{M{=}-L}^{L}Y_{L, M}(\boldsymbol{\Omega})~Y_{L, M}^{*}(\boldsymbol{\Omega^{'}}).
\end{align}
For the $1/r$ Coulomb interaction $v^{(C)}\left(\left|\boldsymbol{\Omega}{-}\boldsymbol{\Omega}^{\prime}\right|\right){=}e^2/\left(\epsilon\ell\sqrt{Q}\left|\boldsymbol{\Omega}{-}\boldsymbol{\Omega}^{\prime}\right|\right)$, where $\epsilon$ is the dielectric constant of the background host material, the corresponding harmonic $v_{L}$ is given by $v^{(C)}_L{=}1/\left(\sqrt{Q}(2L{+1})\right)$~\cite{He94}. All the Coulomb energies in this work are quoted in units of $e^{2}/(\epsilon\ell)$. For parameters appropriate to GaAs, in which the electronic band mass $m_{b}{=}0.067m_{e}$, where $m_{e}$ is the bare electron mass, relative permittivity $\epsilon_{r}{=}\epsilon/\epsilon_{0}{=}12.6$, and Land\'e $g$ factor $g{=}{-}0.44$, the Coulomb energy scale $e^2/(\epsilon\ell){\approx}50\sqrt{B}$ Kelvin (K) and the Zeeman energy $|g|\mu_{B}B_{\rm tot}{\approx}0.3B_{\rm tot}~{\rm K}$, where the perpendicular and total magnetic fields $B$ and $B_{\rm tot}$, respectively, are in Tesla~\cite{Papic22}. Similarly, for a $\mathcal{R}$-ranged Trugman-Kivelson (TK) interaction~\cite{Trugman85} $v^{(\mathcal{R}{-}TK)}(|\boldsymbol{\Omega} {-}\boldsymbol{\Omega^{'}}|){=}(\nabla^{2}_{\boldsymbol{\Omega}})^{\mathcal{R}}\delta(\boldsymbol{\Omega} {-}\boldsymbol{\Omega^{'}})$, the harmonics $v_{L}$ are given by $1/(4\pi)\left[(-L(L{+}1))^{\mathcal{R}}/Q^{\mathcal{R}{+}1}\right ]$~\cite{Dora24}. The summation over $L$ in Eq.~\eqref{eq: normal_ordered_projected_interaction_Hamiltonian} is limited to $2Q$, since $\bar{\rho}^{~I}_{L, M}$ becomes a null operator for $L{>}2Q$, as previously noted. Next, for convenience, we write the normal-ordered operators in Eq.~\eqref{eq: normal_ordered_projected_interaction_Hamiltonian} in terms of the product of projected density operators, i.e.,
\begin{align}
\label{eq: normal_ordered_and_product_density_operator_relation}
    {\colon}~\left[\bar{\rho}^{~I}_{L, M}\right]^{\dagger}\bar{\rho}^{~I}_{L, M}~{\colon}&=\left[\bar{\rho}^{~I}_{L, M}\right]^{\dagger}\bar{\rho}^{~I}_{L, M}~-~\sum_{\tilde{L}{=}0}^{2Q} \mathcal{C}_{\tilde{L},0}^{L, M}~ \bar{\rho}^{~I}_{\tilde{L},0},
\end{align}
where [see also Eq.~\eqref{eq: LL_form_factor}],
\begin{align}
\mathcal{C}_{\tilde{L},0}^{L, M}&=  (-1)^{L+M}~(2\tilde{L}+1)\frac{\left(\mathcal{F}\left(L\right)\right)^2} {\mathcal{F}\left(\tilde{L}\right)}\left(\begin{array}{ccc}
L & L & \tilde{L} \\
-M & M & 0
\end{array}\right)\nonumber\\
&~~~~~~~~~~~~~~~~~~~~~~~\times~\left\{\begin{array}{lll}
L & L & \tilde{L} \\
Q & Q & Q
\end{array}\right\}.
\end{align}
The second term in Eq.~\eqref{eq: normal_ordered_and_product_density_operator_relation} compensates for the self-interaction included in the first term $\left[\bar{\rho}^{~I}_{L, M}\right]^{\dagger}\bar{\rho}^{~I}_{L, M}$. Substituting Eq.~\eqref{eq: normal_ordered_and_product_density_operator_relation} into Eq.~\eqref{eq: normal_ordered_projected_interaction_Hamiltonian}, the Hamiltonian becomes
\begin{align}
    \bar{H} &= \frac{4\pi}{2}\sum_{L{=}0}^{2Q}v_{L}\sum_{M{=}-L}^{L}\left[\bar{\rho}^{~I}_{L, M}\right]^{\dagger}\bar{\rho}^{~I}_{L, M}~-~\bar{H}_{1\rm body}- \mu B \mathbb{S}^{z},
\end{align}
where
\begin{align}
   \bar{H}_{1\rm body}&= \frac{4\pi}{2}\sum_{L{=}0}^{2Q}v_{L}\sum_{M{=}-L}^{L}~\sum_{\tilde{L}{=}0}^{2Q} \mathcal{C}_{\tilde{L},0}^{L, M}~ \bar{\rho}^{~I}_{\tilde{L},0}.
\end{align}
Interestingly, $\bar{H}_{1\rm body}$ is proportional to the identity operator. In other words, using the properties of Wigner $3j$ and $6j$ symbols, $\bar{H}_{1\rm body}$ can be expressed as
\begin{align}
  \bar{H}_{1\rm body}&=\frac{4\pi}{2}\frac{\sqrt{4\pi}}{2Q+1}\left[\sum_{L{=}0}^{2Q} v_{L}~\left[\mathcal{F}\left(L\right)\right]^{2} \right]\bar{\rho}^{~I}_{0,0},
\end{align}
where $\bar{\rho}^{~I}_{0,0}$ acts as an identity operator in both the orbital and spin space [see also Eq.~\eqref{eq: Fock_space_projected_angular_momentum_density_operator}]. Throughout this paper, we set the Zeeman coupling $\mu{=}0$, as we are interested in the contribution to the energies arising purely from the inter-particle interactions. The single-particle Zeeman energies can be readily added to these. To this end, the Hamiltonian we will consider, $\bar{H}$, is
\begin{align}
\label{eq: int_Hamiltonian_without_Zeeman_term}
   \bar{H} &= \frac{4\pi}{2}\sum_{L{=}0}^{2Q}v_{L}\sum_{M{=}-L}^{L}\left[\bar{\rho}^{~I}_{L, M}\right]^{\dagger}\bar{\rho}^{~I}_{L, M}~-~\bar{H}_{1\rm body}.
\end{align}
The energy of a uniform state $\left|\Psi_{0}\right\rangle$ with respect to $\bar{H}$ can be computed from the state's projected static structure factor $\bar{S}^{I}\left(L\right)$ as~\cite{Dora24}
\begin{align}
    \left\langle\bar{H}\right\rangle_{\Psi_{0}}&=\frac{N}{2}\sum_{L{=}0}^{2Q}v_{L}\left(2L+1\right)\left[\bar{S}^{I}\left(L\right)-\frac{4\pi\left(\mathcal{F}\left(L\right)\right)^{2}}{(2Q+1)(2L+1)}\right].
\end{align}
Here, we have used $\left\langle\bar{\rho}^{~I}_{0,0}\right\rangle{=}N/(\sqrt{4\pi})$ and defined $\bar{S}^{I}\left(L\right)$ as follows
\begin{align}
\label{eq: definition_projected_S(L)}
   \bar{S}^{I}\left(L\right)&=\frac{4\pi}{N}\left\langle\Psi_{0}\right|\left[\bar{\rho}^{~I}_{L, M}\right]^{\dagger}\bar{\rho}_{L, M}^{~I}\left|\Psi_{0}\right\rangle.
\end{align}
Directly computing $\bar{S}^{I}\left(L\right)$ in Fock space for FQH states for large system sizes is quite challenging since the Hilbert space grows very quickly with $N$. Instead, $ \bar{S}^{I}\left(L\right)$ can be obtained from the unprojected structure factor $S^{I}\left(L\right)$---which can be efficiently computed for large systems for trial wave functions of our interest using Monte Carlo techniques~\cite{Kamilla97, Balram17}---via the relation~\cite{Dora24}
\begin{align}
\label{eq: relaton_unprojected_projected_structure_factor_fully_polarized_states}
 \bar{S}^{I}\left(L\right)&= S^{I}\left(L\right)-1+\frac{4\pi}{(2L+1)(2Q+1)}\left[\mathcal{F}\left(L\right)\right]^{2}.
\end{align}
An interaction Hamiltonian is often parametrized by a set of Haldane pseudopotentials $\{V_{\mathfrak{m}}\}$~\cite{Haldane83}, where $V_{\mathfrak{m}}$ is the energy of two particles in a relative angular momentum $\mathfrak{m}$ state. To write the Hamiltonian in terms of the product of density operators as in Eq.~\eqref{eq: int_Hamiltonian_without_Zeeman_term},  one can compute appropriate real-space harmonics $v_{L}$ via the relation~\cite{Wooten14}
\begin{align}
\label{eq: inversion_pair_pseudopotential}
v_L & =\frac{1}{(2 Q+1)^2}\left(\begin{array}{ccc}
Q & L & Q \\
-Q & 0 & Q
\end{array}\right)^{-2}~\sum_{\mathfrak{m}=0}^{2 Q}\bigg[(-1)^{\mathfrak{m}}~ V_{\mathfrak{m}}\nonumber \\
& \times \left(2 \left(2Q-\mathfrak{m}\right)+1\right)\left\{\begin{array}{lll}
Q & Q & ~~~~L \\
Q & Q & 2Q-\mathfrak{m}
\end{array}\right\} \bigg],
\end{align}
for a given set of pseudopotentials $\{V_{\mathfrak{m}}\}$.

\subsection{Ground state wave function for fractional quantum Hall states with spin}
\label{eq: GS_WF_with_spin}
Insights into the ground state of the Hamiltonian in Eq.~\eqref{eq: int_Hamiltonian_without_Zeeman_term} for certain values of $v_{L}$, such as the Coulomb interaction in the LLL, can be gained through the CF theory~\cite{Jain89}. The CF theory naturally predicts the possible ground state spin polarizations of electronic states at $\nu{=}n/(2pn{\pm}1)$ and also provides microscopic wave functions that can be used for quantitative tests. The Jain wave function of interacting electrons is constructed from the IQH states of CFs occupying spin-$\uparrow$ and/or spin-$\downarrow$ $\Lambda$Ls. In particular, the ground state wave function with polarization $\gamma{=}(n_{\uparrow}{-}n_{\downarrow})/(n_{\uparrow}{+}n_{\downarrow})$, is given in terms of CFs carrying $2p$ vortices, as~\cite{Jain89}
\begin{align}
\label{eq: CF_GS_WF}
    \Psi_{n/(2pn+1)}^{\left(\gamma\right)}&=\mathcal{P}_{\rm LLL}~\Phi_{1}^{2p}\Phi_{n}\equiv\mathcal{P}_{\rm LLL}~\Phi_{1}^{2p}\Phi_{n_\uparrow}\Phi_{n_\downarrow}.
\end{align}
Here, $n{=}n_{\uparrow}{+}n_{\downarrow}$, where $n_{\uparrow\left(\downarrow\right)}$ is the number of spin-$\uparrow\left(\downarrow\right)$ $\Lambda$Ls filled by CFs in the IQH state $\Phi_{n_{\uparrow}}\left(\Phi_{n_{\downarrow}}\right)$. The Jastrow factor $\Phi_{1}$, which is also the wave function of $\nu{=}1$, is
\begin{align}
\label{eq: Jastrow_factor}
    \Phi_{1}{=}\prod_{1\leq i<j \leq N}\left(u_{i}v_{j}{-}u_{j}v_{i}\right).
\end{align}
Note that in Eq.~\eqref{eq: Jastrow_factor}, the product runs over all the $\uparrow$ and $\downarrow$ electrons. More specfically, $\Phi_{1}$ can be written as
\begin{align}
    \Phi_{1}&=\Phi_{\rm intra}^{\uparrow} \Phi_{\rm intra}^{\downarrow} \Phi_{\rm inter}^{\uparrow,\downarrow},
\end{align}
where
\begin{align}
  \Phi_{\rm intra}^{\uparrow}&=\prod_{1\leq i<j \leq N_{\uparrow}}\left(u^{\uparrow}_{i}v^{\uparrow}_{j}-u^{\uparrow}_{j}v^{\uparrow}_{i}\right),\\
  \Phi_{\rm intra}^{\downarrow}&=\prod_{1\leq i<j \leq N_{\downarrow}}\left(u^{\downarrow}_{N_{\uparrow}+i}v^{\downarrow}_{N_{\uparrow}+j}-u^{\downarrow}_{N_{\uparrow}+j}v^{\downarrow}_{N_{\uparrow}+i}\right),\\
  \Phi_{\rm inter}^{\uparrow,\downarrow}&=\prod_{1\leq i \leq N_{\uparrow}} \prod_{1\leq j \leq N_{\downarrow}}\left(u^{\uparrow}_{i}v^{\downarrow}_{N_{\uparrow}+j}-u^{\downarrow}_{N_{\uparrow}+j}v^{\uparrow}_{i}\right).
\end{align}
Here, $N_{\uparrow}$ and $N_{\downarrow}$ denote the number of $\uparrow$ [labeled $\{1,2,{\cdots}, N_{\uparrow}\}$] and $\downarrow$ [labeled $\{N_{\uparrow}{+}1,N_{\uparrow}{+}2,{\cdots}, N_{\uparrow}{+}N_{\downarrow}{\equiv}N\}$] particles, respectively, with $N{=}N_{\uparrow}{+}N_{\downarrow}$. Throughout this paper, without loss of generality, we consider $n_{\uparrow}{\geq}n_{\downarrow}$ in Eq.~\eqref{eq: CF_GS_WF}. For the case when $n_{\uparrow}{>}0$ and $n_{\downarrow}{=}0$, the state $\Psi$ in Eq.~\eqref{eq: CF_GS_WF} is a fully polarized state. Similarly, $\Psi$ is a spin-singlet state when $n_{\uparrow}{=}n_{\uparrow}$, and to a partially polarized state when $n_{\uparrow}{>}n_{\uparrow}$ with $n_{\uparrow}{\neq}0$. 

At $\nu{=}n/(2np{+}1)$, CFs see an effective positive magnetic field. At fillings $\nu{=}n/(2np{-}1)$, CFs sense an effective negative magnetic field. The Jain CF wave function for these $n/(2np{-}1)$ states is obtained by complex conjugating $\Phi_{n}$ in Eq.~\eqref{eq: CF_GS_WF}, i.e.,
\begin{align}
\label{eq: CF_GS_WF_negative_magnetic_field}
    \Psi_{n/(2pn-1)}^{\left(\gamma\right)}&=\mathcal{P}_{\rm LLL}~\Phi_{1}^{2p}[\Phi_{n_\uparrow}\Phi_{n_\downarrow}]^{*}.
\end{align}
We note that these spinful CF states give a good representation of the Coulomb ground state in the LLL~\cite{Dev92a, Wu93, Jain07, Balram13, Balram15a, Yang19a, Balram20b, Balram21b}.

Besides fermions, the CF theory also provides candidate states for the bosonic FQH effect. An analogous wave function for bosons can be constructed by considering CFs that bind an odd number of vortices. Specifically, by replacing $2p$ with $2p{-}1$ in Eq.~\eqref{eq: CF_GS_WF}, one obtains a bosonic state at $\nu{=}n/\left[(2p{-}1)n{+}1\right]$:
\begin{align}
\label{eq: CF_GS_WF_bosons}
    \Psi_{n/\left[(2p{-}1)n{+}1\right]}^{\left(\gamma\right)}&=\mathcal{P}_{\rm LLL}~\Phi_1^{(2p-1)}\Phi_{n_\uparrow}\Phi_{n_\downarrow}.
\end{align}
We will only consider the bosonic Halperin-$(2p, 2p, 2p{-}1)$ state at $\nu_{b}{=}2/(4p{-}1)$, which corresponds to $n_{\uparrow}{=}n_{\downarrow}{=}1$ in Eq.~\eqref{eq: CF_GS_WF_bosons}, an in particular, focus on the bosonic Halperin-$(2,2,1)$ state at $\nu_{b}{=}2/3$. 

On the sphere, a quantum Hall state of $N$ particles at filling $\nu$ occurs at flux $2Q{=}\nu^{{-}1}N{-}\mathcal{S}$, where $\mathcal{S}$ is the Wen-Zee shift~\cite{Wen92}. The fermionic state in Eq.~\eqref{eq: CF_GS_WF} occurs at $\mathcal{S}{=}2p{+}\mathcal{S}^{\rm IQH}\left(n_{\uparrow},n_{\downarrow}\right)$, while the bosonic state in Eq.~\eqref{eq: CF_GS_WF_bosons} occurs at $\mathcal{S}{=}2p{-}1{+}\mathcal{S}^{\rm IQH}\left(n_{\uparrow},n_{\downarrow}\right)$, where, the shift of the $\left(n_{\uparrow},n_{\downarrow}\right)$ IQH state is 
\begin{align}
  \mathcal{S}^{\rm IQH}\left(n_{\uparrow},n_{\downarrow}\right)&=\frac{(n_{\uparrow})^2 + (n_{\downarrow})^2}{n}.
\end{align}

In this work, we employ the Jain-Kamilla (JK) method~\cite{Jain97, Jain97b, Jain07, Moller05, Davenport12, Balram15a, Gattu24} to project CF wave functions onto the LLL, as it enables us to access fairly large systems that are beyond the reach of exact diagonalization (ED). The JK projection requires at least two Jastrow factors to multiply the IQH state~\cite{Jain07}. Therefore, in this work, all the CF wave functions for the $2/3$ bosonic states are constructed by dividing the corresponding $2/5$ LLL-projected fermionic states by a Jastrow factor~\cite{Chang05b, Liu20}. Note that generally, this procedure results in ill-defined wave functions, since in generic spinful fermionic states, the wave function does not vanish when particles of opposite spin coincide, which then leads to a singularity when the fermionic state is divided by a Jastrow factor~\cite{Wu13}. However, here we will only consider the ground and excited states of the bosonic Halperin-$(2,2,1)$ state, and in the excitations, allow for at most one CF to be in the second Lambda, and for such states, the wave function obtained by dividing the corresponding LLL-projected fermionic ground and excited states of Halperin-$(3,3,2)$ by a Jastrow factor remains well-defined.

For the wave function in Eq.~\eqref{eq: CF_GS_WF}, quantities of interest, such as the structure factor $S_{L}$ and the pair-correlation function $g(r)$, can be computed by carrying out multi-dimensional integrals, which we evaluate using the Metropolis Monte Carlo method~\cite{Binder10}. For reference, to evaluate $S_{L}$ for the Jain wave functions, for the largest systems we accessed ($N{\sim}50$), we run 30 Monte Carlo chains, where in each chain we do $10^{7}$ Monte Carlo iterations. The largest systems considered here should provide a good representative of the thermodynamic limit since the Laughlin and Jain states of our interest have small correlation lengths of the order of a few magnetic lengths~\cite{Balram15b, Balram17, Fulsebakke23}, while the largest systems we consider are much larger, a few tens of magnetic lengths, compared to these correlation lengths.

Before the advent of the CF theory, Halperin generalized the Laughlin state~\cite{Laughlin81} to propose a class of trial wave functions that can incorporate the spin degrees of freedom. These are referred to as the Halperin-$(m,m,n)$ states~\cite{Halperin83}, where $m,~n$ are non-negative integers, and their wave function is given as
\begin{align}
    \label{eq: Halperin_mmn_WF}
    \Psi^{m,m,n}_{\nu=2/(m+n)}&= \left(\Phi_{\rm intra}^{\uparrow}\right)^{m}\left(\Phi_{\rm intra}^{\downarrow}\right)^{m} \left(\Phi_{\rm inter}^{\uparrow,\downarrow}\right)^{n}.
\end{align}
In this paper, we will restrict ourselves to the case $N_{\uparrow}{=}N_{\downarrow}{=}N/2$, for which on the sphere, the Halperin-$(m,m,n)$ state occurs at $\mathcal{S}{=}m$. 
 Exchange symmetry between particles requires $m$ to be odd for fermions and even for bosons. Note that Eq.~\eqref{eq: Halperin_mmn_WF} does not have a definite exchange symmetry when a spin-$\uparrow$ particle is exchanged with a spin-$\downarrow$ particle for $m{>}1$. A properly symmetrized (for even $m$) or antisymmetrized (for odd $m$) version of the Halperin-$(m,m,n)$ wave function is presented in App.~\ref{app: Antisymmetrized_symmetrized_Halperin_states}. Notably, the Halperin-$(m,m,n)$ states with $n{=}m{-}1$ correspond to the Jain state of CFs carrying $(m{-1})$ vortices. In other words, $\Psi^{m,m,m{-}1}_{2/(2m{-}1)}{=}\Phi_1^{(m-1)}\Phi_{1_\uparrow}\Phi_{1_\downarrow}$. For example, the Halperin-$(2,2,1)$ and Halperin-$(3,3,2)$ states are identical to the spin-singlet bosonic $\nu{=}2/3$ [$p{=}1$ and $(n_{\uparrow}, n_{\downarrow}){=}(1,1)$ in Eq.~\eqref{eq: CF_GS_WF_bosons}] and fermionic $\nu{=}2/5$ [$p{=}1$ and $(n_{\uparrow}, n_{\downarrow}){=}(1,1)$ in Eq.~\eqref{eq: CF_GS_WF}] states, respectively. Moreover, when $m{\geq}1,~n{=}m$, the Halperin-$(m,m,n)$ states represent the $\mathbb{S}{=}N/2,~\mathbb{S}^{z}{=}(N_{\uparrow}{-}N_{\downarrow})/2{=}0$ version of the fully polarized $\nu{=}1/m$ Laughlin state, while for $m{\geq}1,~n{=}m{-}1$ these are spin-singlets~\cite{Yoshioka88, Yoshioka89}, i.e., have $\mathbb{S}{=}0$. For other values of $n$, the Halperin-$(m,m,n)$ states are not eigenstates of the $\vec{\mathbb{S}}^{2}$ total spin operator. Furthermore, the Halperin-$(m,m,m{-}1)$ singlet for $m{\geq}2$ is the exact zero-energy highest-density ground state for the $V_{\mathfrak{m}{<}m{-}1}$ pseudopotential Hamiltonian, i.e., $V_{0}{=}1,~V_{1}{=}1,{\cdots},V_{m{-}2}{=}1$ and $V_{m{-}1}{=}0,~V_{m}{=}0,~{\cdots}$~\cite{Yoshioka89}. In particular, the Halperin $(2,2,1)$ and $(3,3,2)$ states are exact zero-energy highest-densty ground states of the $V_{0}$ and $V_{0}{+}V_{1}$ Hamiltonians.

For convenience, we mention here that at $2/3$, we consider two distinct spin-singlet states: the bosonic Jain CF singlet or equivalently the Halperin-$(2,2,1)$ state as discussed above, and the fermionic Jain CF singlet state, described by Eq.~\eqref{eq: CF_GS_WF_negative_magnetic_field} for $n{=}2$ and $p{=}1$.

Finally, we also consider the Haldane-Rezayi state described by the wave function~\cite{Haldane88a, Haldane88b}
\begin{equation}
\label{eq: Haldane_Rezayi_WF}
\Psi_{1/2}{=} {\rm Det} \left( \frac{1}{\left(u^{\uparrow}_{i}v^{\downarrow}_{N_{\uparrow}+j}-u^{\downarrow}_{N_{\uparrow}+j}v^{\uparrow}_{i}\right)^{2}} \right) \times \Psi^{2,2,2}_{1/2}.
\end{equation}
On the sphere, this state occurs at flux $2Q{=}2N{-}4$ and is a spin-singlet. The Haldane-Rezayi state is the exact zero-energy ground state of the hollow-core $V_{1}$ Hamiltonian. This state is believed to be gapless~\cite{Gurarie97, Read00, Seidel11, Crepel19} though its magnetoroton mode is gapped~\cite{Nguyen23}.

In the subsequent sections, we compute different spin excitation gaps for fully polarized primary Jain states at $\nu{=}n/(2n{+}1)\left[p{=}1\right]$, partially polarized states at $\nu{=}3/5$ and $3/7$, fermionic singlet states at $\nu{=}2/5$ and $2/3$, and the bosonic singlet state at $\nu{=}2/3$. The spin excitation gaps of the Haldane-Rezayi spin-singlet at $\nu{=}1/2$ are presented in App.~\ref{app: gaps_Haldane_Rezayi}.

\section{Spin-wave dispersion in fully polarized quantum Hall states}
\label{sec: dispersion_spin_density_mode_polarized_QH}
In this section, we compute the SW dispersion in fully polarized quantum Hall ground states, i.e., those that have $\mathbb{S}{=}\mathbb{S}^{z}{=}N/2$. As mentioned earlier, we focus specifically on the fully polarized primary Jain states at $\nu{=}n/(2n{\pm}1)$ with $n_{\uparrow}{=}n$ and $n_{\downarrow}{=}0$. Below, we consider two different approaches to describe the SW: one based on the density-wave ansatz, and the other using the spin-flip CF exciton wave description. 

\subsection{Density-wave ansatz for the spin-wave}
\label{ssec: Spin-flip density wave ansatz}
We consider the following density-wave ansatz for a SW, obtained by acting the SW operator $\bar{\rho}^{~-}_{L, M}$ [see Eq.~\eqref{eq: up_down_fock_space_SFD}] on $\left|\Psi_{0}\right\rangle$, which creates a spin-flipped excitation with definite angular momentum, i.e.,
\begin{align}
\Psi^{\rm SW}_{L, M}&=\bar{\rho}^{~-}_{L, M}~\left| \Psi_0\right\rangle.
\end{align}
The energy cost to create the $\Psi^{\rm SW}_{L, M}$ excitation is 
\begin{align}
\label{eq: spin_flip_gap_derivation_1}
    \Delta^{\rm SW}\left(L\right)&=\frac{\left\langle\Psi_0\right|\left[\bar{\rho}^{~-}_{L, M}\right]^{\dagger}\bar{H}\bar{\rho}^{~-}_{L, M}\left| \Psi_0\right\rangle}{\left\langle\Psi_0\right|\left[\bar{\rho}^{~-}_{L, M}\right]^{\dagger}\bar{\rho}^{~-}_{L, M}\left| \Psi_0\right\rangle} ~-~\left\langle\Psi_0\right|\bar{H}\left| \Psi_0\right\rangle.
\end{align}
The above expression can be equivalently written in terms of a single commutator as
\begin{align}
\label{eq: spin_flip_gap_derivation_2}
     \Delta^{\rm SW}\left(L\right)&=\frac{\left\langle\Psi_0\right|\left[\bar{\rho}^{~-}_{L, M}\right]^{\dagger}\left[\bar{H},\bar{\rho}^{~-}_{L, M}\right]\left| \Psi_0\right\rangle}{\left\langle\Psi_0\right|\left[\bar{\rho}^{~-}_{L, M}\right]^{\dagger}\bar{\rho}^{~-}_{L, M}\left| \Psi_0\right\rangle}.
\end{align}
To see that Eqs.~\eqref{eq: spin_flip_gap_derivation_1} and~\eqref{eq: spin_flip_gap_derivation_2} are equivalent, one notes that the action of $\left[\bar{\rho}^{~-}_{L, M}\right]^{\dagger}$ on the fully polarized state $\left|\Psi_{0}\right\rangle$ annihilates the state, i.e.,
\begin{align}
\label{eq: sigma_plus_density_on_polarized_ground_state}
    \left[\bar{\rho}^{~-}_{L, M}\right]^{\dagger} ~\left|\Psi_{0}\right\rangle&=(-1)^{M}~\bar{\rho}^{~+}_{L,-M}\left|\Psi_{0}\right\rangle=0,
\end{align}
which straightforwardly follows from the fact that $\sigma^{+}\left|\uparrow\right\rangle{=}0$. This observation then leads to the simplification of the normalization factor $\mathcal{N}^{\rm SW}\left(L\right){=}\left\langle\Psi_0\right|\left[\bar{\rho}^{~-}_{L, M}\right]^{\dagger}\bar{\rho}^{~-}_{L, M}\left| \Psi_0\right\rangle$ as
\begin{align}
\label{eq: normalization_factor_SF}
   \mathcal{N}^{\rm SW}\left(L\right)&=\left\langle\Psi_0\right|\left[\left[\bar{\rho}^{~-}_{L, M}\right]^{\dagger},\bar{\rho}^{~-}_{L, M}\right]\left| \Psi_0\right\rangle\nonumber\\
    &=2\sum_{\tilde{L}{=}0}^{2Q}(-1)^{M}~A_{\tilde{L},0}^{L,L,-M,M}\left\langle\Psi_0\right|\bar{\rho}_{\tilde{L},0}^{z}\left| \Psi_0\right\rangle\nonumber\\
    &=\frac{2N}{\sqrt{4\pi}}(-1)^{M}~A_{0,0}^{L,L,-M,M},
\end{align}
where in the second line of the above equation, we have used Eq.~\eqref{eq: sigma_plus_sigm_minus_density_commutation}, and in the third line, we have used $\left\langle\Psi_0\right|\bar{\rho}_{\tilde{L},0}^{z}\left| \Psi_0\right\rangle{=}\delta_{\tilde{L},0}N/(\sqrt{4\pi})$. In Eq.~\eqref{eq: normalization_factor_SF}, $A_{0,0}^{L,L,-M,M}$ is given by
\begin{align}
\label{eq: }
A_{0,0}^{L,L,-M,M}&=\frac{(-1)^{M}(2Q+1)((2Q)!)^{2}}{\sqrt{\pi}(2Q-L)!~\Gamma(2Q+L+2)}.
\end{align}
Moreover, following a similar line of reasoning, one can express $\left\langle\Psi_0\right|\left[\bar{\rho}^{~-}_{L, M}\right]^{\dagger}\bar{\rho}^{~-}_{L, M}\bar{H}\left| \Psi_0\right\rangle$ as $\left\langle\Psi_0\right|\left[\left[\bar{\rho}^{~-}_{L, M}\right]^{\dagger},\bar{\rho}^{~-}_{L, M}\right]\bar{H}\left| \Psi_0\right\rangle$, and can further show that
\begin{align}
   \label{eq: spin_flip_gap_derivation_3}
 \frac{\left\langle\Psi_0\right|\left[\left[\bar{\rho}^{~-}_{L, M}\right]^{\dagger},\bar{\rho}^{~-}_{L, M}\right]~\bar{H}\left| \Psi_0\right\rangle}{\mathcal{N}^{\rm SW}\left(L\right)}&= \left\langle\Psi_0\right|\bar{H}\left| \Psi_0\right\rangle.
\end{align}
Therefore, Eqs.~\eqref{eq: spin_flip_gap_derivation_1} and~\eqref{eq: spin_flip_gap_derivation_2} are equivalent. In obtaining Eq.~\eqref{eq: spin_flip_gap_derivation_3}, we have employed Eq.~\eqref{eq: sigma_plus_sigm_minus_density_commutation}, and used the fact that $\left\langle\Psi_0\right|\bar{\rho}_{\tilde{L},0}^{\alpha}\bar{H}\left| \Psi_0\right\rangle{=}0$ for $\tilde{L}{\geq}1$, as the rotational invariance of $\bar{H}$ implies that $\bar{H}\left|\Psi\right\rangle$ resides in the $L{=}0$ sector while $\left\langle\Psi_{0}\right|\bar{\rho}^{\alpha}_{\tilde{L}{\geq}1,0}$ lives in a different angular momentum sector, and thus they are orthogonal and the overlap between them is zero. For $L{=}0$, since $\bar{H}\left|\Psi\right\rangle$ preserves the spin of the state $\left|\Psi\right\rangle$ (in particular, its $\mathbb{S}^{z}$ value) and $\bar{\rho}^{~I}_{0,0}{=}\bar{\rho}^{~z}_{0,0}{\propto}\mathbb{I}$, $\Delta^{\rm SW}\left(L{=}0\right){=}0$.

Next, for convenience, we express $\Delta^{\rm SW}\left(L\right)$ [see Eq.~\eqref{eq: spin_flip_gap_derivation_2}] in terms of a double commutator as
\begin{align}
\label{eq: spin_flip_gap_double_Commutator}
     \Delta^{\rm SW}\left(L\right)&=\frac{\left\langle\Psi_0\right|\left[\left[\bar{\rho}^{~-}_{L, M}\right]^{\dagger},\left[\bar{H},\bar{\rho}^{~-}_{L, M}\right]\left| \Psi_0\right\rangle\right]}{\mathcal{N}^{\rm SW}\left(L\right)},
\end{align}
which straightforwardly follows from Eq.~\eqref{eq: sigma_plus_density_on_polarized_ground_state}. Upon evaluating the commutators using Eqs.~\eqref{eq: sigma_minus_sigm_plus_density_commutation}-\eqref{eq: identity_sigm_minus_density_commutation}, $\Delta^{\rm SW}\left(L\right)$ can then be expressed entirely in terms of ground state correlation functions. Due to the rotational invariance of $\bar{H}$, $\Delta^{\rm SW}\left(L\right)$ is independent of the orbital angular momentum's azimuthal quantum number $M$, so we set $M{=}0$ without loss of generality to obtain
\begin{align}
\label{eq: sphere_SF_gap_equation}
    \Delta^{\rm SW}\left(L\right)&=\frac{\bar{F}^{\rm SW}\left(L\right)}{\mathcal{N}^{\rm SW}\left(L\right)},
\end{align}
where the oscillator strength $\bar{F}^{\rm SW}\left(L\right)$ is given by
\begin{align}
\label{eq: oscillator_strength_fully_polarized_states}
 \bar{F}^{\rm SW}\left(L\right)&=4N\sum_{\tilde{L}=0}^{2Q}v_{\tilde{L}}\sum_{\tilde{M}{=}0}^{\tilde{L}}~ \sum_{\tilde{l}{=}|\tilde{L}-L|}^{\tilde{L}+L}\bigg[(-1)^{\tilde{M}}\sqrt{\pi}\left(C_{\tilde{l},\tilde{M}}^{L,\tilde{L},0,\tilde{M}}\right)^{2} \nonumber \\
 & \times A_{0,0}^{\tilde{l},\tilde{l},-\tilde{M},\tilde{M}}-\left(C_{\tilde{L},\tilde{M}}^{L,\tilde{l},0,\tilde{M}}\right)~\left(C_{\tilde{l},\tilde{M}}^{L,\tilde{L},0,\tilde{M}}\right)~\bar{S}^{I}\left(\tilde{L}\right)\bigg].   
\end{align}
Here, $\bar{S}^{I}\left(L\right)$ is the projected static structure factor of the fully polarized ground state [see Eq.~\eqref{eq: definition_projected_S(L)}]. Note that for a fully polarized state, the $z$ and $I$ density [see Eqs.~\eqref{eq: projected_anti_symmetric_density_operator} and \eqref{eq: projected_symmetric_density_operator}] correlation functions are identical, i.e., $\left\langle\left[\bar{\rho}^{~z}_{L, M}\right]^{\dagger}\bar{\rho}^{~z}_{L, M}\right\rangle{=}\left\langle\left[\bar{\rho}^{~I}_{L, M}\right]^{\dagger}\bar{\rho}^{~I}_{L, M}\right\rangle$. For a $\nu{=}1$ IQH state, $\bar{S}^{I}\left(L\right)$ can be exactly computed analytically and is $ \bar{S}^{I}_{\nu{=}1}\left(L\right){=}N\delta_{L,0}$. Thus, for the $\nu{=}1$ IQH state, $\Delta^{\rm SW}\left(L\right)$ simplifies to [note that $C_{\tilde{l},\tilde{M}}^{0,\tilde{L},0,\tilde{M}}{=}0$ in Eq.~\eqref{eq: oscillator_strength_fully_polarized_states}]:

\begin{figure}[tbh!]
        \includegraphics[width=0.99\columnwidth]{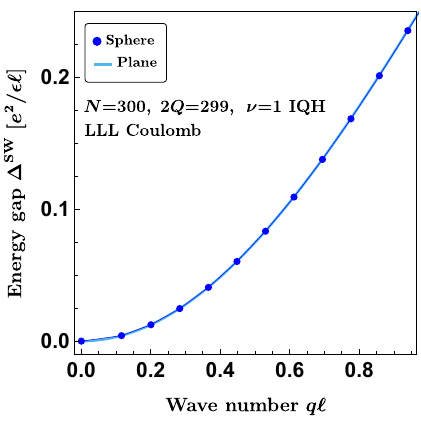}
          \caption{Comparison of the dispersion of the spin-wave mode in the $\nu{=}1$ integer quantum Hall state obtained on the sphere (blue filled dots) and plane (blue shaded line). On the sphere, the spin-wave mode's dispersion is calculated for a finite system of $N{=}300$ electrons, while the planar result is obtained in the thermodynamic limit.}
          \label{fig: SF_dispersion_IQH_state}
        \end{figure}
\begin{align}
\label{eq: SF_gap_nu_1_IQH_state}
    \Delta^{\rm SW}_{\nu{=}1}\left(L\right)&=4\pi\sum_{\tilde{L}={0}}^{2Q}v_{\tilde{L}}\sum_{\tilde{M}{=}0}^{\tilde{L}}~ \sum_{\tilde{l}{=}|\tilde{L}-L|}^{\tilde{L}+L} \Bigg[(-1)^{\tilde{M}}\left(C_{\tilde{l},\tilde{M}}^{L,\tilde{L},0,\tilde{M}}\right)^{2} \nonumber \\
    &~~~~~~~~~~~~~~~~~~~~~~~~~~\qquad \times \frac{A_{0,0}^{\tilde{l},\tilde{l},-\tilde{M},\tilde{M}}}{A_{0,0}^{L,L,0,0}} \Bigg].
\end{align}
This circumvents the rather tedious computation of the SW excitation spectrum of $\nu{=}1$ on the sphere, which was carried out by Nakajima and Aoki~\cite{Nakajima94}. However, unlike our expression, the Nakajima-Aoki computation produces an exact analytic closed-form expression for the SW gap at $\nu{=}1$, in terms of the pseudopotentials $\{V_{\mathfrak{m}}\}$, which is [Eq. (2) of Ref.~\cite{Nakajima94}]:
\begin{align}
\label{eq: SF_gap_Nakajima_Aoki_IQH_state}
    \Delta^{\rm SW, (\rm NA)}_{\nu{=}1}\left(L\right)&=\sum_{\mathfrak{m}{=}0}^{2Q}\Bigg[\left(2\left(2Q-\mathfrak{m}\right)+1\right)(-1)^{\mathfrak{m}}V_{\mathfrak{m}} \\
    &\left[\frac{1}{2Q+1}-(-1)^{\mathfrak{m}}\left\{\begin{array}{lll}
Q & Q & ~~~~L \\
Q & Q & 2Q-\mathfrak{m}
\end{array}\right\}\right]\Bigg].\nonumber
\end{align}
Note that the summation in the above equation is over the relative angular momentum $\mathfrak{m}$. In contrast, Nakajima and Aoki express the summation in terms of the total angular momentum $J$ of two particles. The two quantum numbers $\mathfrak{m}$ and $J$ are related as $\mathfrak{m}{=}2Q{-}J$. We have verified that our result is exactly equivalent to the Nakajima-Aoki result in the sense that the numerical values of the gaps obtained from both expressions match. 

For completeness, we mention below the SW gap equation in the planar (p) geometry, which is the analog of the spherical gap presented in Eq.~\eqref{eq: sphere_SF_gap_equation}~\cite{Rasolt86, Nakajima94}:
\begin{align}
\label{eq: planar_general_SF_gap_equation}
    \Delta^{\rm SW,\left(p\right)}\left(q\right)&=\frac{1}{2\pi}\int_{0}^{\infty} dk ~k v(k)\left[1-J_{0}\left(qk\right)\right]\left[1-S^{I}\left(k\right)\right].
\end{align}
Here, $v\left(k\right)$ is the Fourier component of the real space interaction $v\left(r\right)$ between two particles separated by a distance $r$, $J_{0}\left(x\right)$ denotes the zeroth-order Bessel function of the first kind, and $S^{I}\left(k\right)$ is the planar unprojected structure factor~\cite{Girvin86}. Note that in the above equation, we have set $\ell{=}1$. For FQH states, the unprojected structure factor can be obtained from fitting the pair-correlation function $g(r)$ [see App.~\ref{app: gr_SL_sphere}] obtained from Monte Carlo simulations using the trial wave functions, Fourier transforming it, and fitting it to a particular form, as explained in Refs.~\cite{Girvin86, Fulsebakke23, Dora24}.

At $\nu{=}1$, $S^{I}\left(k\right){=}1{-}e^{-k^2/2}$~\cite{Girvin86, Giuliani08} and consequently, $\Delta^{\rm SW,\left(p\right)}\left(q\right)$ for the Coulomb interaction, for which $v\left(k\right){=}2\pi/k$, reduces to the following result, first derived by Kallin and Halperin~\cite{Kallin84},
\begin{align}
\label{eq: planar_IQH_SF_dispersion}
     \Delta^{\rm SW,\left(p\right)}_{\nu{=}1}\left(q\right)&=\sqrt{\frac{\pi}{2}}\left[1-e^{-q^2/4}I_{0}\left(q^2/4\right)\right],
\end{align}
where $I_{0}\left(x\right)$ is the zeroth-order modified Bessel function of the first kind. 

In Fig.~\ref{fig: SF_dispersion_IQH_state}, we show a comparison of the spherical and planar SW dispersion in the $\nu{=}1$ IQH state for Coulomb-interacting electrons, with the latter corresponding to the thermodynamic limit. The spherical gap is computed for $N{=}300$ electrons using Eq.~\eqref{eq: SF_gap_nu_1_IQH_state} with $v^{(C)}_{\tilde{L}}{=}1/\left(\sqrt{Q}(2\tilde{L}{+1})\right)$~\cite{He94}, or equivalently Eq.~\eqref{eq: SF_gap_Nakajima_Aoki_IQH_state} with the Coulomb pseudopotentials $V_{\mathfrak{m}}{=}\left(2/\sqrt{Q}\right)\left(\binom{2\mathfrak{m}}{\mathfrak{m}}\binom{8Q-2\mathfrak{m}+2}{4Q-\mathfrak{m}+1}/\binom{4Q+2}{2Q+1}^2\right)$~\cite{Fano86}. To facilitate a direct comparison between the two geometries, we have mapped stereographically the angular momentum $L$ on the sphere to the planar linear momentum $q$ as $q\ell{=}\sqrt{L(L{+}1)}/\sqrt{Q}$ (see Apps.~\ref{app: L_to_q} and \ref{app: SF_mode_model_interactions_IQH_state}). The spherical gap closely tracks the planar gap, but eventually, for large enough $L$, there will be a mismatch between the two due to the finite-size curvature effects on the sphere (see Fig.~\ref{fig: SF_dispersion_IQH_state_N_51}). In Secs.~\ref{ssec: spin_wave_dispersion_parallel_vortex_attached_Jain_states} and~\ref{ssec: spin_wave_dispersion_reverse_vortex_attached_Jain_states}, we present a similar comparison for FQH states. 

We conclude this section with two remarks. First, unlike in the GMP gap computations~\cite{Girvin85, Girvin86, Dora24}, the SW gap as computed above is exact, as there is no source of error stemming from expressing Eq.~\eqref{eq: spin_flip_gap_derivation_1} as a double commutator in Eq.~\eqref{eq: spin_flip_gap_double_Commutator}. In particular, arriving at Eq.~\eqref{eq: spin_flip_gap_double_Commutator} does not require $\left|\Psi_{0}\right\rangle$ to be an exact eigenstate of $\bar{H}$---a condition that is necessary in writing the GMP gap in the double commutator form [see also discussion below Eq.~\eqref{eq: AS_density_wave_double_commutator}]~\cite{Girvin85, Girvin86, Dora24}. 

Second, although $\Psi^{\rm SW}_{L, M}$ carries a definite orbital angular momentum quantum number by construction, and has $\mathbb{S}^{z}{=}N/2{-}1$, it \emph{does not} possess a well-defined total spin quantum number for FQH states---except when $L{=}0, 1$. To see this, one can compute the expectation value of the square of the total spin angular momentum operator, $\vec{\mathbb{S}}^{2}$ [see Eq.~\eqref{eq: spin_angular_momentum_components}], for $\Psi^{\rm SW}_{L, M}$, which is given by
\begin{align}
     \frac{\left\langle\Psi^{\rm SW}_{L, M}\right|\boldsymbol{\vec{\mathbb{S}}}^{2}\left|\Psi^{\rm SW}_{L, M}\right\rangle}{\mathcal{N}^{\rm SW}\left(L\right)}&=\mathbb{S}\left(\mathbb{S}-1\right)+ \frac{N~\bar{S}^{I}\left(L\right)}{4\pi~\mathcal{N}^{\rm SW}\left(L\right)}.
 \end{align}
 Here, $\mathbb{S}{=}N/2$ is the total spin of $\left|\Psi_{0}\right\rangle$, as mentioned previously. Noting that $\bar{S}^{I}\left(0\right){=}N$ for FQH states, at $L{=}0$, $\left|\Psi^{\rm SW}_{0, 0}\right\rangle$ carries the same spin $\mathbb{S}{=}N/2$ as the ground state $\left|\Psi_{0}\right\rangle$. At $L{=}1$, since $\bar{S}^{I}\left(1\right){=}0$ [this follows from the fact that $\bar{\rho}^{I}_{1, M}$ annihilates $\left|\Psi_{0}\right\rangle$], the state $\left|\Psi^{\rm SW}_{1,M}\right\rangle$ has a definite spin quantum number $\mathbb{S}{=}(N/2){-1}$. On the other hand, the states $\left|\Psi^{\rm SW}_{L{>}1, M}\right\rangle$ do not carry a definite $\mathbb{S}$ due to a finite offset from the value $(N/2)(N/2-1)$. Interestingly, for the $\nu{=}1$ IQH state, as $\bar{S}^{I}\left(L\right){=}N\delta_{L,0}$, the states $\Psi^{\rm SW,\nu{=}1}_{L, M}$ have a definite $\mathbb{S}$ for all $L$. In particular, the state $\left|\Psi^{\rm SW,\nu{=}1}_{0,0}\right\rangle$ has $\mathbb{S}{=}N/2$, while the states $\left|\Psi^{\rm SW,\nu{=}1}_{L{\geq}1,0}\right\rangle$ have $\mathbb{S}{=}(N/2){-}1$. Accordingly, for FQH states, it is meaningful to compare the gap of only $\left|\Psi^{\rm SW}_{1, M}\right\rangle$ with the corresponding gap of states obtained from ED~\cite{Rezayi87} and CF theory, as these states have a definite $\mathbb{S}$ for all $L$ by construction, whereas $\left|\Psi^{\rm SW}_{L{>}1, M}\right\rangle$ do not. 
 
\subsection{Spin-wave dispersion from spin-flip composite fermion excitons in primary Jain states}
\label{ssection: spin_flip_CF_exciton_wave}
\begin{figure*}[tbh!]
        \includegraphics[width=0.93\textwidth]{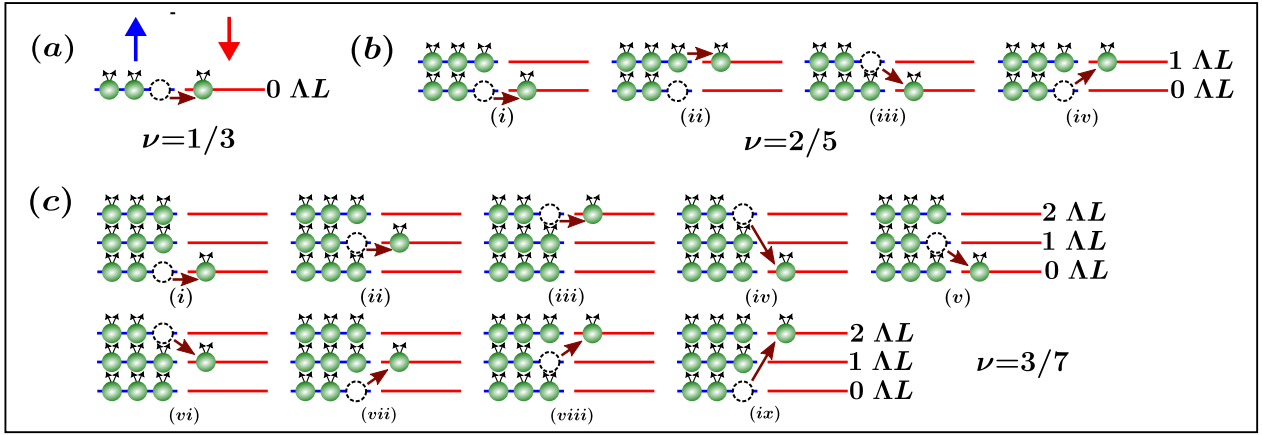}
          \caption{Schematic of the spin-flip CF excitons included in the CF-diagonalization-based spin-flip CF exciton gap computation in $(a)$ $\nu{=}1/3$ Laughlin, $(b)$ $\nu{=}2/5$ Jain, and $\nu{=}3/7$ Jain states. The blue horizontal lines denote the spin-$\uparrow$ $\Lambda$Ls, while the red horizontal lines denote the spin-$\downarrow$ $\Lambda$Ls. }
          \label{fig: SF_gap_schematic}
        \end{figure*} 
In the CF theory, various spin-flip exciton states can be constructed by exciting a CF from an occupied $\uparrow$-${\Lambda}L$ to an empty $\downarrow$-${\Lambda}L$ in the fully polarized state at $\nu{=}n/(2n{\pm}1)$. The gapless SW excitation of the fully polarized state corresponds to the excitons $i_{\uparrow}{\rightarrow}i_{\downarrow}$, with $i{=}0,1,{\cdots},(n{-}1)$. However, for the $n/(2n{\pm}1)$ states with $n{\geq}2$, it has been previously found that a lower energy, hence improved, SW excitation can be obtained by mixing the above $\Lambda$L conserving spin-flipped excitons with additional spin-flipped excitons involving inter-$\Lambda$L transitions~\cite{Mandal01, Wurstbauer11}, $i_{\uparrow}{\rightarrow}j_{\downarrow}$, where $j{<}i$, since in these states the CF cyclotron energy of the exciton is lower than that in the ground state. The range of angular momenta for these spin-flip CFEs can be worked out by the addition of the angular momentum of the constituent CF hole and spin-flipped CF particle~\cite{Jain07, Balram16d}.  

In general, for the $n/(2n{\pm}1)$ Jain states, one can consider $n^2$ spin-flipped CF exciton basis states, where a CF from the $i^{\rm th}$ occupied $\uparrow$-$\Lambda$L is moved to the $j^{\rm th}$ empty $\downarrow$-$\Lambda$L, with $i,j{=}0,1,2,{\cdots},(n{-}1)$. One then performs CF diagonalization (CFD)~\cite{Mandal02, Jain07, Balram15}, i.e., diagonalizes the interaction of interest (for example, the LLL Coulomb interaction) within this set of basis states, and determines the SW dispersion from the resulting lowest-energy eigenvalue at each orbital angular momentum. A discussion on the construction of the spin-flipped CF basis states and the CFD can be found in Refs.~\cite{Mandal01, Majumder14, Makki25}, which we do not repeat here. A schematic illustration of the $n^2$ CF basis states used to compute the CF SW dispersion in the primary Jain states at $\nu{=}n/(2n{+1})$, for $n{=}1,2,3$, is presented in Fig.~\ref{fig: SF_gap_schematic}. We note that for the $\nu{=}1/3~[n{=}1]$ Laughlin state, this amounts to considering solely the spin-flipped CF excitation $0_{\uparrow}{\rightarrow}0_{\downarrow}$ that resides entirely in the lowest $\Lambda$L. Therefore, the wave function of the spin-flipped CF excitation at $1/3$ is simply the wave function of the $\nu{=}1$ SW state times $\Phi_1^2$. The spin-flip CFE excitation for Laughlin states ranges from $L{=}1$ to $L{=}N{-}1$.

At Laughlin fillings $\nu{=}1/(2p{+}1)$, $p{\geq}1$, Nakajima and Aoki computed the SW dispersion using the CF framework, albeit without explicitly employing the CF wave functions~\cite{Nakajima94}. They noted the fact that attaching $2p$ vortices to electrons maps a pair of electrons at relative angular momentum $\mathfrak{m}$ to a pair of CFs at relative angular momentum $\mathfrak{m}{-}2p$~\cite{Nakajima94}. At the mean-field level, where CFs form a $\nu^{\star}{=}1$ IQH state, they approximated the SW dispersion using the same expression presented for electrons at $\nu{=}1$ in Eq.~\eqref{eq: SF_gap_Nakajima_Aoki_IQH_state}, but now appropriately modified for CFs. Since CFs sense an effective reduced flux $2Q^{\star}{=}2Q{-}2p(N{-}1)$, one replaces $2Q$ in Eq.~\eqref{eq: SF_gap_Nakajima_Aoki_IQH_state} with $2Q^{\star}$. Thus, the mean-field CF SW gap obtained by Nakjima and Aoki is~\cite{Nakajima94}
\begin{align}
\label{eq: SF_gap_NAkajima_Aoki_Laughlin_state}
    \Delta^{\rm SW, (\rm NA)}_{\nu{=}1/(2p{+1})}\left(L\right)&=\sum_{\mathfrak{m}{=}0}^{2Q^{*}}\Bigg[\left(2\left(2Q^{*}-\mathfrak{m}\right)+1\right)(-1)^{\mathfrak{m}}\tilde{V}_{\mathfrak{m}}\nonumber\\
    &\left[\frac{1}{2Q^{*}+1}-(-1)^{\mathfrak{m}}\left\{\begin{array}{lll}
Q^{*} & Q^{*} & ~~~~L \\
Q^{*} & Q^{*} & 2Q^{*}-\mathfrak{m}
\end{array}\right\}\right]\Bigg].
\end{align}
Here, $\tilde{V}_{\mathfrak{m}}$ is the pseudopotential sensed by CFs, which one relates to the electron pseudopotential $V_{\mathfrak{m}{+2p}}$ as $ \tilde{V}_{\mathfrak{m}}/\tilde{\ell}{=}V_{\mathfrak{m}{+2p}}/\ell$, where $\tilde{\ell}{=}\sqrt{2p{+}1}\ell$ is the effective magnetic length of CFs~\cite{Nakajima94}. 

Finally, we point out that for $L{=}1$, the SW ansatz and the CF SW in the Laughlin states are identical. To see this, let us denote the Laughlin state as $\left|\Psi_{1/m}^{\rm Laughlin}\right\rangle{=}\left(\Phi_1\right)^{2m} {\otimes}\left|\uparrow_1\uparrow_2{\cdots}\uparrow_N\right\rangle$, and the $L{=}1$, $M{=}1$ SW ansatz as $\left|\Psi_{1,1}^{\rm SW,\rm Laughlin}\right\rangle{=}\bar{\rho}_{1,1}^{-}\left|\Psi_{1/m}^{\rm Laughlin}\right\rangle$. Using $\bar{\rho}_{1,1}^{-}{=}\sum_{i{=}1}^{N} \sigma_{i}^{-}{\otimes}u_i\partial/\partial v_i$~\cite{Pu23}, one can straightforwardly show that $\left|\Psi_{1,1}^{\rm SW,\rm Laughlin}\right\rangle{=}\Phi_{1}^{m{-}1} \bar{\rho}_{1,1}^{-}\left[\Phi_1 {\otimes}\left|\uparrow_1\uparrow_2{\cdots}\uparrow_N\right\rangle\right]$, which is exactly the CF SW state since $\left|\Phi_{1,1}^{\rm SW,~\nu{=}1~\rm IQH}\right\rangle{=}\bar{\rho}_{1,1}^{-}\left[\Phi_1 {\otimes}\left|\uparrow_1\uparrow_2{\cdots}\uparrow_N\right\rangle\right]$. 

In the next two sections, we compare the CF SW dispersion with that obtained from the SW density ansatz for the primary Jain states at $\nu{=}n/(2n{\pm}1)$.

\subsection{Results: Spin-wave in parallel-vortex attached primary Jain states}
\label{ssec: spin_wave_dispersion_parallel_vortex_attached_Jain_states}

\begin{figure*}[tbh!]
        \includegraphics[width=0.66\columnwidth]{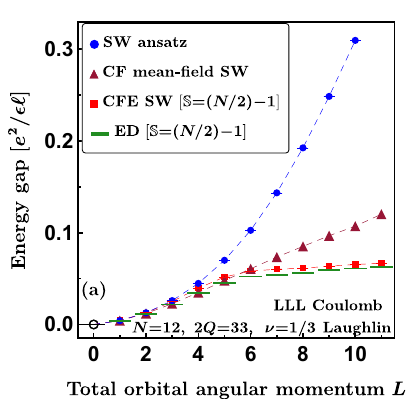}
        \includegraphics[width=0.66\columnwidth]{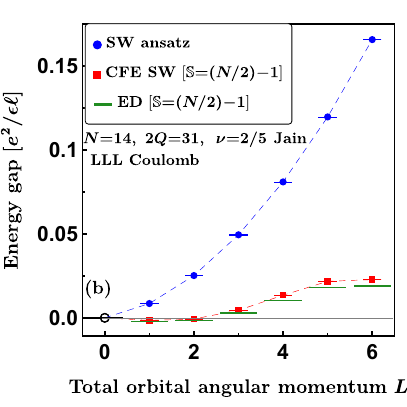}
         \includegraphics[width=0.66\columnwidth]{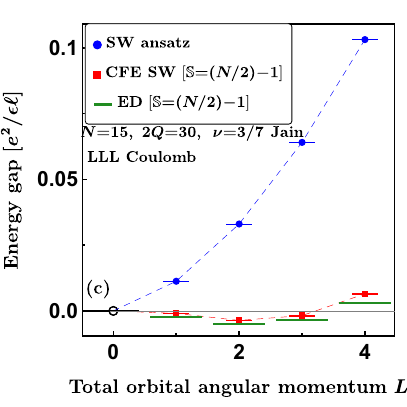} \\
         \includegraphics[width=0.66\columnwidth]{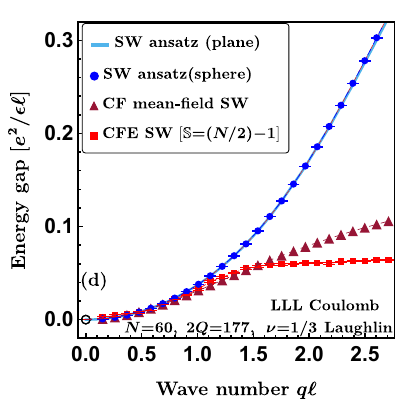}
        \includegraphics[width=0.66\columnwidth]{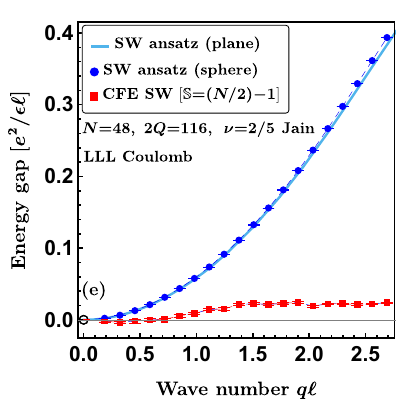}
         \includegraphics[width=0.66\columnwidth]{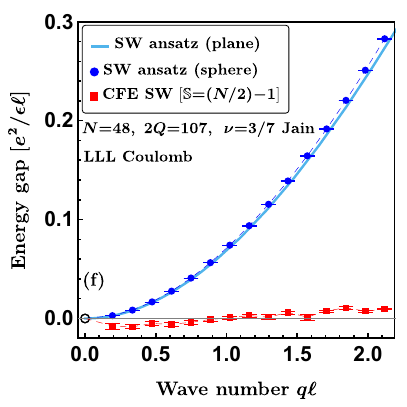}
          \caption{Top panels: Comparison of the dispersion of the spin-wave mode in primary Jain states computed following different methods for small system sizes on the sphere. Here, ED denotes the spin-wave gap obtained from exact diagonalization, results for which are taken from Ref.~\cite{Balram15c}. The bottom panels show a comparison of the dispersion of the sphere spin-wave mode in primary Jain states, computed for larger systems, along with a comparison with planar geometry results. }
          \label{fig: SF_dispersion_primary_Jain_states_1_3_2_5_3_7}
        \end{figure*} 

The SW dispersion, computed following the methods outlined in Secs.~\ref{ssec: Spin-flip density wave ansatz} and~\ref{ssection: spin_flip_CF_exciton_wave}, for the LLL Coulomb interaction, is presented in Fig.~\ref{fig: SF_dispersion_primary_Jain_states_1_3_2_5_3_7} for the first three primary Jain states in the $n/(2n{+}1)$ sequence. We have also computed the SW mode using ED for small systems, shown by green lines in the top panels of Fig.~\ref{fig: SF_dispersion_primary_Jain_states_1_3_2_5_3_7}. As evident from Figs.~\ref{fig: SF_dispersion_primary_Jain_states_1_3_2_5_3_7}$(a{-}c)$, the exact dispersion is captured remarkably well by the CF SW across all allowed angular momenta $L$, in sharp contrast to the density-wave ansatz for the SW. Notably, in the $\nu{=}1/3$ Laughlin state, the density ansatz for the SW lies very close to the CF SW at $L{=}1$, suggesting that the two are identical, as expected; however, this is not the case in the $\nu{=}2/5$ and $3/7$ Jain states. Consistent with previous studies~\cite{Mandal01, Majumder14}, we find that the CF SW dispersion for the $2/5$ and $3/7$ Jain states [see red filled squares in Fig.~\ref{fig: SF_dispersion_primary_Jain_states_1_3_2_5_3_7}$(b)$ and Fig.~\ref{fig: SF_dispersion_primary_Jain_states_1_3_2_5_3_7}$(c)$] exhibits a negative energy roton minimum at small angular momenta. This feature is absent in the dispersion of the density mode for the SW, which increases monotonically. The CF SW dispersion for comparatively bigger systems, presented in Figs.~\ref{fig: SF_dispersion_primary_Jain_states_1_3_2_5_3_7}$(e{-}f)$, further corroborates the presence of a roton minimum at small wavenumbers in the $2/5$ and $3/7$ Jain states. The negative energy roton minima indicate that, for the LLL Coulomb interaction, the fully polarized Jain states at $2/5$ and $3/7$ [in general, at $\nu{=}n/(2n{+}1)$ for $n{>}1$] become unstable toward a partially polarized or unpolarized state as the Zeeman energy is lowered. In contrast, the fully polarized state at $\nu{=}1/3$ remains stable down to zero Zeeman energy, as evidenced by the absence of a roton minimum in its CF SW dispersion [see Fig.~\ref{fig: SF_dispersion_primary_Jain_states_1_3_2_5_3_7}$(c)$ and Fig.~\ref{fig: SF_dispersion_primary_Jain_states_1_3_2_5_3_7}$(d)$]. Interestingly, resonant inelastic light scattering (RILS) measurements have observed a gapless SW mode at long wavelengths, which has a gap exactly equal to the Zeeman energy, as mandated by Larmor's theorem, in primary Jain states~\cite{Pinczuk93}. Furthermore, the negative energy roton minima in non-Laughlin primary Jain states manifest as intensity peaks at energies lower than the Zeeman energy in RILS measurements~\cite{Kang00, Dujovne03b, Dujovne05, Gallais06, Gallais06a, Wurstbauer11}. 

In Figs.~\ref{fig: SF_dispersion_primary_Jain_states_1_3_2_5_3_7}$(e{-}f)$, we also show the dispersion of the density-mode for the SW obtained in the planar geometry, using Eq.~\eqref{eq: planar_general_SF_gap_equation}, to facilitate its comparison with our results on the sphere. At small wave numbers, the planar gap closely matches the spherical one. However, as the wave number increases, the curvature effects on the finite sphere become important. Moreover, the mapping $q\ell{=}\sqrt{L(L{+}1)}/\sqrt{Q}$ also becomes less accurate as $L$ increases [see App.~\ref{app: L_to_q}]. This results in a noticeable difference between the finite spherical and thermodynamic planar gaps for the SW density-mode as $q\ell$ increases, as is also seen for the $\nu{=}1$ IQH state [see Fig.~\ref{fig: SF_dispersion_IQH_state_N_51}]. 

For comparison, we also show the Nakajima-Aoki mean-field CF SW gap at $\nu{=}1/3$ [see Eq.~\eqref{eq: SF_gap_NAkajima_Aoki_Laughlin_state}] in Figs.~\ref{fig: SF_dispersion_primary_Jain_states_1_3_2_5_3_7}$(a)$ and~\ref{fig: SF_dispersion_primary_Jain_states_1_3_2_5_3_7}$(d)$. As evident from the figures, Nakajima-Aoki's mean-field CF SW dispersion [deep-red triangles] provides a better description than the density-wave ansatz for the SW [blue filled circles], remaining valid for up to larger wave numbers; however, eventually at large enough wave numbers, it too deviates from the CF SW gap [see Fig.~\ref{fig: SF_dispersion_primary_Jain_states_1_3_2_5_3_7}$(d)$]. 

\begin{figure}[tbh!]
\centering
\begin{tabular}{cc}
        \includegraphics[width=0.499\columnwidth]{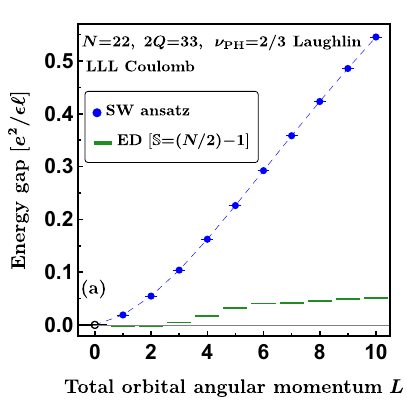}&
        \includegraphics[width=0.499\columnwidth]{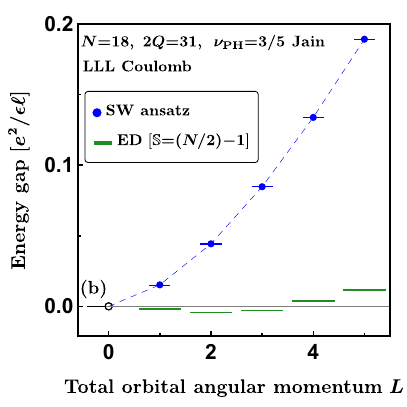}\\
         \includegraphics[width=0.499\columnwidth]{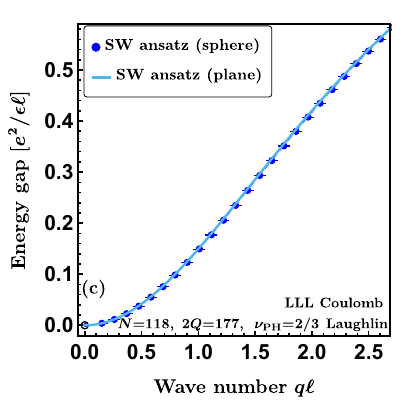}&
         \includegraphics[width=0.499\columnwidth]{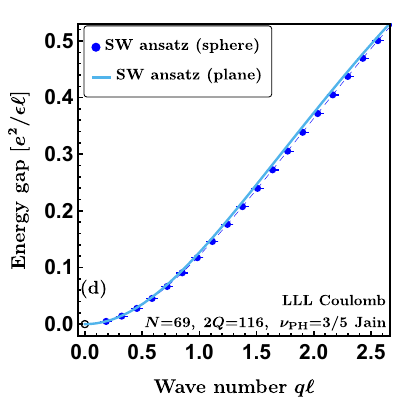}
         \end{tabular}
          \caption{Spin-wave dispersion in the $\nu{=}2/3$ [left panels] and $3/5$ [right panels] fractional quantum Hall states modeled as particle-hole conjugates of the $1/3$ Laughlin and $2/5$ Jain states, respectively. Panels $(a)$ and $(b)$ compare the spin-wave dispersion obtained using the gap equation with the corresponding exact diagonalization results for small systems on the sphere. Panels $(c)$ and $(d)$ show a comparison between the spin-wave dispersion computed on the sphere for a larger system and that obtained on the planar geometry.}
          \label{fig: SF_dispersion_primary_Jain_states_2_3_and_3_5}
        \end{figure} 
        
\subsection{Results: Spin-wave in reverse-vortex attached primary Jain states}
\label{ssec: spin_wave_dispersion_reverse_vortex_attached_Jain_states}
In this section, we present the LLL Coulomb SW dispersion for reverse-vortex attached fully polarized states at $\nu{=}2/3$ and $3/5$, which belong to the $\nu{=}n/(2n{-}1)$ Jain sequence [see Eq.~\eqref{eq: CF_GS_WF_negative_magnetic_field}]. In the presence of an active spin degree of freedom, the states at $\nu{=}2/3$ and $3/5$ are not related to the states at $\nu{=}1/3$ and $2/5$, respectively, by a particle-hole (PH) transformation. In general, when spin is in play, PH transformation maps $\nu{=}n/(2n{-}1)$ to $\nu{=}2{-}n/(2n{-}1)$, in contrast to the spin-frozen case, where states at $\nu{=}n/(2n{-}1)$ and $(n{-}1)/[2(n{-}1){+}1]$ are related by a PH transformation. Thus, although the fully polarized ground states at $2/3$ and $3/5$ are related to those at $1/3$ and $2/5$, respectively, by a PH-transformation, their SW dispersions are expected to differ. Indeed, as evident from the ED results shown in Fig.~\ref{fig: SF_dispersion_primary_Jain_states_2_3_and_3_5}$(a)$, the SW dispersion at $\nu{=}2/3$ has a negative energy roton, which is absent at $\nu{=}1/3$. This negative energy spin-flip roton at $\nu{=}2/3$ can be understood using CF theory as arising from the CF-cyclotron energy lowering transition of a CF from the occupied $1_{\uparrow}$-$\Lambda$L to the empty $0_{\downarrow}$-$\Lambda$L. In contrast, at $\nu{=}1/3$, the spin-flip inter-$\Lambda$L transitions either increase or leave constant the CF cyclotron energy; hence, no negative energy spin-flip rotons exist for it. Similarly, at $\nu{=}3/5$, the CF cyclotron energy lowering transitions---such as $2_{\uparrow}{\rightarrow}0_{\downarrow}$, $2_{\uparrow}{\rightarrow}1_{\downarrow}$ and $1_{\uparrow}{\rightarrow}0_{\downarrow}$---manifest as negative energy rotons in the spin-flip dispersion seen in ED results presented in Fig.~\ref{fig: SF_dispersion_primary_Jain_states_2_3_and_3_5}$(b)$. These results, along with the $\Lambda$L occupancy of CFs in the ground state, suggest that the spin-polarized state at $\nu{=}2/3$ can undergo a transition to an unpolarized state as the Zeeman energy is lowered. Similarly, the spin-polarized $3/5$ Jain state can transition to a partially polarized state as the Zeeman energy decreases~\cite{Kukushkin99}.

For comparison with ED results, we have also presented the dispersion of the density-wave ansatz for the SW [see blue filled circles] in Figs.~\ref{fig: SF_dispersion_primary_Jain_states_2_3_and_3_5}$(a{-}b)$ for small systems. As is evident from the figures, the density-wave ansatz for the SW fails to provide an accurate description of the SW dispersion at any angular momenta. 

In Figs.~\ref{fig: SF_dispersion_primary_Jain_states_2_3_and_3_5}$(c{-}d)$, we show the dispersion of the density-wave ansatz for the SW computed for larger systems on the sphere and compare it with that obtained in the planar geometry. The computation of the density-wave ansatz for the SW gap on both the sphere and plane requires the input of the projected static structure factor of only the fully polarized ground state, as is evident from Eqs.~\eqref{eq: sphere_SF_gap_equation} and~\eqref{eq: planar_general_SF_gap_equation}, respectively. This allows us to leverage some results of the fully polarized ground state. In particular, the projected structure factor $\bar{S}^{I}_{\nu}$ of the reverse-vortex attached fully polarized Jain state at $\nu$ can be obtained, to an excellent approximation~\cite{Balram21b}, from the projected structure factor $\bar{S}^{I}_{1{-}\nu}$ of the fully polarized PH conjugate Jain state at $1{-}\nu$. In the planar geometry, $\bar{S}^{I}_{\nu}$ and  $\bar{S}^{I}_{1{-}\nu}$ are related as~\cite{Balram15b, Nguyen17}:
\begin{align}
\label{eq: relation_S(q)_between_PH_conjuagate_states_plane}
\nu\bar{S}^{I}_{\nu}\left(q\right)&=\left(1-\nu\right)\bar{S}^{I}_{1{-}\nu}\left(q\right).
\end{align}
An analogous relation on the sphere can be obtained by noting that on a finite sphere $\nu$ is given by $\nu{=}N/(2Q{+}1)$, where $N$ is the number of particles in the state at filling $\nu$. Substituting $\nu{=}N/(2Q{+}1)$ in the above Eq.~\eqref{eq: relation_S(q)_between_PH_conjuagate_states_plane} yields:
\begin{align}
\label{eq: relation_S(q)_between_PH_conjuagate_states_sphere}
    N\bar{S}^{I}_{\nu}\left(L\right)&=N_{1-\nu}~\bar{S}^{I}_{1{-}\nu}\left(L\right)~~\forall ~L>0,
\end{align}
where $N_{1{-}\nu}{=}2Q{+1}{-}N$ is the number of particles in the state at filling $1{-}\nu$ (or equivalently, the number of holes at $\nu$). Here, we define $\bar{S}^{I}_{\nu}\left(L{=}0\right){=}N$ and $\bar{S}^{I}_{1{-}\nu}\left(L{=}0\right){=}N_{1{-}\nu}$. The structure factor used in obtaining the planar and spherical dispersions of the density-wave ansatz for the SW at $2/3$ and $3/5$, shown in Fig.~\ref{fig: SF_dispersion_primary_Jain_states_2_3_and_3_5}, were obtained using the structure factors of the hole-conjugate $1/3$ Laughlin and $2/5$ Jain states, respectively, using Eqs.~\eqref{eq: relation_S(q)_between_PH_conjuagate_states_plane} and \eqref{eq: relation_S(q)_between_PH_conjuagate_states_sphere}.

\section{Excitation gaps in partially polarized and unpolarized states}
\label{sec: Spin_gap_unpolarized__partially_polarized_states}
This section extends our discussion of the excitation gaps to non-fully polarized states, including singlet states with $\mathbb{S}{=}0$, partially polarized (PP) states with $0{<}\mathbb{S}{<}N/2$, and fully polarized states with $\mathbb{S}{=}N/2$ but $|\mathbb{S}^{z}|{<}N/2$, which we refer to as FP-$N/2$-multiplet since these states are part of the $\mathbb{S}^{z}$ multiplet of the fully polarized $\mathbb{S}{=}N/2$ state. Here, we will consider both spin-flip and spin-conserving excitation gaps. For the fully polarized states, the spin-conserving gaps were already considered in our previous work~\cite{Dora24}. 

\subsection{Spin-flip gap from density wave ansatz}
\label{ssec: Spin_gap_from_density_wave_ansatz_wave_functions}
As mentioned previously, various ansatz states for spin excitations can be obtained by acting with spin-density operators---such as $\bar{\rho}_{L, M}^{~+}$, $\bar{\rho}_{L, M}^{~-}$, and $\bar{\rho}_{L, M}^{~z}$---on the ground state $\left|\Psi_{0}\right\rangle$. For a spin-singlet ground state $\left|\Psi_{0}^{\mathbb{S}{=}0}\right\rangle$, spin-density wave states $\bar{\rho}_{L, M}^{~+}\left|\Psi_{0}^{\mathbb{S}{=}0}\right\rangle$, $\bar{\rho}_{L, M}^{~z}\left|\Psi_{0}^{\mathbb{S}{=}0}\right\rangle$,  $\bar{\rho}_{L, M}^{~-}\left|\Psi_{0}^{\mathbb{S}{=}0}\right\rangle$, all possess $\mathbb{S}{=}1$ for $L{\geq}1$, but differ in their azimuthal spin-quantum numbers $\mathbb{S}^{z}{=}1,0,{-}1$, respectively. As mentioned previously, this follows from the singlet nature of the ground state and the fact that $\bar{\rho}_{L, M}^{~+}$, $\bar{\rho}_{L, M}^{~-}$, and $\bar{\rho}_{L, M}^{~z}$ are spin $\mathbb{S}{=}1$ operators with $\mathbb{S}^{z}{=}1,0,{-}1$, respectively. Since the corresponding density wave states are just different $\mathbb{S}^{z}$-multiplets of the same $\mathbb{\vec{S}}^{2}$-eigenstate, they have the same energy for a spin-rotation preserving interaction such as Coulomb. Thus, for a spin-singlet ground state, it suffices to just consider one of these excitations, and here we only compute the dispersion of the ADW mode corresponding to the state $\bar{\rho}_{L, M}^{~z}\left|\Psi_{0}^{\mathbb{S}{=}0}\right\rangle$. 

The action of the operators $\bar{\rho}_{L, M}^{~+}$, $\bar{\rho}_{L, M}^{~-}$, and $\bar{\rho}_{L, M}^{~z}$, for $L{>}0$, on a PP ground state $\left|\Psi_{0}^{\rm PP}\right\rangle$ generally leads to excitations that are not spin-eigenstates. Interestingly, for a PP state with $\mathbb{S}{=}S_{0}{<}N/2$ and $\mathbb{S}^{z}{=}S_{0}$, $\bar{\rho}_{L, M}^{~+}$ generates a density-wave state with definite spin $S_{0}{+}1$ [see App.~\ref{ssec: spin_density_wave_PP}]. In this section, we present the dispersion of the ADW mode for the PP ground states. The interaction gap of states $\bar{\rho}_{L, M}^{~\alpha}\left|\Psi_{0}^{\rm PP}\right\rangle$, with $\alpha{=}z,{+},{-}$, generally differ from each other, as these states are not related to each other by spin raising and lowering operators---unlike in the case of a spin singlet ground state. We have not considered here the interaction gap of excitations created by $\bar{\rho}_{L, M}^{\pm}$, such as $\bar{\rho}_{L, M}^{\pm}\left|\Psi_{0}^{\rm PP}\right\rangle$, as they require evaluating correlation functions like $\left\langle\bar{\rho}_{L,-M}^{+}~\bar{\rho}_{L, M}^{-}\right\rangle_{\rm PP}$, which are challenging to compute for large system sizes. 

Except for at $\nu{=}1$, the spin-density wave excitations in the FP-$N/2$ multiplet, obtained from the action of $\bar{\rho}_{L, M}^{~\alpha}$ on a fully polarized ground state with $|\mathbb{S}^{z}|{<}N/2$, generically, do not possess definite spin quantum numbers for all $L$. The various ground states in the FP-$N/2$ multiplet can be constructed from the successive application of $\mathbb{S}^{-}$ on the fully polarized state with $\mathbb{S}^{z}{=}N/2$, i.e., $\left|\Psi_{\nu}^{N/2, N/2-\mathfrak{n}}\right\rangle{=}\left(\mathbb{S}^{-}\right)^{\mathfrak{n}}\left|\Psi_{\nu}^{N/2, N/2}\right\rangle$, where $\mathfrak{n}{=}1,2,{\cdots},N$. At $\nu{=}1$, the application of $\bar{\rho}_{L, M}^{~\alpha}$ with $\alpha{=}z,{+},{-}$ on these ground states result in spin excitations with $\mathbb{S}{=}N/2{-}1$ for $L{\geq}1$, except for the states $\bar{\rho}_{L, M}^{~+}\mathbb{S}^{-}\left|\Psi_{\nu{=}1}^{N/2, N/2}\right\rangle$ and $\bar{\rho}_{L, M}^{~+}\left|\Psi_{\nu{=}1}^{N/2, N/2}\right\rangle$, which are annihilated. This is because [see App.~\ref{ssec: spin_states_in_the_maximal_spin_multiplet}]
\begin{align}
 \label{eq: annihilated_states} 
 \bar{\rho}_{L, M}^{~+}\mathbb{S}^{-}\left|\Psi_{\nu{=}1}^{N/2, N/2}\right\rangle&=\left[\mathbb{S}^{-}\bar{\rho}_{L, M}^{~+}{+}\bar{\rho}_{L, M}^{~z}\right]\left|\Psi_{\nu{=}1}^{N/2, N/2}\right\rangle,
\end{align}
and, both $\bar{\rho}_{L, M}^{+}$ and $\bar{\rho}_{L{\geq}1,M}^{z}$ annihilate $\left|\Psi_{\nu{=}1}^{N/2, N/2}\right\rangle$. Except for the cases where $\alpha{=}+$ with $\mathfrak{n}{=}0$ or $\mathfrak{n}{=}1$, all spin excitation states generated by $\bar{\rho}_{L, M}^{~\alpha}$ differ only in their $\mathbb{S}^{z}$ quantum number. Consequently, for a $SU(2)$ invariant interaction, they all share the same interaction gap. For example, spin excitations obtained from $\bar{\rho}_{L, M}^{z}$, $\bar{\rho}_{L, M}^{+}$, and $\bar{\rho}_{L, M}^{-}$ in $\left|\Psi_{\nu{=}1}^{N/2,0}\right\rangle$ or equivalently, the Halperin-$(1,1,1)$ state [see Eq.~\eqref{eq: Halperin_mmn_WF}]---the $\mathbb{S}^{z}{=}0$ version of the $\nu{=}1$ fully polarized state $\left|\Psi_{\nu{=}1}^{N/2, N/2}\right\rangle$---all exhibit the same gap as in $\left|\Psi_{\nu{=}1}^{N/2, N/2}\right\rangle$, at each $L$. 

Moving to FQH states, the spin-density wave excitations, $\bar{\rho}_{L{\geq1}, M}^{\alpha}\left|\Psi_{\nu}^{N/2, N/2-\mathfrak{n}}\right\rangle$ carry a definite spin only at $L{=}1$ [see App.~\ref{ssec: spin_states_in_the_maximal_spin_multiplet}]. Consequently, the dispersion of $\bar{\rho}_{L{>}1,M}^{\alpha}\left|\Psi_{\nu}^{N/2, N/2-\mathfrak{n}}\right\rangle$ built atop an $\mathfrak{n}$ spin-flipped FQH ground state $\left|\Psi_{\nu}^{N/2, N/2-\mathfrak{n}}\right\rangle$ differs for each $\alpha{=}z,+,-$, and for each $\mathfrak{n}$. However, at $L{=}1$, the interaction gap of all of these modes are identical, as they differ only in their $\mathbb{S}^{z}$ quantum number---similar to the case at $\nu{=}1$ [Note that for $\alpha{=}+$ with $\mathfrak{n}{=}0$ or $\mathfrak{n}{=}1$, the states $\bar{\rho}_{L{=}1, M}^{\alpha}\left|\Psi_{\nu}^{N/2, N/2-\mathfrak{n}}\right\rangle$ are annihilated, for the same reason as in the $\nu{=}1$ case, see Eq.~\eqref{eq: annihilated_states}, with a modification that here only $\bar{\rho}_{1, M}^{z}\left|\Psi_{\nu}^{N/2, N/2}\right\rangle{=}0$ ($\bar{\rho}_{1, M}^{+}\left|\Psi_{\nu}^{N/2, N/2}\right\rangle{=}0$ simply from the fully polarized nature of $\left|\Psi_{\nu}^{N/2, N/2}\right\rangle$)]. As an example, the fully polarized $1/3$ Laughlin state and its $\mathbb{S}^{z}{=}0$ counterpart, the Halperin-$(3,3,3)$ state, exhibit different spin-density wave dispersions for $\bar{\rho}_{L, M}^{\alpha}$ with $\alpha{=}z,+,-$; however, their gaps at $L{=}1$ are identical. In App.~\ref{app: spin_quantum_number_spin_density_states}, we provide a detailed derivation to infer the spin quantum numbers of different spin-density wave states.  

Next, we compute the dispersion of the ADW mode. This is the energy required to excite the state $\bar{\rho}_{L, M}^{~z}\left|\Psi_{\nu}\right\rangle$ relative to the ground state $\left|\Psi_{\nu}\right\rangle$, as a function of $L$. In other words, 
\begin{align}
    \Delta^{\rm ADW}\left(L\right)&=\frac{\left\langle\Psi_{\nu}\right|\left[\bar{\rho}^{~z}_{L, M}\right]^{\dagger}\bar{H}\bar{\rho}^{~z}_{L, M}\left| \Psi_{\nu}\right\rangle}{\left\langle\Psi_{\nu}\right|\left[\bar{\rho}^{~z}_{L, M}\right]^{\dagger}\bar{\rho}^{~z}_{L, M}\left| \Psi_{\nu}\right\rangle} ~-~\left\langle\Psi_{\nu}\right|\bar{H}\left| \Psi_{\nu}\right\rangle.
\end{align}
The gap $\Delta^{\rm ADW}\left(L\right)$ can be determined from the ground state two-point correlation functions, which can be made evident by expressing the above equation in terms of a double commutator. In doing so, we assume that $\bar{H}\left| \Psi_{\nu}\right\rangle{=}E_{0}\left| \Psi_{\nu}\right\rangle$, i.e., $\left| \Psi_{\nu}\right\rangle$ is the exact ground state of $\bar{H}$ with energy $E_{0}$. [Even if $\left| \Psi_{\nu}\right\rangle$ is not an exact ground state of $\bar{H}$, the error due to this assumption remains small, provided that $\left| \Psi_{\nu}\right\rangle$ is a good variational state~\cite{Dora24}.] Consequently, one obtains 
\begin{align}
\label{eq: AS_density_wave_double_commutator}
    \Delta^{\rm ADW}\left(L\right)&=\frac{1}{2}\frac{\left\langle\Psi_{\nu}\right|\left[\left[\bar{\rho}^{~z}_{L, M}\right]^{\dagger},\left[\bar{H},\bar{\rho}^{~z}_{L, M}\right]\right]\left| \Psi_{\nu}\right\rangle}{\left\langle\Psi_{\nu}\right|\left[\bar{\rho}^{~z}_{L, M}\right]^{\dagger}\bar{\rho}^{~z}_{L, M}\left| \Psi_{\nu}\right\rangle}.
\end{align}
In obtaining the above equation, we have used the identities $\left[\bar{\rho}^{z}_{L, M}\right]^{\dagger}{=}(-1)^{M}\bar{\rho}^{z}_{L, {-}M}$, and $\left\langle\Psi_{\nu}\right|\left[\bar{\rho}^{z}_{L, {-}M}\right]^{\dagger} \bar{H}\bar{\rho}^{z}_{L, {-}M}\left| \Psi_{\nu}\right\rangle{=}\left\langle\Psi_{\nu}\right|\left[\bar{\rho}^{z}_{L, M}\right]^{\dagger} \bar{H}\bar{\rho}^{z}_{L, M}\left| \Psi_{\nu}\right\rangle$, which follows from the rotational invariance of $\bar{H}$ in real space. Next, we define the projected total $\mathbb{S}^{z}$-structure factor $\bar{S}^{z}\left(L\right)$ as
\begin{align}
    \bar{S}^{z}\left(L\right)&=\frac{4\pi}{N}\left\langle\Psi_{\nu}\right|\left[\bar{\rho}^{~z}_{L, M}\right]^{\dagger}\bar{\rho}^{~z}_{L, M}\left| \Psi_{\nu}\right\rangle. 
\end{align}
Similar to Eq.~\eqref{eq: relaton_unprojected_projected_structure_factor_fully_polarized_states}, $\bar{S}^{z}\left(L\right)$ is related to its unprojected counterpart $S^{z}\left(L\right)$ as
\begin{align}
\label{eq: relaton_total_Sz_unprojected_projected_structure_factor}
 \bar{S}^{z}\left(L\right)&= S^{z}\left(L\right)-1+\frac{4\pi}{(2L+1)(2Q+1)}\left[\mathcal{F}\left(L\right)\right]^{2}.
\end{align}
Without loss of generality, we fix $M{=}0$, as $\Delta^{\rm ADW}\left(L\right)$ does not depend on $M$ due to the rotation invariance of $\bar{H}$. The commutators in $\Delta^{\rm ADW}\left(L\right)$ [see Eq.~\eqref{eq: AS_density_wave_double_commutator}] can be readily evaluated from Eq.~\eqref{eq: spinful_operator_commutation_algebra}, allowing it to be expressed in terms of ground state correlation functions as
    \begin{align}
\label{eq: AS_density_wave_gap_equation}
  \Delta^{\rm ADW}\left(L\right)&=\frac{4\pi}{2\bar{S}^{z}(L)}\sum_{\tilde{L}=0}^{2Q}\frac{v_{\tilde{L}}}{2}\sum_{\tilde{M}=0}^{\tilde{L}}~\sum_{\lambda{=}\left|\tilde{L}-L\right|}^{\left|\tilde{L}+L\right|}\bigg[4\left(\alpha_{\lambda}^{(\tilde{L},L,\tilde{M},0)}\right)^{2}\nonumber\\
  &~~~~~~\times\bar{S}^{z}(\lambda)+ 4\alpha_{\lambda}^{(\tilde{L},L,\tilde{M},0)}~\alpha_{\tilde{L}}^{(L,\lambda,0,\tilde{M})}~\bar{S}^{I}(\tilde{L})\bigg].
\end{align}
We conclude this section by discussing the dispersion of the SDW mode obtained from acting $\bar{\rho}_{L, M}^{~I}$ on the ground state. The operator $\bar{\rho}_{L, M}^{~I}$ acts as the identity operator in the spin space. Therefore, the state $\bar{\rho}_{L, M}^{~I}\left|\Psi_{\nu}\right\rangle$ has the same spin as that of the ground state $\left|\Psi_{\nu}\right\rangle$, and thus results in a spin-conserving density-wave excitation, the analog of the GMP mode~\cite{Girvin85, Girvin86} for spinful states. The dispersion of the SDW mode can be obtained in a similar way to the ADW mode, and is given by
    \begin{align}
\label{eq: SS_density_wave_gap_equation}
  \Delta^{\rm SDW}\left(L\right)&=\frac{4\pi}{2\bar{S}^{I}(L)}\sum_{\tilde{L}=0}^{2Q}\frac{v_{\tilde{L}}}{2}\sum_{\tilde{M}=0}^{\tilde{L}}~\sum_{\lambda{=}\left|\tilde{L}-L\right|}^{\left|\tilde{L}+L\right|}\bigg[4\left(\alpha_{\lambda}^{(\tilde{L},L,\tilde{M},0)}\right)^{2}\nonumber\\
  &~~~~~~\times\bar{S}^{I}(\lambda)+ 4\alpha_{\lambda}^{(\tilde{L},L,\tilde{M},0)}~\alpha_{\tilde{L}}^{(L,\lambda,0,\tilde{M})}~\bar{S}^{I}(\tilde{L})\bigg].
\end{align}
\begin{figure*}[tbh!]
        \includegraphics[width=0.97\textwidth]{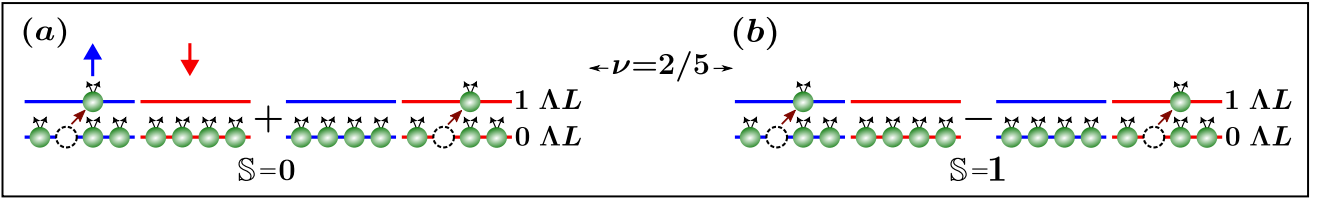}
          \caption{Schematic of the $(a)$ symmetric and $(b)$ antisymmetric CF-basis states deployed in the computation of the spin gap in the $2/5$ spin-singlet Jain state.}
          \label{fig: symmetric_antisymmetric_CF_spin_gap_schematic}
        \end{figure*} 

\begin{figure*}[tbh!]
        \includegraphics[width=0.97\textwidth]{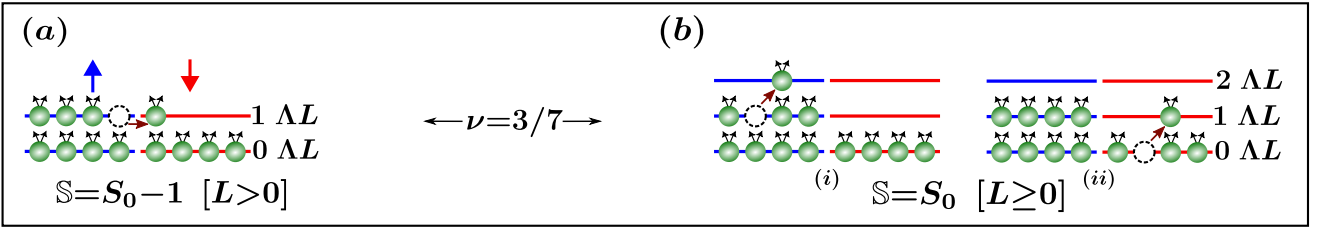}
          \caption{Schematic of $(a)$ the spin-flip CFE state and $(b)$ the spin-conserving CFE basis states used to compute the spin-flip and spin-conserving gaps in the Jain $3/7$ partially polarized state. The spin-flip basis states have spin $\mathbb{S}{=}S_{0}-1$ for $L{>}0$, whereas the spin-conserving CFE basis states have spin $\mathbb{S}{=}S_{0}$ for all allowed $L$, where $S_{0}$ is the spin of the ground state. }
          \label{fig: CFE_basis_states_3_7_PP_Jain_state}
        \end{figure*}         
    
\subsection{Excitation gaps from composite fermion excitons}
\label{ssec: CFE_gaps}
In this section, we employ the CF theory to compute the spin-flip and spin-conserving excitation gaps for spin-singlet and PP FQH states. In the subsequent sections, we compare this result with that obtained from the density-wave ansatzes presented earlier. 

In the CF theory, singlet states are formed when equal numbers of $n_{\uparrow}$ and $n_{\downarrow}$ $\Lambda$Ls are filled, i.e., $n_{\uparrow}{=}n_{\downarrow}$. The lowest energy spin excitations are obtained from CF basis states where the CF hole and CF particle of the constituent CFE occupy the topmost filled and the bottommost empty $\Lambda$Ls, respectively. Specifically, at each $L$, the CF basis states are constructed from: (i) $\mathbb{S}^{z}$ conserving, i.e., $\Delta\mathbb{S}^z{=}0$, CFEs corresponding to $\left(n_{\uparrow}{-}1\right){\rightarrow}n_{\uparrow}$ and $\left(n_{\downarrow}{-}1\right){\rightarrow}n_{\downarrow}$; (ii) spin-flip CFEs with $\Delta\mathbb{S}^{z}{=}{+}1$ from  $\left(n_{\downarrow}{-}1\right){\rightarrow}n_{\uparrow}$; and (iii) spin-flip CFEs with $\Delta\mathbb{S}^z{=}{-}1$ from $\left(n_{\uparrow}{-}1\right) {\rightarrow} n_{\downarrow}$. For a spin-rotation-invariant interaction, basis states differing in $\mathbb{S}^{z}$ do not mix; therefore, the basis states (i), (ii), and (iii) do not mix under a spin-preserving interaction. The CFE basis states (ii) and (iii), which describe the CF SW, belong to the $\mathbb{S}{=}1$ multiplet with $\mathbb{S}^{z}{=}{+}1$ and $\mathbb{S}^{z}{=}{-}1$, respectively, and therefore result in identical dispersions for a spin-invariant interaction. On the other hand, the interaction can couple the two CFE basis states in (i), i.e., the off-diagonal matrix elements of the interaction in the basis of $\left(n_{\uparrow}{-}1\right){\rightarrow}n_{\uparrow}$ and $\left(n_{\downarrow}{-}1\right){\rightarrow}n_{\downarrow}$ are non-zero. Nevertheless, at each $L$, the antisymmetric superposition of these basis states results in $\mathbb{S}{=}1$ while the symmetric superposition results in $\mathbb{S}{=}0$. Thus, we use this basis of antisymmetric and symmetric combinations to decouple the interaction. A schematic of these antisymmetric and symmetric CFE states is depicted in Fig.~\ref{fig: symmetric_antisymmetric_CF_spin_gap_schematic}, for the Halperin-$(3,3,2)$ states [A similar schematic figure, with CFs carrying a single vortex, can be envisioned for the bosonic Halperin-$\left(2,2,1\right)$ state.]. The antisymmetric CFE (A-CFE) mode shares the same dispersion as the above-discussed CF SW modes with $\mathbb{S}^{z}{=}1$ [(ii)] and $ \mathbb{S}^{z}{=}{-}1$ [(iii)] in the $\mathbb{S}{=}1$ sector. In other words, A-CFE mode corresponds to the $\mathbb{S}^{z}{=}0$ component of the $\mathbb{S}{=}1$ SW multiplet. Therefore, we compute only the dispersion of the A-CFE mode and compare it with the earlier discussed ADW dispersion. The symmetric CFE (S-CFE) mode describes the CF spin-conserving density wave excitation and will be compared with the SDW dispersion. 

In the CF theory, for a PP ground state with $n_{\uparrow}{>}n_{\downarrow}$, the SW is modeled by taking a basis of the spin-flipped CFEs corresponding to the transition
\begin{itemize}
    \item $\left(n_{\uparrow}{-}1\right) {\rightarrow} \left(n{-}n_{\downarrow}{-}1\right)$, $\left(n_{\uparrow}{-}1\right) {\rightarrow} \left(n{-}n_{\downarrow}{-}2\right)$,${\cdots}$, $\left(n_{\uparrow}{-}1\right) {\rightarrow} \left(n_{\downarrow}\right)$
    \item $\left(n_{\uparrow}{-}2\right) {\rightarrow} \left(n{-}n_{\downarrow}{-}2\right)$, $\left(n_{\uparrow}{-}2\right) {\rightarrow} \left(n{-}n_{\downarrow}{-}3\right)$,${\cdots}$, $\left(n_{\uparrow}{-}2\right) {\rightarrow} \left(n_{\downarrow}\right)$
    \item ${\vdots}$
    \item $\left(n{-}n_{\uparrow}[{=}n_{\downarrow}]\right) {\rightarrow} \left(n_{\downarrow}\right)$.
\end{itemize}
When $n_{\uparrow}{>}n_{\downarrow}{+}1$ (for example, the partially polarized state at $4/7$ and $4/9$), the SW has a roton, analogous to the spin-roton observed in the fully polarized states at $n/(2n{\pm}1)$ with $n{\geq}2$. 

Here, we will restrict ourselves to the partially polarized states at $3/7$ and $3/5$ for which $(n_{\uparrow}, n_{\downarrow}){=}(2,1)$ and thus, $n_{\uparrow}{=}n_{\downarrow}{+}1$. For these fractions, the CF SW is composed of only the spin-flip CFE $\left(n_{\uparrow}{-}1\right) {\rightarrow} \left(n_{\downarrow}\right)$ [see Fig.~\ref{fig: CFE_basis_states_3_7_PP_Jain_state}$(a)$]. For $L{=}0$, this state has the same  $\mathbb{S}$ as the ground state [with $\mathbb{S}^{z}$ one lower than that of the ground state] while for $L{\geq}1$ it has one spin lower than the $\mathbb{S}$ of the ground state, and thus forms the SW. Similar to the fully polarized states, owing to Larmor's theorem, the SW dispersion should be gapless in the long-wavelength limit for PP states too. The spin-flip CFE $\left(n_{\uparrow}{-}1\right) {\rightarrow} \left(n_{\downarrow}\right)$ has the same CF cyclotron energy as the ground state, and as we will see below, produces a gapless SW at $3/7$ and $3/5$, consistent with Larmor's theorem.

The spin-conserved CF exciton is obtained by performing the CFD within the subspace of CFE basis states $\left(n_{\uparrow}{-}1\right) {\rightarrow} n_{\uparrow}$ and $\left(n_{\downarrow}{-}1\right) {\rightarrow} n_{\downarrow}$ [note that both these states satisfy Fock's cyclic conditions~\cite{Hamermesh62}, and therefore, have the same $\mathbb{S}$ as the ground state since the spins cannot be raised further as the corresponding positions in the $\uparrow$-$\Lambda$L are occupied] and selecting the lowest-energy state. A schematic of the spin-conserving CFE basis states is illustrated in Fig.~\ref{fig: CFE_basis_states_3_7_PP_Jain_state}$(b)$.
        
\begin{figure*}[tbh!]
        \includegraphics[width=0.66\columnwidth]{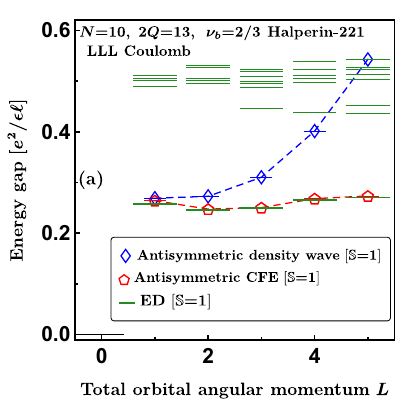}
        \includegraphics[width=0.66\columnwidth]{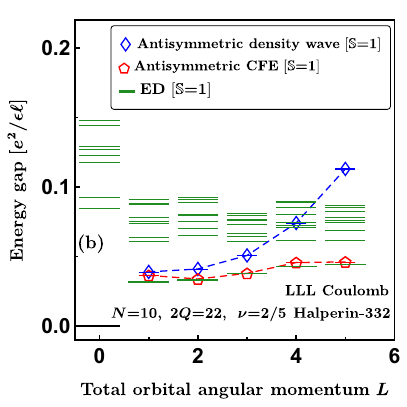}
        \includegraphics[width=0.66\columnwidth]{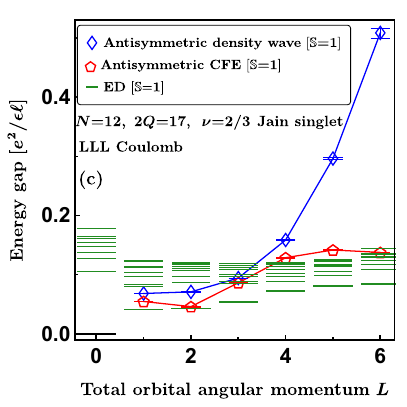}\\
         \includegraphics[width=0.66\columnwidth]{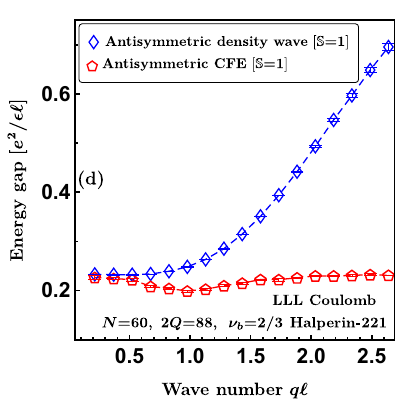}
        \includegraphics[width=0.66\columnwidth]{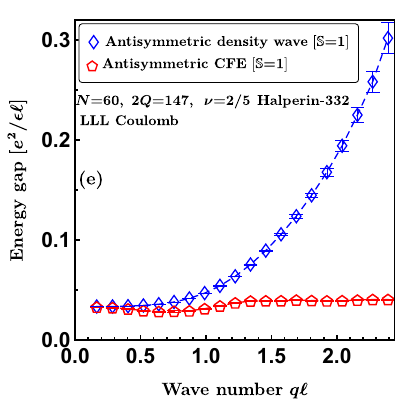}
         \includegraphics[width=0.66\columnwidth]{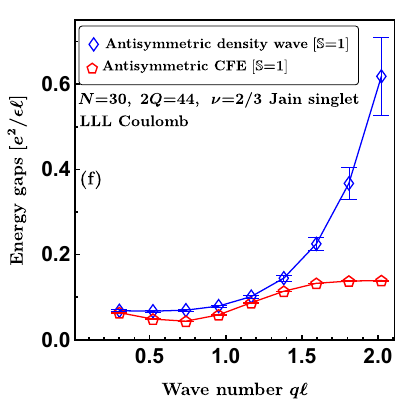}\\
     \caption{Spin-flip dispersion for the LLL Coulomb interaction in the bosonic Halperin-$\left(2,2,1\right)$, fermionic Halperin-$\left(332\right)$, and fermionic Jain $2/3$ spin-singlet states. Panels $\left(a\right)$, $\left(b\right)$, and $\left(c\right)$ present a comparison between the antisymmetric density wave and the antisymmetric CFE gaps with the corresponding ED spin excitation gaps. Panels $\left(e\right)$, $\left(f\right)$, and $\left(g\right)$ compare the antisymmetric density wave and CFE dispersions for large system sizes.}
          \label{fig: ADW_dispersion_Halperin_221_332_Jain_ss_2_3}
        \end{figure*}
\begin{figure}[tbh!]
\begin{tabular}{cc}
        \includegraphics[width=0.499\columnwidth]{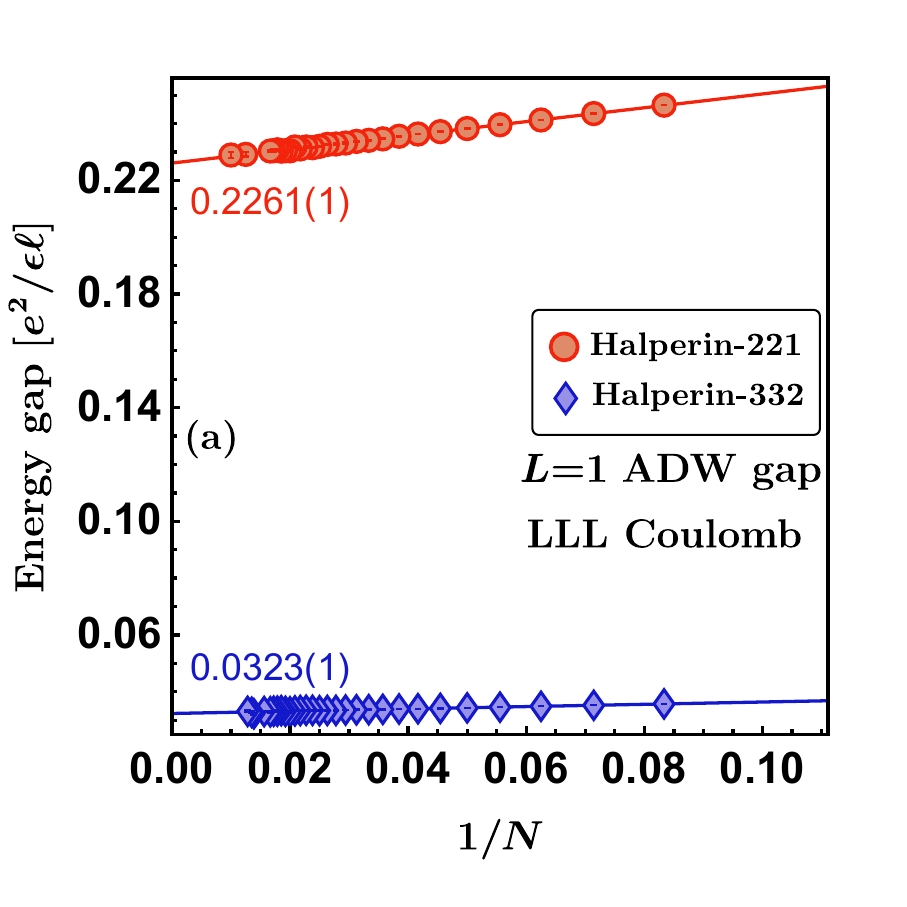}&
         \includegraphics[width=0.499\columnwidth]{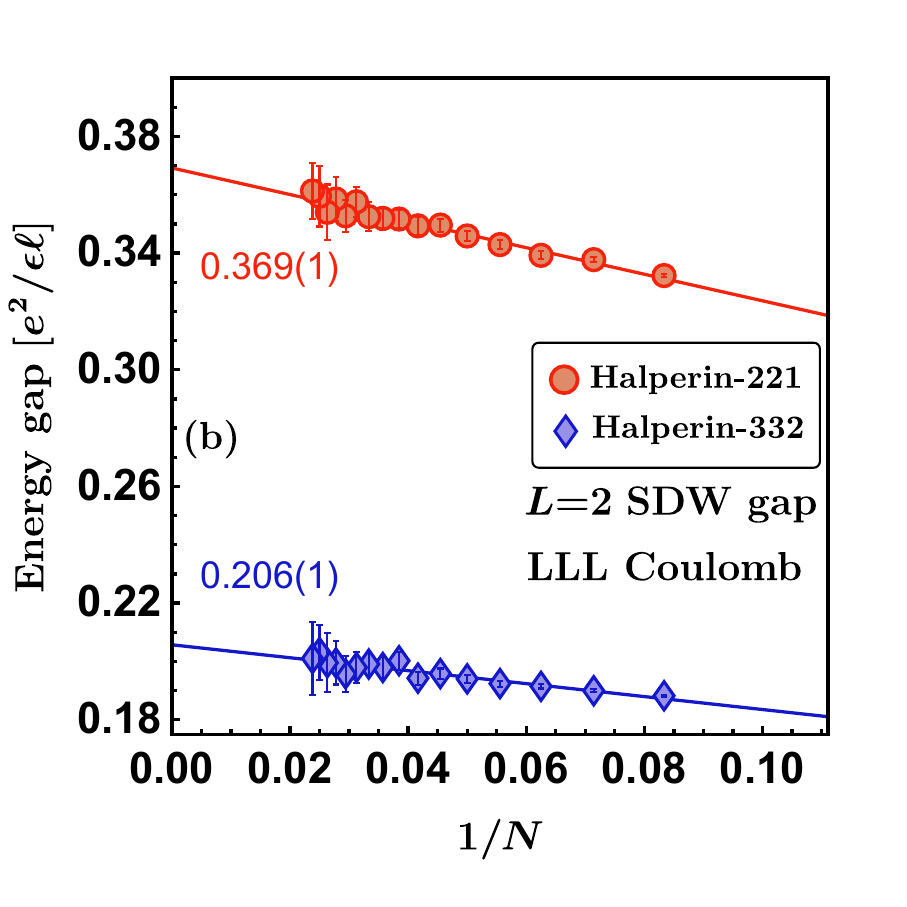}\\
         \includegraphics[width=0.499\columnwidth]{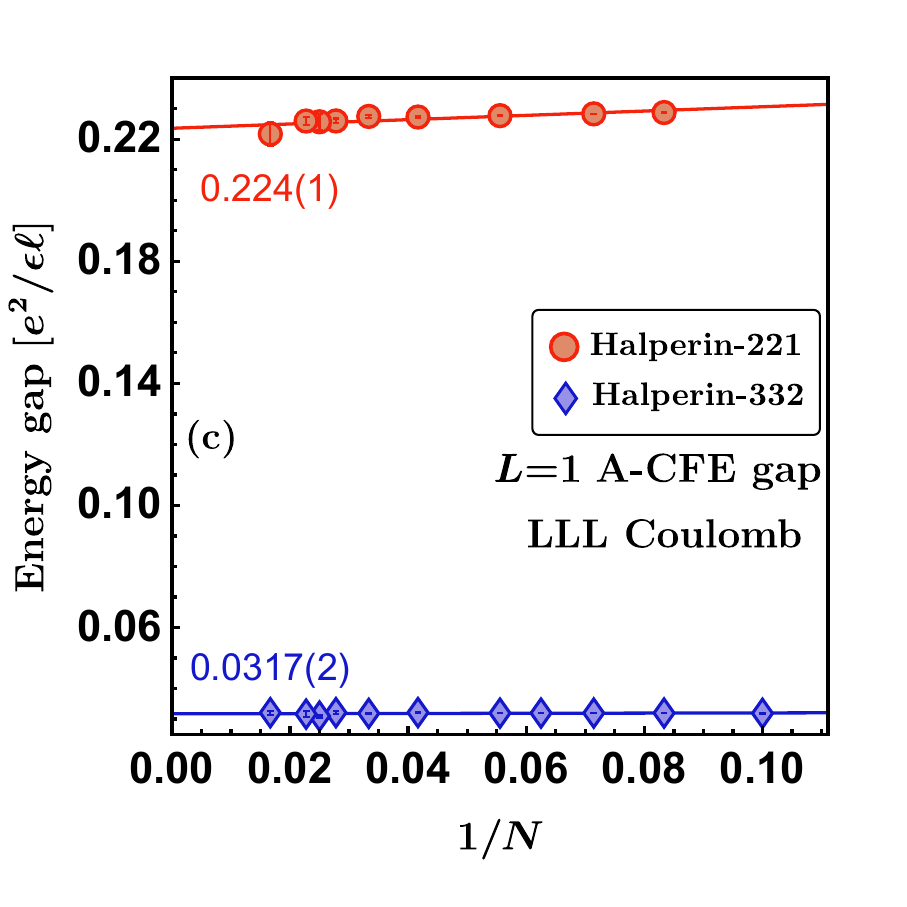}&
         \includegraphics[width=0.499\columnwidth]{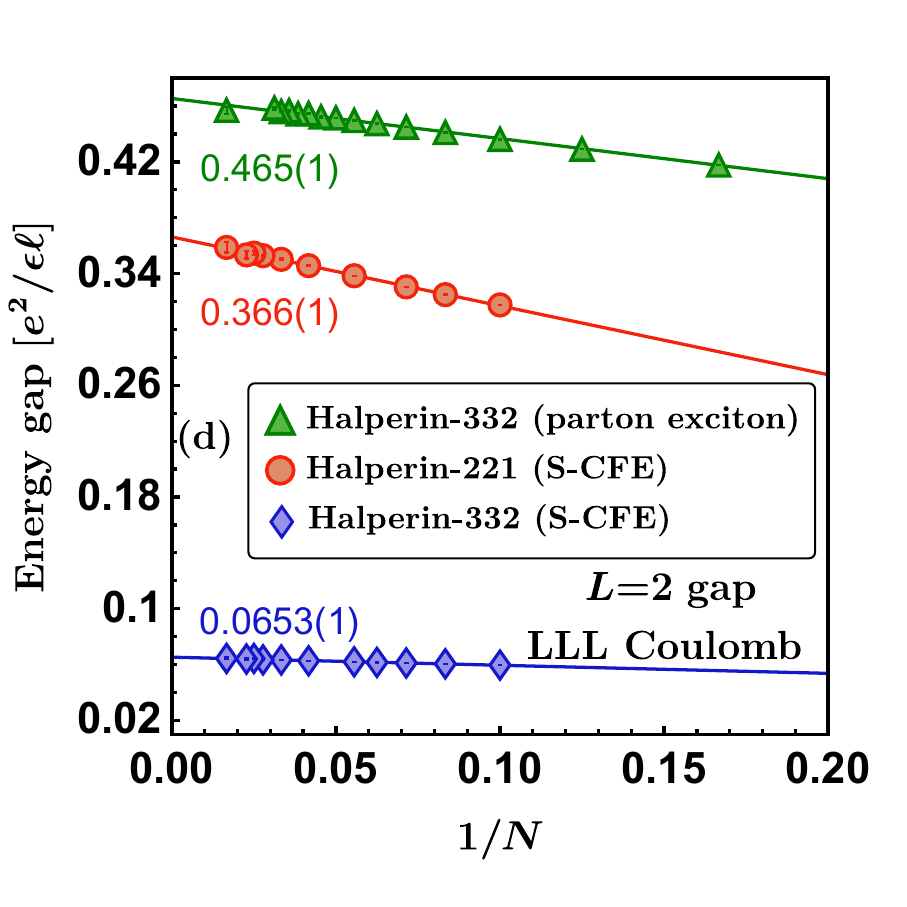}
         \end{tabular}
          \caption{Thermodynamic extrapolation of the density-corrected~\cite{Morf86b} Coulomb gaps of $(a)$ the $L{=}1$ antisymmetric density wave, $(b)$ the $L{=}2$ symmetric density wave (equivalently, $L{=}2$ Girvin-MacDonald-Platzman state), $(c)$ the $L{=}1$ antisymmetric composite fermion exciton (CFE) state, and $(d)$ $L{=}2$ symmetric composite fermion exciton state for the Halperin-$(2,2,1)$ and Halperin-$(3,3,2)$ states obtained from a linear fit in $1/N$ of finite-system results on the sphere. Panel $(d)$ shows the density-corrected~\cite{Morf86b} Coulomb gap of the $L{=}2$ state of the parton mode [see Eq.~\eqref{eq: parton_mode_Halperin_332}], equivalently the parton graviton, in the Halperin-$(3,3,2)$ state. The number in the parentheses of the extrapolated gap is the error in the intercept of the linear fit, and the error bar on each point is the statistical uncertainty stemming from its Monte Carlo evaluation. Some of the results for the Halperin-$(3,3,2)$ state are taken from Ref.~\cite{Liu20}.}
          \label{fig: thermodynamic_extrapolation_L_1_ADW_L_2_SDW_gaps}
        \end{figure}        
        
\subsection{Results: Spin-flip gaps in singlet and partially polarized states}
\label{ssec: Spin_flip_gaps_in_singlet_partially_polarized_states}

\subsubsection{Singlet states}
The SW dispersion of Coulomb-interacting particles in various singlet states, computed using the density wave ansatz (ADW mode), CF theory (A-CFE mode), and ED, is shown in Fig.~\ref{fig: ADW_dispersion_Halperin_221_332_Jain_ss_2_3}. Unlike in the fully polarized states, the SW is gapped in spin-singlet states. The low-lying branch $\mathbb{S}{=}1$ excitations, i.e., the SW mode, is well-separated from the other excitations. Remarkably, in the Halperin-$(2,2,1)$ and Halperin-$(3,3,2)$ states [Figs.~\ref{fig: ADW_dispersion_Halperin_221_332_Jain_ss_2_3}$(a)$ and~\ref{fig: ADW_dispersion_Halperin_221_332_Jain_ss_2_3}$(b)$], the exact dispersion is well-captured by the A-CFE mode, whereas the ADW mode is accurate only in the long-wavelength limit, i.e., at $L{=}1$. The corresponding dispersions for larger systems are shown in Figs.~\ref{fig: ADW_dispersion_Halperin_221_332_Jain_ss_2_3}$(d)$ and~\ref{fig: ADW_dispersion_Halperin_221_332_Jain_ss_2_3}$(e)$, where the long-wavelength gaps of the ADW and A-CFE mode are nearly identical. See Fig.~\ref{fig: thermodynamic_extrapolation_L_1_ADW_L_2_SDW_gaps}$(a)$ and~\ref{fig: thermodynamic_extrapolation_L_1_ADW_L_2_SDW_gaps}$(c)$ for the thermodynamic limit extrapolations of the $L{=}1$ ADW  and A-CFE gaps, respectively. As the density on a finite sphere differs from that in the thermodynamic limit, we first density-correct the finite size gaps by multiplying a factor of $\sqrt{2Q\nu/N}$, and then extrapolate the density-corrected gaps to the thermodynamic limit~\cite{Morf86b}. The nearly identical energies suggest that the A-CFE and ADW modes could be identical at $L{=}1$ for the Halperin-$(m,m,m{-}1)$ singlet states. It is useful to note that the $L{=}1$ ADW state in the  Halperin-$(m,m,m{-}1)$ states with $m{>}2$ can be written in terms of the ADW state in the Halperin-$(2,2,1)$ state as
\begin{align}
    \label{eq: L_1_ADW_Halperin}
    \bar{\rho}^{z}_{1,1}\Psi^{m,m,m{-}1}_{\nu{=}2/(2m{-}1)}&=(m-1)\Phi_{1}^{m-2}~\bar{\rho}^{z}_{1,1}\Psi^{2,2,1}_{\nu_{b}{=}2/3}.
\end{align}
Following Ref.~\cite{Kamilla96b}, it can be shown that, in the thermodynamic limit, the A-CFE and the ADW states in the Halperin-$(m,m,m{-}1)$ states with $m{\geq}2$ are identical at long wavelengths. In other words, provided that ADW state in the Halperin-$(m,m,m{-}1)$ state can be written as $\bar{\rho}^{z}_{q} \Psi^{m,m,m{-}1}_{\nu{=}2/(2m{-}1)}{=}\mathcal{P}_{\rm LLL}(\Phi_1)^{m{-}1}\rho_{q}^{z}\Psi_{\nu{=}2}^{1,1,0}$, and noticing that the planar unprojected spin-flip density operator $\rho_{q{\to}0}^{z}$ is equivalent to the A-CFE operator $\rho_{q}^{\rm A-CFE,0{\to}1}$, which creates the anti-symmetric CF exciton [the single-vortex analog of Fig.~\ref{fig: symmetric_antisymmetric_CF_spin_gap_schematic}(b), that is appropriate for bosons at $\nu_{b}{=}2/3$], it follows that the A-CFE and ADW states are identical as $q{\to}0$ for the Halperin states. 

In the Jain $2/3$ singlet state, the A-CFE mode does not accurately capture the exact dispersion, in particular, with significant deviations at larger angular momenta, as seen in Fig.~\ref{fig: ADW_dispersion_Halperin_221_332_Jain_ss_2_3}$(c)$. This mismatch can be attributed to the use of the JK projection method to obtain the LLL wave function of a spinful CF state. The JK projection allows opposite-spin electrons to come arbitrarily close, thereby increasing the energy of the CF state. This discrepancy can be resolved by performing the ``hard-core" projection (HCP)~\cite{Wu93}, wherein a Jastrow factor is kept outside the LLL-projection operator, which ensures that even opposite-spin electrons are excluded from occupying the same state. However, unlike the JK projection, the HCP can be implemented only for small system sizes, which limits its applicability for obtaining results in the thermodynamic limit. To this end, we note that, despite being computed using the JK projection scheme, the A-CFE mode in the Halperin-$(2,2,1)$ and Halperin-$(3,3,2)$ states captures the exact dispersion surprisingly well. This issue has been pointed out in the literature~\cite{Balram15a}, where the JK projection is found to yield better results for parallel vortex attached non–fully-polarized states at $\nu{=}n/(2n{+}1)$ than their reverse vortex attached counterparts at $\nu{=}n/(2n{-}1)$. Similar to the Halperin states, the energies of the ADW and A-CFE modes in the Jain $2/3$ singlet state are nearly equal in the long-wavelength limit, as seen from the Figs.~\ref{fig: ADW_dispersion_Halperin_221_332_Jain_ss_2_3}$(c)$ and~\ref{fig: ADW_dispersion_Halperin_221_332_Jain_ss_2_3}$(f)$.
\begin{figure}[tbh!]
\centering
\begin{tabular}{cc}
        \includegraphics[width=0.499\columnwidth]{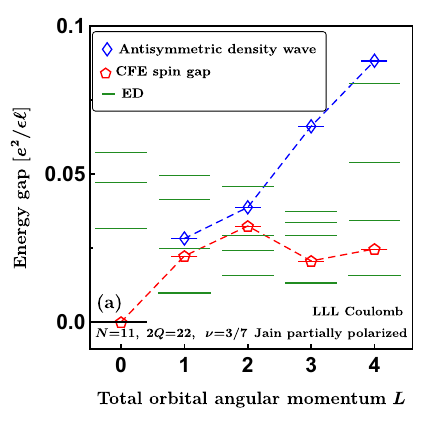}&
        \includegraphics[width=0.499\columnwidth]{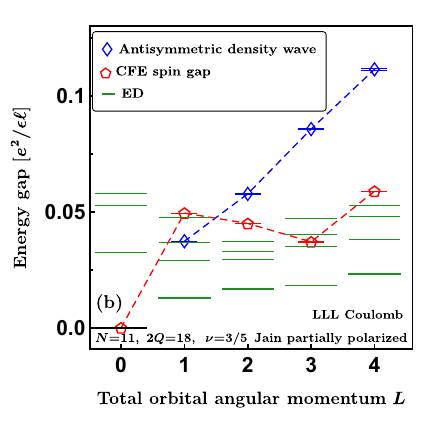}\\
          \includegraphics[width=0.499\columnwidth]{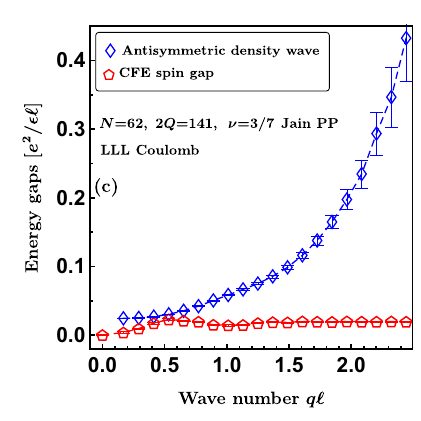}&
          \includegraphics[width=0.499\columnwidth]{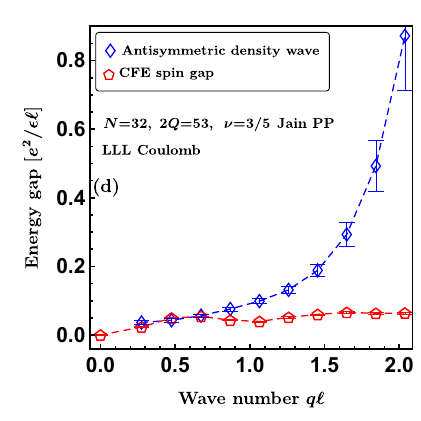}
        \end{tabular}
          \caption{LLL Coulomb spin-wave dispersion in the partially polarized Jain $3/7$ and $3/5$ states. The top panels $(a)$ and $(b)$ show a comparison between the spin-wave dispersion from ED and CF theory for small system sizes. The bottom panels $(c)$ and $(d)$ present results from CF theory for larger systems. For reference, we have also included the antisymmetric density wave dispersion (blue empty pentagons), which does not describe a gapless spin-wave mode [see text].}
          \label{fig: ADW_dispersion_primary_Jain_states_pp_3_7_and_3_5_pp}
        \end{figure} 

\subsubsection{Partially polarized states}
The exact SW mode in the Jain $3/7$ and $3/5$ PP states is presented in Figs.~\ref{fig: ADW_dispersion_primary_Jain_states_pp_3_7_and_3_5_pp}$(a)$ and~\ref{fig: ADW_dispersion_primary_Jain_states_pp_3_7_and_3_5_pp}$(b)$, respectively. We find that the spin-flip CF excitons, corresponding to the transition $1_{\uparrow} {\rightarrow} 1_{\downarrow}$, do not model the exact dispersion very accurately. As was the case for the 2/3 spin-singlet Jain state, here too, JK projection does not produce a very good variational state for the spin-flip CF excitons. Here, too, a better version of the CF SW mode can be obtained by performing the HCP, which is expected to provide a good description of the actual gapless SW. However, since this approach is limited to small system sizes, we have not pursued it further. For completeness, in Figs.~\ref{fig: ADW_dispersion_primary_Jain_states_pp_3_7_and_3_5_pp}$(c)$ and~\ref{fig: ADW_dispersion_primary_Jain_states_pp_3_7_and_3_5_pp}$(d)$, we show the dispersion of the CF SW obtained via the JK projection for large systems. The CF SW mode goes gapless in the long-wavelength limit, as one would anticipate in PP states. 

Our results [see Figs.~\ref{fig: ADW_dispersion_primary_Jain_states_pp_3_7_and_3_5_pp}$(c)$ and~\ref{fig: ADW_dispersion_primary_Jain_states_pp_3_7_and_3_5_pp}$(d)$] suggest that the ADW mode remains gapped in the long-wavelength limit and therefore does not correspond to the gapless SW of PP states. In general, unlike in singlet states, the ADW mode, and SW modes, those obtained from acting $\bar{\rho}_{L, M}^{~+}$ or $\bar{\rho}_{L, M}^{~-}$ on the ground state, are all distinct in PP states, since these modes are not related by $\mathbb{S}^{\pm}$ operators. The gapless SW mode in PP states arises from the density-wave ansatz for the SW obtained from $\bar{\rho}_{L, M}^{~-}$, as it smoothly connects to the ground state multiplet at $L{=}0$ with $\mathbb{S}^{z}{=}\mathbb{S}_{0}{-}1$. In contrast, $\bar{\rho}_{0,0}^{+}$ annihilates the ground state with $\mathbb{S}{=}\mathbb{S}^{z}{=}\mathbb{S}_{0}$.   

\begin{figure*}[tbh!]
         \includegraphics[width=0.66\columnwidth]{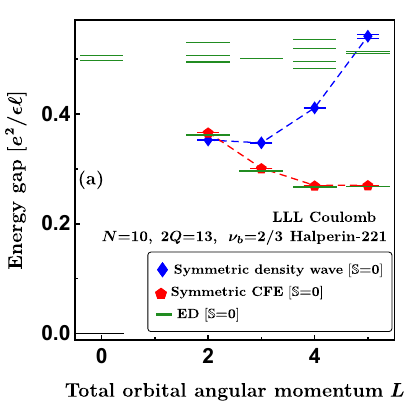}
        \includegraphics[width=0.66\columnwidth]{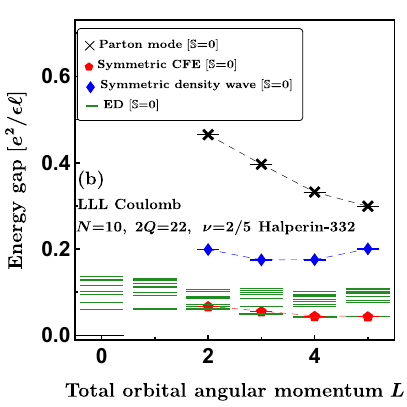}
      \includegraphics[width=0.66\columnwidth]{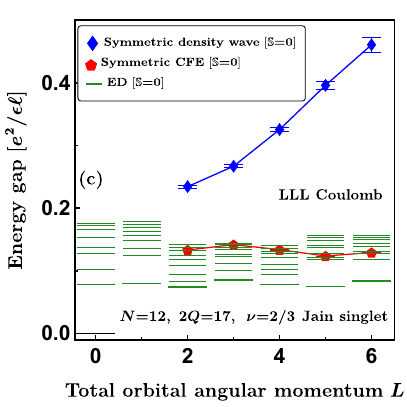}\\
       \includegraphics[width=0.66\columnwidth]{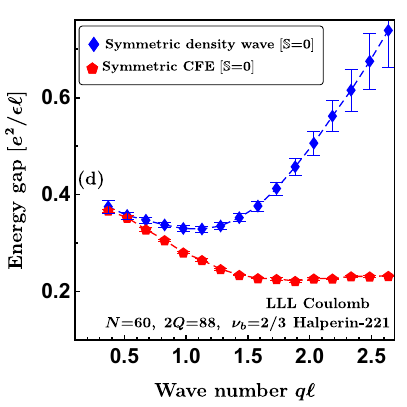}
      \includegraphics[width=0.66\columnwidth]{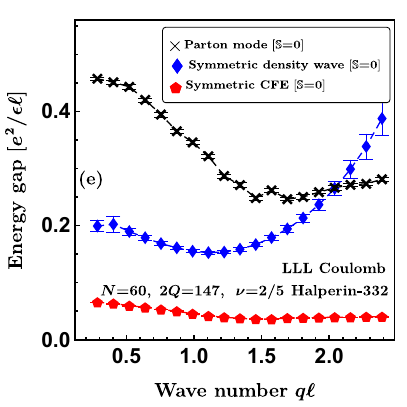}
      \includegraphics[width=0.66\columnwidth]{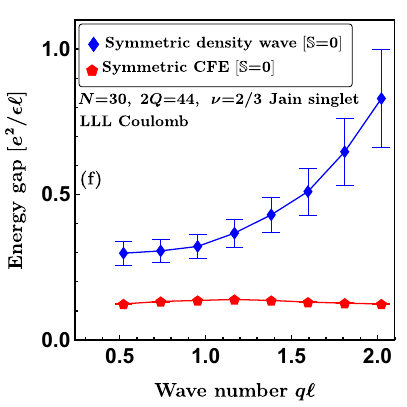} \\
        
       \caption{LLL Coulomb dispersion of the spin-conserving charge-neutral mode in the bosonic $(2,2,1)$, fermionic $(3,3,2)$ Halperin, and spin-singlet Jain $2/3$ states. The top panels present the comparison of the symmetric density wave and symmetric CFE mode with the exact dispersion for small systems. The bottom panels present the dispersion of the symmetric density wave and the symmetric CFE mode for bigger systems. The extra high-energy parton mode in the Halperin-$(3,3,2)$ state is indicated by black crosses. } 
          \label{fig: symmetric_SW_dispersion_small_large_system_size}
        \end{figure*}
        
\subsection{Results: Spin-conserving gaps in unpolarized states}
\label{ssec: spin_conserving_gaps_unpolarized_states}

\subsubsection{Singlet states: Prediction of a parton mode in the Halperin-$\left(m,m,m-1\right)$ states}
This section presents results on the dispersion of spin-conserving neutral excitations in singlet states for the LLL Coulomb interaction, computed following the methods discussed in the previous sections. The SDW dispersion, also referred to as the GMP mode, is obtained from Eq.~\eqref{eq: SS_density_wave_gap_equation}, and is compared with the S-CFE mode in Fig.~\ref{fig: symmetric_SW_dispersion_small_large_system_size}. For comparison, we have also computed the exact dispersion of spin-conserving neutral excitations, referred to as the magnetoroton mode, which corresponds to the branch of the lowest energy excited states in the exact spectra. As evident from Figs.~\ref{fig: symmetric_SW_dispersion_small_large_system_size}$(a)$ and~\ref{fig: symmetric_SW_dispersion_small_large_system_size}$(b)$, the magnetoroton mode is well described by the S-CFE mode in the Halperin-$(2,2,1)$~\cite{Wu13} and $(3,3,2)$ singlet states, while the GMP mode fails to capture it accurately. Note that the S-CFE mode starts from $L{=}2$, as the $L{=}1$ S-CFE state is annihilated upon projection to the LLL~\cite{Liu20}. Similarly, since $\bar{\rho}_{1, M}^{~I}$ annihilates the ground state, the GMP mode also starts from $L{=}2$. Surprisingly, the $L{=}2$ GMP state behaves distinctly in the  Halperin-$(2,2,1)$ and Halperin-$(3,3,2)$ states. Specifically, in the  Halperin-$(2,2,1)$ state [see Figs.~\ref{fig: symmetric_SW_dispersion_small_large_system_size}$(a)$ and~\ref{fig: symmetric_SW_dispersion_small_large_system_size}$(d)$], the GMP mode is close in energy to the S-CFE mode in the long-wavelength limit (meaning at $L{=}2$ for finite systems on the sphere). It therefore provides a good description of the long-wavelength magnetoroton mode, consistent with previous studies~\cite{Liu20}, and as has been seen in the spin-polarized primary Jain states~\cite{Dora24}. In contrast, the long-wavelength GMP mode differs significantly in energy from the S-CFE mode in the Halperin-$(3,3,2)$ state as seen in Figs.~\ref{fig: symmetric_SW_dispersion_small_large_system_size}$(b)$ and~\ref{fig: symmetric_SW_dispersion_small_large_system_size}$(e)$. Thermodynamic extrapolation of the $L{=}2$ SDW and S-CFE gaps in the  Halperin-$(2,2,1)$ and Halperin-$(3,3,2)$ states are shown in Fig.~\ref{fig: thermodynamic_extrapolation_L_1_ADW_L_2_SDW_gaps}$(b)$ and~\ref{fig: thermodynamic_extrapolation_L_1_ADW_L_2_SDW_gaps}$(d)$, respectively.

The magnetoroton mode seen in the exact Coulomb spectrum for the spin-singlet bosonic state at $\nu_{b}{=}2/3$ [see Fig.~\ref{fig: symmetric_SW_dispersion_small_large_system_size}$(a)$], clearly starts from $L{=}2$, and this branch of excitations is well-separated from other excitations. In contrast, curiously, there is a low-lying excitation at $L{=}1$ in the exact Coulomb spectra of the spin-singlet state of fermions at $\nu{=}2/3$ and $\nu{=}2/5$, but we believe this low-lying $L{=}1$ state arises from two CF excitons, and is not part of the magnetoroton mode.

This peculiar behavior of the GMP mode is reminiscent of that seen in fully-polarized primary and secondary Jain states. Unlike in primary Jain states, in secondary Jain states, such as $\nu{=}2/7$ and $2/9$, the GMP mode is not accurate even in the long-wavelength. The $L{=2}$ GMP state, referred to as the GMP graviton, splits into a low-energy CF graviton and a high-energy parton graviton~\cite{Balram21d, Nguyen22, Wang22, Balram24}. Following this logic, the large mismatch between the long-wavelength gaps of GMP and S-CFE modes in the Halperin-$(3,3,2)$ state suggests the presence of an additional parton mode in it. Encouragingly, one can indeed write down the wave function of this parton mode as follows. We first note that the Halperin-$(m,m,m{-}1)$ states are made of two types of parton: one forming a $\Phi_1$ state (multiple partons of this type exist when $m{>}2$) and the other forming a $\Phi_{1,1}$ state (this parton is unique). We then suggestively write the Halperin-$(3,3,2)$ state as
\begin{equation}
\label{eq: Halperin_332}
\Psi^{3,3,2}_{\nu{=}2/5}{=}\Phi_1^{2}\Phi_{1_\uparrow}\Phi_{1_\downarrow}{\equiv}\Phi_1^{2}\Phi_{1,1}.
\end{equation}
The wave function of the parton mode is then given by 
\begin{equation}
\label{eq: parton_mode_Halperin_332}
\Psi^{\rm parton}_{2/5} = \mathcal{P_{\rm LLL}}\Phi_{1,1} \Phi_{1} \Phi^{\rm exciton}_{1}  {\sim} \Phi_{1,1} \Psi^{\rm CFE}_{1/2}{\equiv}\Phi_{1,1} \frac{\Psi^{\rm CFE}_{1/3}}{\Phi_1}.
\end{equation}
Here, $\Phi_{1,1}{\equiv}\Psi^{1,1,0}_{\nu=2}$ is the $\nu{=}2$ Halperin-$(1,1,0)$ spin-singlet state. The $\sim$ sign in the above equation indicates that the projection to the LLL is carried out in a particular manner to facilitate the evaluation of the wave function for large systems, i.e., only the rightmost side of Eq.~\eqref{eq: parton_mode_Halperin_332} is amenable to JK projection. We expect that such details of the projection result in only minor quantitative differences~\cite{Balram15a, Balram16b, Anand22}. In general, other Halperin-$(m,m,m{-1})$ states, described by the wave function,
\begin{equation}
\label{eq: Halperin_mmm_1}
\Psi^{m,m,m{-}1}_{2/(2m{-}1)}{=}\Phi_1^{(m-1)}\Phi_{1,1},
\end{equation}
admit a parton mode for $m{\geq}3$, the wave function for which is
\begin{equation}
\label{eq: parton_mode_Halperin_m_m_mm1}
\Psi^{\rm parton}_{2/(2m-1)} = \Phi_{1,1} \Psi^{\rm CFE}_{1/(m-1)},
\end{equation}
where $\Psi^{\rm CFE}_{1/(m{-}1)}$ is the CFE of the $1/(m{-}1)$ Laughlin state, i.e., $\Psi^{\rm CFE}_{1/(m{-}1)}{=}\mathcal{P}_{\rm LLL} \Phi^{m{-}2}_{1}\Phi^{\rm exciton}_{1}$. The state in Eq.~\eqref{eq: parton_mode_Halperin_m_m_mm1} has $\mathbb{S}{=}0$ since it is a fully symmetric in spin-state $\Psi^{\rm CFE}_{1/(m-1)}$ [that has $\mathbb{S}{=}N/2$] multiplying a spin-singlet state $\Phi_{1,1}$~\cite{Hamermesh62}. For convenience, we mention below the wave function of the low-energy S-CFE mode in the Halperin-$(m,m,m{-1})$ states, whose construction is discussed in Sec.~\ref{ssec: CFE_gaps},
\begin{equation}
\label{eq: CFE_mode_Halperin_mmn}
\Psi^{\rm S-CFE}_{2/(2m-1)} = \mathcal{P_{\rm LLL}}\Phi_{1}^{m-1} \Phi^{S,\rm exciton}_{1,1}.
\end{equation}
Here, $ \Phi^{S,\rm exciton}_{1,1}$ is the symmetric linear combination of the CF exciton states corresponding to the transitions $0_{\uparrow}{\rightarrow}1_{\uparrow}$ and $0_{\downarrow}{\rightarrow}1_{\downarrow}$ in $\Phi_{1,1}$. A schematic representation of the state in Eq.~\eqref{eq: CFE_mode_Halperin_mmn} can  be envisioned from  Fig.~\ref{fig: symmetric_antisymmetric_CF_spin_gap_schematic}$(a)$ with CFs carrying $(m{-1})$ vortices. A HCP version of Eq.~\eqref{eq: CFE_mode_Halperin_mmn} is 
\begin{align}
\label{eq: hard_core_projected_CFE_mode_Halperin_mmn}
\Psi^{\rm HCP-CFE}_{2/(2m-1)} &=\Phi_{1}^{m-2}\mathcal{P_{\rm LLL}}\Phi_{1} \Phi^{\rm S-exciton}_{1,1}\nonumber\\
    &=\Phi_{1}^{m-2}\Psi^{\rm (2,2,1)~S-CFE}_{2/3},
\end{align}
which, for $m{=}3$ is 
\begin{equation}
\label{eq: hard_core_projected_CFE_mode_Halperin_332}
\Psi^{\rm HCP-CFE}_{2/5} {=} \Phi_{1}\mathcal{P_{\rm LLL}}\Phi_{1} \Phi^{\rm S-exciton}_{1,1}{=} \Phi_{1}\Psi^{\rm (2,2,1)~S-CFE}_{2/3},
\end{equation}
which vanishes when an up and down electron is brought together, i.e., $\langle V_{0}\rangle_{\Psi^{\rm HCP-CFE}_{2/5}}{\equiv}\langle V_{0}^{\rm inter}\rangle_{\Psi^{\rm HCP-CFE}_{2/5}}{=}0$; however, this wave function is not amenable to large-scale numerical calculations. In general, for the wave function in Eq.~\eqref{eq: hard_core_projected_CFE_mode_Halperin_mmn}, $\langle V_{\mathfrak{m}{\leq}m{-}3}\rangle{=}0$. In contrast, the wave function of Eq.~\eqref{eq: parton_mode_Halperin_332} has $\langle V_{0}\rangle_{\Psi^{\rm parton}_{2/5}}{\equiv}\langle V_{0}^{\rm inter}\rangle_{\Psi^{\rm parton}_{2/5}}{\neq}0$, which suggests that the state $\Psi^{\rm parton}_{2/5}$ cannot be obtained via a HCP of an excitation of the spin-singlet state at $2/5$.

Interestingly, one can also infer the possibility of a parton mode in the Halperin-$(m,m,m{-}1)$ states from the corresponding wave function of the GMP state at $L{=}2$. We note that the GMP state in the Halperin-$(m,m,m{-}1)$ state is given by
\begin{align}
\label{eq: GMP_Halperin_mmn}
    \Psi^{\rm GMP}_{L, M}&=\bar{\rho}_{L, M}^{~I}\left[\Psi^{m,m,m{-}1}_{2/(2m{-}1)}\right]=\bar{\rho}_{L, M}^{~I}\left[ \Phi_{1}^{m-1}\Phi_{1,1}\right].
\end{align}
In real space, $\bar{\rho}_{L, M}^{~I}$ at $M{=}L$ admits a simple form given by $\bar{\rho}_{L, M}^{~I}{=}\sum_{i}u_{j}^L(\partial/\partial v_j)^L$~\cite{Pu23}. Consequently, the GMP state at $L{=}2$, equivalently the GMP graviton, can be simplified by using the chain rule of differentiation as 
\begin{align}
    \label{eq: GMP_graviton}
    \Psi^{\rm GMP-graviton}_{2/(2m-1)}&=\sum_{j{=}1}^{N}u_{j}^{2}\frac{\partial^2} {\partial v_{j}^{2}}\left(\Phi_{1}^{m-1}\Phi_{1,1}\right)
     \nonumber\\
    &= (m-1)\Phi_{1}^{m-2}\sum_{j{=}1}^{N} u_{j}^2\frac{\partial^2} {\partial v_{j}^{2}} \left(\Phi_1\Phi_{1,1}\right)   \nonumber\\
    & + \Phi_{1,1}\sum_{j{=}1}^{N}u_{j}^2\frac{\partial^2} {\partial v_{j}^{2}}\Phi_{1}^{m-1} \nonumber\\
    &= \Psi^{\rm CF-graviton}_{2/(2m-1)} + \Psi^{\rm p-graviton}_{2/(2m-1)},
\end{align}
where, 
\begin{align}
\label{eq: CF_graviton}
    \Psi^{\rm CF-graviton}_{2/(2m-1)}&=(m-1)\Phi_{1}^{m-2}\bar{\rho}_{2,2}\Phi_1\Phi_{1,1} \\
    &\equiv (m-1)\Phi_{1}^{m-2} \Psi^{\rm HCP-CF-graviton}_{2/3},\nonumber
\end{align}
and
\begin{align}
\label{eq: parton_graviton}
   \Psi^{\rm p-graviton}_{2/(2m-1)}&= \Phi_{1,1}\bar{\rho}_{2,2}\Phi_{1}^{m-1} \\
   &\propto \Phi_{1,1}(\Phi_1)^{m-3}\sum_{j{=}1}^{N}\left(u_j\frac{\partial}{\partial v_j}\Phi_1\right)^2,~m{\geq}3. \nonumber
\end{align} 
In obtaining Eq.~\eqref{eq: GMP_graviton}, we have used the fact that $\bar{\rho}_{2,2}$ annihilates the IQH states $\Phi_{1,1}$ and $\Phi_{1}$, i.e., $\bar{\rho}_{2,2}\Phi_{1,1}{=}0{=}\bar{\rho}_{2,2}\Phi_{1}$. 

For $m{=}2$, i.e., the Halperin-$(2,2,1)$ state, $\Psi^{\rm p-graviton}_{2/3}$ vanishes identically; consequently, $\Psi^{\rm CF-graviton}_{2/3}{=}\bar{\rho}_{2,2}\Phi_1\Phi_{1,1}$ is exactly the same as $\Psi^{\rm GMP-graviton}_{2/3}$. Since the energy of $\Psi^{\rm GMP-graviton}_{2/3}$ is close to the energy of the CF graviton in the Halperin-$(2,2,1)$ state [see Fig.~\ref{fig: symmetric_SW_dispersion_small_large_system_size}($a$) and~\ref{fig: symmetric_SW_dispersion_small_large_system_size}($d$)], we identify $\Psi^{\rm GMP-graviton}_{2/3}$ in Eq.~\eqref{eq: GMP_graviton} with the $L{=}2,~M{=}2$ state of $\Psi^{\rm HCP-CFE}_{2/3}$ [see Eq.~\eqref{eq: hard_core_projected_CFE_mode_Halperin_mmn}], denoted as $\Psi^{\rm HCP-CF-graviton}_{2/3}$. Therefore, $\Psi^{\rm CF-graviton}_{2/3}$ is nearly equivalent to the $L{=}2,~M{=}2$ state of $\Psi^{\rm HCP-CFE}_{2/3}$. More generally, multiplying both $\Psi^{\rm HCP-CF-graviton}_{2/3}$ and the $L{=}2,~M{=}2$ state of $\Psi^{\rm HCP-CFE}_{2/3}$ by $\Phi_1^{m-2}$, we arrive at the conclusion that $\Psi^{\rm CF-graviton}_{2/(2m-1)}$ is nearly equivalent to the $L{=}2,~M{=}2$ state of $\Psi^{\rm HCP-CFE}_{2/(2m-1)}$. For practical purposes, we shall further identify $\Psi^{\rm CF-graviton}_{2/(2m-1)}$ with the $L{=}2,~M{=}2$ state of $\Psi^{\rm S-CFE}_{2/(2m-1)}$ defined in Eq.~\eqref{eq: CFE_mode_Halperin_mmn} since, unlike $\Psi^{\rm HCP-CFE}_{2/(2m-1)}$, $\Psi^{\rm S-CFE}_{2/(2m-1)}$ is computable for large systems.

Similarly, for $m{\geq}3$, $\Psi^{\rm p-graviton}_{2/(2m-1)}$ [see Eq.~\eqref{eq: parton_graviton}] is exactly equivalent to the graviton state of the parton mode $\Psi^{\rm parton}_{2/(2m-1)}$ [see Eq.~\eqref{eq: parton_mode_Halperin_m_m_mm1}]. This is because, $\bar{\rho}^{I}_{2,2}\Phi_{1}^{m}{=}\mathcal{P}_{\rm LLL} \Phi_{1}^{m{-}1}\rho^{I}_{2,2}\Phi_1$, where $\rho_{2,2}$ is the unprojected total density operator. In the $q{\to}0$ long-wavelength limit, the GMP and CFE excitations are identical for the $\Phi_1^{m}$ Laughlin states at $\nu{=}1/m$ since $\rho_{q{\to}0}\Phi_{n}{=}\lim_{q{\to}0}\Phi^{n{\to}n{+}1~{\rm exciton}}_{n}(q)$~\cite{Kamilla96b}, and thus, the operator $\rho_2$ corresponds to the lowest-energy CFE operator for Laughlin states in the $Q{\to}\infty$ limit. For a system at finite flux $Q$, $\rho_{2,2}\Phi_1$ results in two kinds of inter LL terms: $Q{\rightarrow}Q{+}1$ transition from the LLL to the second LL and $Q{\rightarrow}Q{+}2$ transition from the LLL to the third LL. Surprisingly, even for finite systems, explicit calculations show that once projected, the above two inter-LL terms for Laughlin states become identical to each other~\cite{Balram13, Balram24}. Thus, the GMP and CF graviton wave functions for the Laughlin states are identical for even finite systems~\cite{Pu23, Pu24}. Therefore, multiplying the GMP and CF graviton wave functions of the Laughlin state by $\Phi_{1,1}$, we conclude that $\Psi^{\rm p-graviton}_{2/(2m-1)}$ is identical to the $L{=}2$ state of $\Psi^{\rm parton}_{2/(2m-1)}$. 

To summarize, the GMP graviton of the $m{\geq}3$ Halperin-$(m,m,m{-}1)$ state is a linear combination of the HCP-CFE graviton and the parton graviton. In other words, the GMP graviton splits into the HCP-CF and parton gravitons, and is likely fully exhausted by them~\cite{Balram24} (This is because, as elucidated above, in the parton construction, only two possibilities exist: one could either create an exciton in the $\Phi_1$ state or in the $\Phi_{1,1}$ state.). Next, to infer the energy scale of these parton and CFE modes, we discuss their clustering properties for the $m{=}3$ Halperin-$(3,3,2)$ state of our interest. As noted above, the HCP-CFE wave function given in Eq.~\eqref{eq: hard_core_projected_CFE_mode_Halperin_332} vanishes when an up and down electron are brought to the same point, and thus has $\langle V_0 \rangle{=}0$, while that is not the case for the parton mode's wave function given in Eq.~\eqref{eq: parton_mode_Halperin_332}. In the language of conformal Hilbert spaces~\cite{Wang22}, the HCP-CFE lives within the Hilbert space defined by the zero-modes of $V_{0}$, while the parton-mode lies outside this Hilbert space. For the LLL Coulomb interaction, since $V_0$ is dominant compared to other pseudopotentials, the parton mode has a higher energy than the HCP-CFE mode. Even when the CFE is not HCP, this appears to be the case as can be seen in Figs.~\ref{fig: symmetric_SW_dispersion_small_large_system_size}$(b)$ and~\ref{fig: symmetric_SW_dispersion_small_large_system_size}$(e)$, presumably since the CFE and HCP-CFE have good overlaps with each other. In general, since the $L{=}2$ state of Eq.~\eqref{eq: parton_mode_Halperin_m_m_mm1} and Eq.~\eqref{eq: parton_graviton} are equivalent, the parton graviton, for $m{\geq}3$, has $\left\langle V_{\mathfrak{m}{<}m{-}3}\right\rangle{=}0$. Moreover, since in the HCP-CF graviton, $\left\langle V_{\mathfrak{m}{<}m{-}2}\right\rangle{=}0$, the GMP graviton has $\left\langle V_{\mathfrak{m}{<}m{-}3}\right\rangle{=}0$, for $m{\geq}3$. The wave functions of the CFE and parton modes, in particular, their gravitons, are not orthogonal to each other, and one should diagonalize the Coulomb interaction in the subspace of these two modes~\cite{Balram24, Bose25} to obtain the mode dispersions. We leave a detailed exploration of this matter to future work. In Fig.~\ref{fig: thermodynamic_extrapolation_L_1_ADW_L_2_SDW_gaps}$(d)$, we show the density-corrected~\cite{Morf86b} gap of the parton graviton in the thermodynamic limit.  

Other singlet states, such as the fermionic $2/3$ Jain state, or the bosonic and fermionic Jain states at $2/7$ and $2/9$, can also host these parton modes, and their wave functions can be constructed analogously to the ones given above for $2/5$. The CF graviton for the spin-singlet Halperin-$(m,m,m{-}1)$ state has the same chirality as the graviton of the $1/3$ Laughlin state~\cite{Liou19, Balram21d}, and so does the parton graviton, when it exists for $m{\geq}3$. In contrast, the CF graviton for the spin-singlet Jain state at $\nu{=}2/3$ has opposite chirality to that of the graviton of the $1/3$ Laughlin state. These modes can be detected via circularly polarized light-scattering measurements~\cite{Liang24}. 

Analogous to the parton mode of Eq.~\eqref{eq: parton_mode_Halperin_m_m_mm1}, one can potentially construct an additional spin-flip mode, which exists only for $m{\geq}2$, described by the wave function
\begin{equation}
\label{eq: SF_parton_mode_Halperin_m_m_mm1}
\Psi^{\rm parton-SW}_{2/(2m-1)} = \Phi_{1,1} \Psi^{\rm CFE-SW}_{1/(m-1)},
\end{equation}
where $\Psi^{\rm CFE-SW}_{1/(m{-}1)}$ is the CFE SW in the $\mathbb{S}_{z}{=}0$ version of the $1/(m{-}1)$ Laughlin state, i.e., $\Psi^{\rm CFE-SW}_{1/(m{-}1)}{=}\Phi^{m{-}2}_{1}\Phi^{\rm SW-exciton}_{1}$. However, CFE SW state has $\mathbb{S}^{z}{=}{+}1$ and that value is incompatible with the $\mathbb{S}^{z}{=}0$ value of $\Phi_{1,1}$ thereby precluding the construction of the wave function given in Eq.~\eqref{eq: SF_parton_mode_Halperin_m_m_mm1}. Another possibility is to consider 
\begin{equation}
\label{eq: anti_symmetric_spin_parton_mode_Halperin_m_m_mm1}
\Psi^{\rm parton-ADW}_{2/(2m-1)} = \Phi_{1,1} \bar{\rho}^{z}\Psi^{m{-}1, m{-}1, m{-}1}_{1/(m-1)}.
\end{equation}
However, this state does not carry $\mathbb{S}{=}1$, and thus the mode does not have definitive spin, since $\bar{\rho}^{z}\Psi^{m{-}1, m{-}1, m{-}1}_{1/(m-1)}$ for $L{\geq}2$ has no definite spin. At $L{=}1$ $\bar{\rho}^{z}\Psi^{m{-}1, m{-}1, m{-}1}_{1/(m-1)}$ has $\mathbb{S}{=}N/2{-}1,~\mathbb{S}^{z}{=}0$, which is not fully symmetric in the spin-space, and therefore, when multiplied by the singlet $\Phi_{1,1}$ does not yield a state with a definitive spin quantum number. The ADW and A-CFE energies are close in the long-wavelength limit [see Fig.~\ref{fig: ADW_dispersion_Halperin_221_332_Jain_ss_2_3}], suggesting that the latter fully exhausts the former in the $q{\to}0$ limit and no additional spin-flip parton modes are present in the small wave number limit. 

\subsubsection{Partially polarized states}
\begin{figure}[tbh!]
\centering
\begin{tabular}{cc}
        \includegraphics[width=0.499\columnwidth]{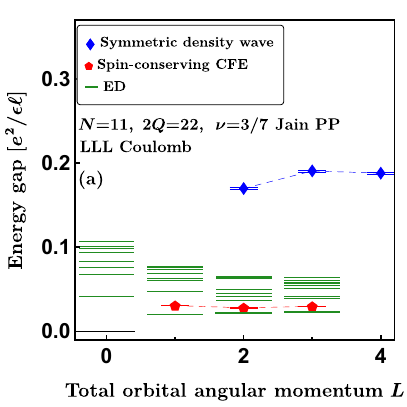}
        \includegraphics[width=0.499\columnwidth]{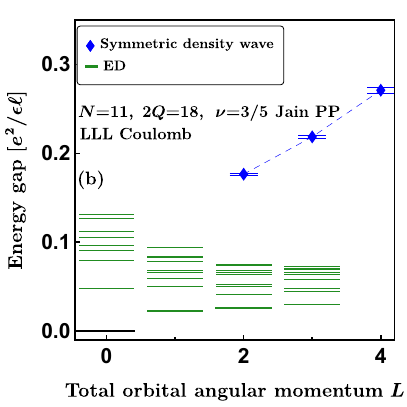}\\
          \includegraphics[width=0.499\columnwidth]{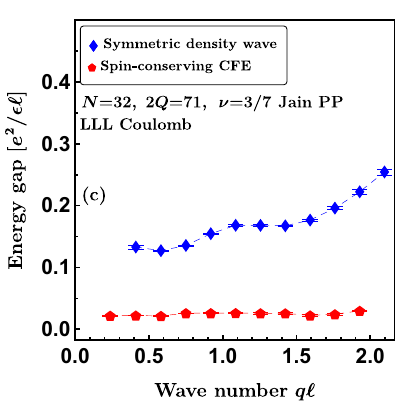}
          \includegraphics[width=0.499\columnwidth]{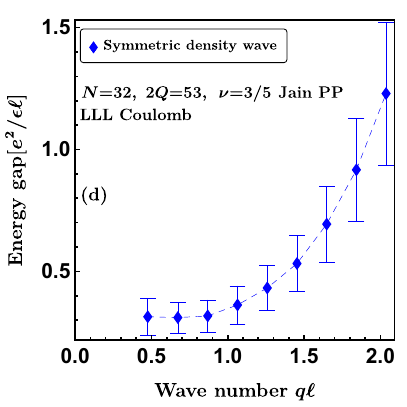}
         \end{tabular}
          \caption{Spin-conserving charge neutral LLL Coulomb gap in the partially polarized Jain $3/7$ and $3/5$ states. Panels $(a)$ and $(c)$ show the dispersions of the symmetric density wave and the spin-conserving CFE mode in the Jain $3/7$ PP state. Similarly, panels  $(b)$ and $(d)$ present the symmetric density wave gap in the Jain $3/5$ PP state. Results from ED for small system sizes are also presented in the top panel.}
        \label{fig: SS_density_wave_dispersion_primary_Jain_states_pp_3_7_and_3_5}
\end{figure} 

Strikingly, unlike in the fully polarized and singlet states, the spin-conserved CFE mode starts from $L{=}1$ in PP states [see Fig,~\ref{fig: SS_density_wave_dispersion_primary_Jain_states_pp_3_7_and_3_5}($a$)], consistent with a low-lying exact state at that angular momentum. Unlike for the spin-singlet states, in the CF basis states forming the spin-conserved CFE mode [see Fig.~\ref{fig: CFE_basis_states_3_7_PP_Jain_state}(b)], the spin-up and spin-down FQH fluids are excited or deformed differently in partially polarized states, allowing for a state at $L{=}1$ to occur. Note that since the JK-projection is less accurate for the reverse vortex attached PP states, we have not computed the spin-conserved CFE mode in the PP $3/5$ Jain state. The GMP mode cannot capture this $L{=}1$ state since it only starts from $L{=}2$. Moreover, similar to the singlet states, our results for the PP states, shown in Fig.~\ref{fig: SS_density_wave_dispersion_primary_Jain_states_pp_3_7_and_3_5}, suggest that they also host a high-energy parton mode. Specifically, the SDW dispersion, equivalently the GMP mode, lies far in energy from the magnetoroton mode, even in the long-wavelength limit, i.e., at $L{=}2$, as seen from Figs.~\ref{fig: SS_density_wave_dispersion_primary_Jain_states_pp_3_7_and_3_5}$(a)$ and~\ref{fig: SS_density_wave_dispersion_primary_Jain_states_pp_3_7_and_3_5}$(b)$. The wave function of the parton mode in the PP Jain $3/7$ state, described by the wave function
\begin{align}
    \label{eq: Jain_3_7_pp}
\Psi^{\rm Jain-PP}_{3/7}{=}\Phi_1^{2}\Phi_{2_\uparrow}\Phi_{1_\downarrow}{\equiv}\Phi_1^{2}\Phi_{2,1},
\end{align}
can only be constructed by brute-force projection as
\begin{align}
    \label{eq: parton_mode_Jain_3_7_pp}
\Psi^{\rm parton}_{3/7}&=\mathcal{P_{\rm LLL}}\Phi_{2,1} \Phi_{1} \Phi^{\rm exciton}_{1}.
\end{align}
This is because writing it as $\Psi^{\rm Jain-PP}_{3/7}\Psi^{\rm CFE}_{1/2}/\Phi_{1}^{2}$, to enable the use of JK-projection, makes the wave function ill-defined when two electrons, separated by a distance $r$, approach each other since the numerator only vanishes as $r$ while the denominator vanishes as $r^2$. An analogous wave function of the parton mode in other PP primary Jain states can also be written. As these wave functions are not amenable to JK projection, we have not pursued computing their dispersions. For the PP secondary Jain states, for example, at $3/13$, the parton mode wave function can be obtained via JK-projection as $\Psi^{\rm Jain-PP}_{3/7}\Psi^{\rm CFE}_{1/2}$. This idea suggests that the $L{=}1$ state of the parton mode for PP Jain states is annihilated upon projection to the LLL, just like the $L{=}1$ GMP state.  

To summarize this subsection, for the singlet and PP primary Jain states at $\nu{=}n/(2n{\pm}1)$, the GMP mode fails to capture their dynamics in the spin-conserving sector even in the long-wavelength limit. This should be contrasted with the case of fully polarized primary Jain states, where the GMP mode gives an accurate description of the spin-preserving lowest-lying collective neutral mode in the small wavenumber limit~\cite{Dora24, Balram24}.

\section{Conclusion}
\label{sec: conclusion}
We computed the dispersion of neutral excitations, including both spin-flip and spin-conserving ones, on a Haldane sphere, for spinful quantum Hall states belonging to the fermionic $n/(2n{\pm}1)$ and bosonic $n/(n{\pm}1)$ primary Jain sequence described by the composite fermion theory. The dispersions are evaluated using two approaches---density-wave ansatzes and composite fermion excitons. To compute the dispersion of the density wave states, we generalized the algebra of spinless projected density operators on a sphere~\cite{Dora24} to the spinful case. Along with this algebra, the only input required to compute the density-wave gaps is the ground-state static structure factor, which we evaluate using Monte Carlo methods.

Comparing these dispersions against the results obtained from exact diagonalization on small systems, we find that the composite fermion theory gives an accurate description of the low-lying neutral collective modes. In contrast, generically, the density-wave ansatz fails to capture the collective modes. Nevertheless, the density-wave states for spin-flip excitations are accurate in the long-wavelength limit in certain cases. Specifically, for fully polarized states, the density ansatz for the spin-wave results in a gapless (up to the single-particle Zeeman gap) dispersion in the long-wavelength limit, as is expected for states with non-zero polarizations. Similarly, the antisymmetric density wave dispersion lies close to the antisymmetric composite fermion exciton mode in the long-wavelength limit for spin-singlet bosonic and fermionic Jain states at $2/3$ and $2/5$, respectively. However, the density-wave ansatz for the spin-wave fails to capture the roton minimum seen in the exact spin-wave dispersion of fully polarized Jain states at $\nu{=}n/(2n{\pm}1)$, $n{>}1$. Notably, both the gapless mode and the roton minimum have been observed experimentally in fully polarized Jain states~\cite{Kang00, Dujovne03b, Dujovne05, Gallais06, Gallais06a}. 

Unlike their spin-flip counterparts, the spin-conserving density wave excitations do not provide an accurate description, even in the long-wavelength limit, for generic spin-singlet and partially polarized states. In other words, the symmetric density wave, equivalently the GMP mode, lies far away from the actual low-lying spin-conserving mode [that is well captured by the symmetric composite fermion exciton]. However, the GMP mode does work accurately in the long-wavelength limit for the specific case of the bosonic Halperin-$(2,2,1)$ state. The failure of the GMP mode to capture the excitation even in the long-wavelength limit---whether for the $2/5$ spin-singlet, or more generally the spin-singlet states at $\nu{=}2/(2m{-}1)$ with $m{\geq}3$, or the partially polarized primary Jain states---suggests the intriguing possibility that these states support a high-energy parton mode alongside the low-energy symmetric composite fermion exciton mode. To summarize, our results show that the GMP mode captures the spin-conserving dynamics of only fully polarized primary Jain states, and that too only in the long-wavelength limit~\cite{Dora24, Balram24}. 
 
We end with some pointers to possible extensions of our work. Although we focused primarily on the LLL Coulomb interaction, the framework we presented can readily incorporate, via tweaking of the pseudopotentials, the Coulomb interactions corresponding to other LLs, such as higher LLs or the various LLs of relativistic electrons arising in graphene and its multi-layer incarnations~\cite{Dora23}. By tuning the interaction, one can see if the spinful bosonic Halperin-$(2,2,1)$ state is susceptible to a nematic instability~\cite{Yang20, Pu24, Dora24}, which occurs when its GMP mode closes in the long-wavelength limit. In the future, it would be useful to generalize our results to double-layer systems~\cite{Liu19, Li19, Faugno20, Liu20} or hetero-orbital settings~\cite{Huang25} where the $SU(2)$ symmetry of the interaction is broken. If the $SU(2)$ symmetry of the interaction is only weakly broken, the ground and excited state wave functions considered here can still provide a good starting point for that scenario. The dispersion of the spinless neutral collective modes for many partonic FQH states~\cite{Jain89b} was studied recently~\cite{Dora24, Bose25}. Our work can be readily extended to evaluate the dispersion of the spinful neutral collective modes in those and other partonic FQH states.

\begin{acknowledgments}
We acknowledge valuable discussions with Mytraya Gattu, Jainendra Jain, Prashant Kumar, Zlatko Papi\'c, and Arkadius\'z W\'ojs. The work was made possible by financial support from the Science and Engineering Research Board (SERB) of the Department of Science and Technology (DST) via the Mathematical Research Impact Centric Support (MATRICS) Grant No. MTR/2023/000002. Computational portions of this research work were conducted using the Nandadevi and Kamet supercomputers maintained and supported by the Institute of Mathematical Sciences' High-Performance Computing Center. Some numerical calculations were performed using the DiagHam package~\cite{diagham}, for which we are grateful to its authors.
\end{acknowledgments}

\appendix
\section{Fully antisymmetrized (symmetrized) wave functions of fermionic (bosonic) Halperin states}
\label{app: Antisymmetrized_symmetrized_Halperin_states}
In this appendix, we present the expressions of the fully antisymmetrized (symmetrized) Halperin-$(m,m,n)$ wave functions, appropriate for describing fermionic (bosonic) states for odd (even) $m$. The full Halperin-$(m,m,n)$ wave function describing bosons or fermions can be obtained by explicitly including the spin degrees of freedom in Eq.~\eqref{eq: Halperin_mmn_WF}, and subsequently symmetrizing or anti-symmetrizing as
\begin{align}
    \label{eq: Halperin_mmn_WF_symmetrized_anti_symmetrized}
   \tilde{\Psi}^{m,m,n}_{\nu=2/(m+n)}&\equiv \mathcal{A}_{\pm}\left[\Psi^{m,m,n}_{\nu=2/(m+n)} \otimes\right.\nonumber\\
   &\left.\left |\uparrow_{1}\uparrow_{2}{\cdots}\uparrow_{N_{0}}\downarrow_{N_0+1}\downarrow_{N_0+2}{\cdots}\downarrow_{2N_{0}}\right\rangle \right].
\end{align}
Here, $\mathcal{A}_{+}$ and $\mathcal{A}_{-}$ denote the symmetrizing and antisymmetrizing operators that symmetrize and anti-symmetrize over the set of reduced $(1/2)\binom{N}{N/2}$ permutations [since all the up spins have the right symmetry, all the down spins have the right symmetry, the (anti)symmetrization requires only exchanges between the up and down spins need to be carried out], respectively. For ease of notation, we have defined $N_{\uparrow}{=}N_{\downarrow}{=}N/2{\equiv}N_{0}$. 

Let us first consider the simplest case with $n{=}m$. Since the state $\Psi^{m,m,m}_{\nu{=}1/m}$, for odd (even) $m$, is fully antisymmetric (symmetric) under the exchange of spatial coordinates of two particles, the spin part has to be fully symmetric for fermions (odd $m$) and bosons (even $m$). Therefore, in this case, the spin part decouples from the spatial wave function. As there are equal numbers of spin-$\uparrow$ and spin-$\downarrow$ particles, the spin part belongs to the maximally symmetric multiplet $\mathbb{S}{=}N/2$ with $\mathbb{S}^{z}{=}0$. Thus, the total wave function of the Halperin-$(m,m,m)$ state is given by
\begin{align}
\label{eq: full_Halperin_mmm_WF}
\tilde{\Psi}^{m,m,m}_{\nu=1/m}=\Psi^{m,m,m}_{\nu=1/m}\times\left(\mathbb{S}^{-}\right)^{N_0}\left| \uparrow_{1}\uparrow_{2}{\cdots}\uparrow_{N_{0}}{\cdots}\uparrow_{2N_0} \right\rangle.
\end{align}
In the above equation, there are $\binom{2N_0}{N_0}$ number of distinct spin configurations, which are generated by the action of $\left(\mathbb{S}^{-}\right)^{N_0}$ on the fully polarized state. 

For $n{\neq}m$ and odd $m$, one can antisymmetrize Eq.~\eqref{eq: Halperin_mmn_WF_symmetrized_anti_symmetrized} to obtain a fermionic wave function as
\begin{widetext}
\begin{align}
\label{eq: expansion_anti_symmetrized_Halperin_mmn_state}
\tilde{\Psi}^{m,m,n}_{\nu=2/(m+n)}&=\overbrace{\underbrace{\Psi^{m,m,n}_{\nu=2/(m+n)}\left(\mathfrak{u}_{1},\mathfrak{u}_{2},{\cdots},\mathfrak{u}_{N_0};\mathfrak{u}_{N_0{+}1},\mathfrak{u}_{N_0{+}2},{\cdots},\mathfrak{u}_{2N_0}\right)\otimes\bigg[\left|\boldsymbol{\Uparrow}\boldsymbol{\Downarrow}\right\rangle \pm \left|\boldsymbol{\Downarrow}\boldsymbol{\Uparrow}\right\rangle\bigg]}_{\Psi_{\rm base}}}^{\rm no-exchange}-\sum_{i=1}^{N_0}\sum_{\tilde{i}{=}1}^{N_0}\overbrace{\Psi_{\rm base}\left(i\leftrightarrow N_0+\tilde{i}\right)}^{\rm 1-pair-exchange}\nonumber\\[1.4 ex]
   &+\sum_{1\leq i<j\leq N_0}~\sum_{1\leq \tilde{i}<\tilde{j}\leq N_0} \overbrace{\Psi_{\rm base}\left(\left(i,j\right)\leftrightarrow \left(N_0+\tilde{i},N_0+\tilde{j}\right)\right)}^{\rm 2-pairs-exchange}-\overbrace{\cdots\cdots\cdots\cdots\cdots\cdots\cdots}^{\left([N_0]-1\right)-\text{pairs-exchange}}\nonumber\\[1 ex]
  & +(-1)^{[N_0]} \sump_{1\leq i<j<\cdots <r\leq N_0}~\sum_{1\leq \tilde{i}<\tilde{j}<\cdots <\tilde{r}\leq N_0}\overbrace{\Psi_{\rm base}\left(\overbrace{\left(i,j,\cdots r\right)}^{[N_0] ~\text{tuple}}\leftrightarrow \overbrace{\left(N_0+\tilde{i},N_0+\tilde{j},\cdots N_0+\tilde{r}\right)}^{[N_0] ~\text{tuple}}\right)}^{[N_0]-\text{pairs-exchange}}.
\end{align}
\end{widetext}
In the above equation, we have defined 
\begin{align}
    \big|\boldsymbol{\Uparrow}\boldsymbol{\Downarrow}\big\rangle \pm \big|\boldsymbol{\Downarrow}\boldsymbol{\Uparrow}\big\rangle&\equiv\bigg|\uparrow_{1}\uparrow_{2}{\cdots}\uparrow_{N_{0}}\downarrow_{N_0+1}\downarrow_{N_0+2}{\cdots}\downarrow_{2N_{0}} \bigg\rangle\nonumber\\
    &\pm\bigg|\downarrow_{1}\downarrow_{2}{\cdots}\downarrow_{N_{0}}\uparrow_{N_0+1}\uparrow_{N_0+2}{\cdots}\uparrow_{2N_{0}}\bigg\rangle,
\end{align}
and
\begin{align}
\label{eq: Even_odd_cases}
    [ N_0 ]& \equiv \left\lfloor\dfrac{N_0}{2} \right\rfloor=
\begin{cases}
\dfrac{N_0}{2}, & \text{if $N_0$ is even}, \\[6pt]
\dfrac{N_0 - 1}{2}, & \text{if $N_0$ is odd,}
\end{cases} ~~~ ~~ \\[1.5ex]
&\text{and}~~
\label{eq: prime_indicates_factor_1_2}
\sump_{i,j,\cdots}=
\begin{cases}
\dfrac{1}{2}\sum_{i,j,\cdots}, & \text{if $N_0$ is even}, \\[6pt]
\sum_{i,j,\cdots}, & \text{if $N_0$ is odd}
\end{cases},
\end{align}
where $\lfloor x \rfloor$ is the greatest integer${\leq}x$. Here, $\mathfrak{u}{=}(u,v)$, and the symbol $\left(i,j,{\cdots}, r\right){{\leftrightarrow}}\left(N_0{+}\tilde{i}, N_0{+}\tilde{j},{\cdots} N_0{+}\tilde{r}\right)$ denotes the exchange of the $i^{\rm th}$ particle's position coordinate and spin with that of the $\left(N_0{+}\tilde{i}\right)^{\rm th}$ particle and similarly for others. The first term in Eq.~\eqref{eq: expansion_anti_symmetrized_Halperin_mmn_state} represents the base particle configuration, with no exchange, denoted by $\Psi_{\rm base}$. The ${+}$ sign in $\Psi_{\rm base}$ is for even $N_0$ while the ${-}$ sign is for odd $N_{0}$. The second term corresponds to exchanging both the position and spin of one electron with those of another electron of opposite spin. The third term represents a similar exchange, but involving a cluster of two electrons of the same spin with the two electrons of opposite spin, and so on for other higher terms. In Eq.~\eqref{eq: expansion_anti_symmetrized_Halperin_mmn_state}, there are at most $[N_0]$ clusters of spin-$\uparrow$ electrons that are exchanged with the corresponding $[N_0]$ cluster of spin-$\downarrow$ electrons. Moreover, the prime on the summation in the last term of Eq.~\eqref{eq: expansion_anti_symmetrized_Halperin_mmn_state} indicates multiplication by a factor of $1/2$ when $N_0$ is even, to compensate for over-counting done by summing over all exchanges which are not distinct [see also below]

The number of distinct configurations for exchanging coordinates of different $j$-clusters of $\uparrow$-spin electrons with that of different $j$-clusters of $\downarrow$-spin particles is $\binom{N_0}{j}^2$.
The combinatorial factor $\binom{N_0}{j}$ counts the distinct ways of clustering $j$ like-spin electrons from the total $N_0$ electrons of that spin. Consequently, $\binom{N_0}{j}^2$ counts the distinct ways of pairing different $j{-}$clustered electrons of one spin with $j{-}$electron clusters of opposite spin for exchange. An exception occurs at $j{=}[N_0]$ with even $N_0$, where only $(1/2)\binom{N_0}{j}^2$ distinct exchanges exist, and for this reason there is a factor of $1/2$ in the last term of Eq.~\eqref{eq: expansion_anti_symmetrized_Halperin_mmn_state} [see also  Eq.~\eqref{eq: prime_indicates_factor_1_2}]. The total number of distinct exchanges, or equivalently, the total number of different spin configurations $N_{\rm spin-config}$ in Eq.~\eqref{eq: expansion_anti_symmetrized_Halperin_mmn_state} is
\begin{widetext}
\begin{align}
\label{eq: even_N_0_spin_configurations}
  N_{\rm spin-config}&=2\left[\binom{N_0}{0}^2+\binom{N_0}{1}^2+\binom{N_0}{2}^2+\cdots+\binom{N_0}{[N_0]-1}^2\right] +\binom{N_0}{[N_0]}^2 =\binom{2N_0}{N_0},~~\text{for even $N_0$},\\[2ex]
  \label{eq: odd_N_0_spin_configurations}
  N_{\rm spin-config}&=2\left[\binom{N_0}{0}^2+\binom{N_0}{1}^2+\binom{N_0}{2}^2+\cdots+\binom{N_0}{[N_0]}^2\right]  =\binom{2N_0}{N_0},~~\text{for odd $N_0$},
\end{align}
\end{widetext}
as expected. In the above equations, the factor of two multiplying the terms enclosed in the square brackets arises from the two different spin configurations in each term of Eq.~\eqref{eq: expansion_anti_symmetrized_Halperin_mmn_state}. In contrast, for even $N_0$, the factor of two that would have appeared in the last term of Eq.~\eqref{eq: even_N_0_spin_configurations} is canceled by the compensating factor of $1/2$ in Eq.~\eqref{eq: expansion_anti_symmetrized_Halperin_mmn_state} that comes from Eq.~\eqref{eq: Even_odd_cases}. 

Similarly, for even $m$, a fully symmetrized Halperin-$(m,m,n)$ state can be obtained by replacing all minus signs with plus signs in Eq.~\eqref{eq: expansion_anti_symmetrized_Halperin_mmn_state}. For convenience, we provide below an example of the antisymmetrized $\tilde{\Psi}^{1,1,0}_{\nu=2}$ state for a few particles. For $N{=}2,~N_{0}{=}1$, there are no terms with exchange of particles since $[N_{0}]{=}0$, and the wave function is given by (up to a normalization factor):
\begin{align}
\label{eq: antisymmetrized_110_N_2}
\tilde{\Psi}^{1,1,0}_{\nu=2, N{=}2}&= \left|\uparrow_{1}\downarrow_2\right\rangle- \left|\downarrow_{1}\uparrow_2\right\rangle .
\end{align}
The spatial part in the above equation is unity, as there are not enough like-spin particles to have intra-particle correlations. Similarly, for $N{=}4,~N_{0}{=}2$, $[N_0]{=}1$, the wave function $\tilde{\Psi}^{1,1,0}_{\nu=2, N_0{=}2}$ involves exchange of at most one pair of opposite spin electrons, and is given by (up to a normalization factor):
\begin{widetext}
\begin{align}
\label{eq: antisymmetrized_110_N_4}
    \tilde{\Psi}^{1,1,0}_{\nu=2, N{=}4} &=\underbrace{\left(u_1v_2{-}u_2v_1\right)\left(u_3v_4{-}u_4v_3\right)\left[\left|\right\uparrow_1\uparrow_2\downarrow_3\downarrow_4\rangle+\left|\right\downarrow_1\downarrow_2\uparrow_3\uparrow_4\rangle\right]}_{\Psi_{\rm base}^{1,1,0}}~ -\frac{1}{2}\sum_{i=1}^{2}\sum_{\tilde{i}{=}1}^{2}\Psi_{\rm base}^{1,1,0}\left(i\leftrightarrow \left(2+\tilde{i}\right)\right)\nonumber\\[1.4ex]
    &=\left(u_1v_2{-}u_2v_1\right)\left(u_3v_4{-}u_4v_3\right)\left[\left|\right\uparrow_1\uparrow_2\downarrow_3\downarrow_4\rangle+\left|\right\downarrow_1\downarrow_2\uparrow_3\uparrow_4\rangle\right]-\left(u_3v_2{-}u_2v_3\right)\left(u_1v_4{-}u_4v_1\right)\left[\left|\right\downarrow_1\uparrow_2\uparrow_3\downarrow_4\rangle+\left|\right\uparrow_1\downarrow_2\downarrow_3\uparrow_4\rangle\right]\nonumber\\[1.1ex]
    &-\left(u_4v_2{-}u_2v_4\right)\left(u_3v_1{-}u_1v_3\right)\left[\left|\right\downarrow_1\uparrow_2\downarrow_3\uparrow_4\rangle+\left|\right\uparrow_1\downarrow_2\uparrow_3\downarrow_4\rangle\right].
\end{align}
\end{widetext}
In certain cases, it is useful to interpret the wave function of the Halperin-$(m,m,m{-}1)$ state as
\begin{align}
    \tilde{\Psi}^{m,m,m-1}_{\nu=2/(2m-1)}&= \Phi_1^{m-1} \tilde{\Psi}^{1,1,0}_{\nu=2},
\end{align}
where all the spin-coordinates are contained within the spin-singlet $\nu{=}2$ IQH state, $\tilde{\Psi}^{1,1,0}_{\nu=2}$. Since $\Phi_1^{m-1}$ [equivalently, the wave function of the Halperin-$(m{-}1,m{-}1,m{-}1)$ state] is fully symmetric in the spin-space, i.e., has $\mathbb{S}{=}N/2$, any definite spin-$\mathbb{S}$ state (ground or excitation) of the spin-singlet $\nu{=}2$ IQH state, upon multiplication by $\Phi_1^{m-1}$, is a definite spin-$\mathbb{S}$ state of the spin-singlet $\nu{=}2/(2m{-}1)$ FQH state. This is the reasoning behind why the A-CFE and S-CFE modes of the Halperin-$(m,m,m{-}1)$ states carry definite spin. 

Next, following Fock's cyclic condition~\cite{Hamermesh62, Jain07}, we discuss whether the state $\tilde{\Psi}^{m,m,n}_{\nu=2/(m+n)}$ possesses a well-defined spin quantum number for $n{\neq}m$. According to the Fock's cyclic condition, the state $\tilde{\Psi}^{m,m,n}_{\nu=2/(m+n)}$ has definite spin $\mathbb{S}{=}\mathbb{S}^{z}$ if and only if the anti-symmetrization (symmetrization) of any $k^{\rm th}$ spin-$\uparrow$ particle with respect to all spin-$\downarrow$ particles (or vice-versa, since $N_{\uparrow}{=}N_{\downarrow}$) in the spatial wave function $\Psi^{m,m,n}_{\nu=2/(m+n)}$ vanishes, for odd (even) $m$, i.e.,
\begin{equation}
   \label{eq: Fock's_cyclic_condition}
    \Psi^{m,m,n}_{\nu=2/(m+n)} \pm \sum_{i{=}1}^{N_{\downarrow}} \Psi^{m,m,n}_{\nu=2/(m+n)}\left(k\leftrightarrow N_{\uparrow}+i\right)=0. 
\end{equation}
Here ${+}$ stands for even $m$ and ${-}$ for odd $m$. The Halperin-$(m,m,n)$ states, for $n{=}m{-}1$, satisfy Fock's cyclic condition and are therefore spin-singlet states, i.e, $\mathbb{S}{=}\mathbb{S}^{z}{=}0$. In particular, for the Halperin-$(1,1,0)$ state for small particle numbers, for which the wave functions are given in Eqs.~\eqref{eq: antisymmetrized_110_N_2} and~\eqref{eq: antisymmetrized_110_N_4}, the singlet property can be readily verified. In contrast, for $n{\neq}m{-}1$, the Halperin-$(m,m,n)$ states do not satisfy Fock's cyclic condition and are therefore not singlets; in fact, they are not definite spin eigenstates~\cite{Yoshioka89}. Note that in this paper, we have only considered the Halperin-$(m,m,n)$ states with $n{\leq}m$, since for $n{>}m$, the wave function describes a state with stronger inter-species repulsion than intra-species repulsion and will thus phase separate~\cite{Gail08, Simon25}.
 
We conclude this appendix by noting that the various physical quantities of interest, such as, the average energy of a spin-rotation and particle-index-permutation invariant interaction, the pair correlation function, or the structure factor, can be computed directly from $\Psi^{m,m,n}_{\nu=2/(m+n)}$, without explicitly constructing the fully exchange-symmetric wave function $\tilde{\Psi}^{m,m,n}_{\nu=2/(m+n)}$. For example, consider a spin-rotation and particle-index-permutation invariant interaction $V\left(\{\boldsymbol{\mathfrak{u}}_{i}\}\right)$. Its expectation value with respect to $\tilde{\Psi}^{m,m,n}_{\nu=2/(m+n)}\left(\{\boldsymbol{\mathfrak{u}}_{i}\}\right)$ is given by
\begin{align}
\label{eq: averge_energy_exchange_symmetric_WF}
    \left\langle V\right\rangle_{\tilde{\Psi}}&=\frac{\int d^{2}\boldsymbol{\mathfrak{u}}_{1}d^{2}\boldsymbol{\mathfrak{u}}_{2}{\cdots}d^{2}\boldsymbol{\mathfrak{u}}_{N} V\left(\{\boldsymbol{\mathfrak{u}}_{i}\}\right)\left|\tilde{\Psi}^{m,m,n}_{\nu=2/(m+n)}\left(\{\boldsymbol{\mathfrak{u}}_{i}\}\right)\right|^2}{\int d^{2}\boldsymbol{\mathfrak{u}}_{1}d^{2}\boldsymbol{\mathfrak{u}}_{2}{\cdots}d^{2}\boldsymbol{\mathfrak{u}}_{N}\left|\tilde{\Psi}^{m,m,n}_{\nu=2/(m+n)}\left(\{\boldsymbol{\mathfrak{u}}_{i}\}\right)\right|^2}.
\end{align}
Noting that $\tilde{\Psi}^{m,m,n}_{\nu=2/(m+n)}\left(\{\boldsymbol{\mathfrak{u}}_{i}\}\right)$ is a superposition of orthogonal spin configurations, each weighted by a particle-index-permuted $\Psi^{m,m,n}_{\nu=2/(m+n)}$ wave function, we obtain:
\begin{align}
  \left|\tilde{\Psi}^{m,m,n}_{\nu=2/(m+n)}\left(\{\boldsymbol{\mathfrak{u}}_{i}\}\right)\right|^2&=\left|\Psi^{m,m,n}_{\nu=2/(m+n)}\left(\{\boldsymbol{\mathfrak{u}}_{i}\}\right)\right|^2\\
  &+\sum_{P} \left|\Psi^{m,m,n}_{\nu=2/(m+n)}\left(\{P\boldsymbol{\mathfrak{u}}_{i}\}\right)\right|^2. \nonumber
\end{align}
Here, $P$ represents permutations of particle indices that differ from the base configuration in $\Psi^{m,m,n}_{\nu=2/(m+n)}\left(\{\boldsymbol{\mathfrak{u}}_{i}\}\right)$. Next, we substitute the above equation in Eq.~\eqref{eq: averge_energy_exchange_symmetric_WF}, and use the fact that we can redefine the integration variables, which are dummy variables since they are integrated over, to bring back $\Psi^{m,m,n}_{\nu=2/(m+n)}\left(\{P\boldsymbol{\mathfrak{u}}_{i}\}\right)$ to $\Psi^{m,m,n}_{\nu=2/(m+n)}\left(\{\boldsymbol{\mathfrak{u}}_{i}\}\right)$ as $V\left(\{\boldsymbol{\mathfrak{u}}_{i}\}\right)$ is permutation-invariant, i.e., $V\left(\{\boldsymbol{\mathfrak{u}}_{i}\}\right){=}V\left(\{P\boldsymbol{\mathfrak{u}}_{i}\}\right)$. Consequently, we obtain,
\begin{align}
\label{eq: averge_energy_without_exchange_symmetric_WF}
    \left\langle V\right\rangle_{\tilde{\Psi}} &=\frac{\int d^{2}\boldsymbol{\mathfrak{u}}_{1}d^{2}\boldsymbol{\mathfrak{u}}_{2}{\cdots}d^{2}\boldsymbol{\mathfrak{u}}_{N} V\left(\{\boldsymbol{\mathfrak{u}}_{i}\}\right)\left|\Psi^{m,m,n}_{\nu=2/(m+n)}\left(\{\boldsymbol{\mathfrak{u}}_{i}\}\right)\right|^2}{\int d^{2}\boldsymbol{\mathfrak{u}}_{1}d^{2}\boldsymbol{\mathfrak{u}}_{2}{\cdots}d^{2}\boldsymbol{\mathfrak{u}}_{N}\left|\Psi^{m,m,n}_{\nu=2/(m+n)}\left(\{\boldsymbol{\mathfrak{u}}_{i}\}\right)\right|^2}\nonumber\\
    &=\left\langle V\right\rangle_{\Psi}.
\end{align}
Therefore, the average energy is identical whether computed with $\tilde{\Psi}^{m,m,n}_{\nu=2/(m+n)}$ or with $\Psi^{m,m,n}_{\nu=2/(m+n)}$. A similar argument applies to the computation of the pair correlation function and the structure factor.

\section{Spin quantum numbers of spin-flip density-wave states in various polarized ground states}
\label{app: spin_quantum_number_spin_density_states}
In this appendix, we present a derivation to see whether different spin-flip density-wave states carry definite spin quantum numbers. For this purpose, it is useful to note the following commutators:
\begin{align}
\label{eq: S_square_rho_z_commutator}
    \left[\vec{\mathbb{S}}^{2},\bar{\rho}_{L, M}^{~z}\right]&=\bar{\rho}_{L, M}^{~-}\mathbb{S}^{+} - \bar{\rho}_{L, M}^{~+}\mathbb{S}^{-} +2\bar{\rho}_{L, M}^{~z},\\
    \label{eq: S_square_rho_plus_commutator}
    \left[\vec{\mathbb{S}}^{2},\bar{\rho}_{L, M}^{~+}\right]&=2\bar{\rho}_{L, M}^{~+}\mathbb{S}^{z} - 2\bar{\rho}_{L, M}^{~z}\mathbb{S}^{+}+2\bar{\rho}_{L, M}^{~+},\\
    \label{eq: S_square_rho_minus_commutator}
     \left[\vec{\mathbb{S}}^{2},\bar{\rho}_{L, M}^{~-}\right]&=2\bar{\rho}_{L, M}^{~z}\mathbb{S}^{-} - 2\bar{\rho}_{L, M}^{~-}\mathbb{S}^{z} +2\bar{\rho}_{L, M}^{~-}.
\end{align}
These commutators can be straightforwardly derived by considering corresponding unprojected density operators given in Eq.~\eqref{eq: first_quantized_angular_momentum_space_density_operator}. This follows from the fact that $\vec{\mathbb{S}}^{2}$ has no orbital part (equivalently, the orbital part is identity), consequently, it has the same commutation with both $\rho_{L, M}^{\alpha}$ and $\bar{\rho}_{L, M}^{~\alpha}$, where $\alpha{=}z,+,-$. One can further make use of the identities
\begin{align}
    \vec{\mathbb{S}}^{2}&=\sum_{j{=}1}^{N}\vec{s}_{j}^{~2}+\sum_{1\leq j<k\leq N}2~\vec{s}_{j}\cdot \vec{s}_{k},
\end{align}
and
\begin{align}
    \vec{s}_{j}\cdot \vec{s}_{k}&=\left(\frac{s_{j}^{+}+s_{j}^{-}}{2}\right)\left(\frac{s_{k}^{+}+s_{k}^{-}}{2}\right) \nonumber\\
    &+\left(\frac{s_{j}^{+}-s_{j}^{-}}{2\iota}\right)\left(\frac{s_{k}^{+}-s_{k}^{-}}{2\iota}\right) + s_{j}^{z}s_{k}^{z}.
\end{align}
It is also useful to note:
\begin{align}
\label{eq: rho_plus_minus_S_z}
\left[\bar{\rho}_{L, M}^{~\pm},\mathbb{S}^{z}\right]&=\mp\bar{\rho}_{L, M}^{~\pm},\\[6pt]
\label{eq: S_plus_minus_rho_minus_plus}
\left[\bar{\rho}_{L, M}^{~\pm},\mathbb{S}^{\mp}\right]&=\pm2\bar{\rho}_{L, M}^{~z},\\[6pt]
\label{eq: rho_z_S_plus_minus}
\left[\bar{\rho}_{L, M}^{~z},\mathbb{S}^{\pm}\right]&=\pm\bar{\rho}_{L, M}^{~\pm}.
\end{align}
These commutators can be derived from Eq.~\eqref{eq: spinful_operator_commutation_algebra} by noting that $\mathbb{S}^{\alpha}{=}\bar{\rho}_{0,0}^{\alpha}$. Alternatively, one can also evaluate these commutators from the corresponding unprojected density operators, as mentioned earlier.

\subsection{States in the maximal spin $\mathbb{S}{=}N/2$ multiplet}
\label{ssec: spin_states_in_the_maximal_spin_multiplet}

Here, we demonstrate that the spin-flip density waves obtained from the IQH states in the maximal spin $\mathbb{S}{=}N/2$ multiplet have a definite spin at all $L$, whereas for FQH states in the $\mathbb{S}{=}N/2$ multiplet, the spin-density waves possess a definite spin only at $L{=}1$.

Let us begin by considering the state $\left|\Psi_{\nu}^{N/2, N/2}\right\rangle$ with $\mathbb{S}{=}\mathbb{S}^{z}{=}N/2$. The action of $\bar{\rho}^{~z}_{L, M}$ on this state is identical to that of $\bar{\rho}^{~I}_{L, M}$, and thus leaves the spin of $\left|\Psi_{\nu}^{N/2, N/2}\right\rangle$ unaltered; moreover, it does not create a spin-flip density wave. In contrast, $\bar{\rho}^{~+}_{L, M}$ annihilates the ground state $\left|\Psi_{\nu}^{N/2, N/2}\right\rangle$. Only non-trivial spin-flip density wave is created by $\bar{\rho}^{~-}_{L, M}$. To infer the spin of the resulting state, we act with the operator $\vec{\mathbb{S}}^{2}$ to obtain 
\begin{align}
\label{eq: spin_FP_N_2_N_2}
  \vec{\mathbb{S}}^{2} \bar{\rho}^{~-}_{L, M} \left|\Psi_{\nu}^{N/2, N/2}\right\rangle&= S_{0}(S_0-1)\bar{\rho}^{~-}_{L, M}\left|\Psi_{\nu}^{N/2, N/2}\right\rangle\nonumber\\
 & +2\mathbb{S}^{-}\bar{\rho}_{L, M}^{~z}\left|\Psi_{\nu}^{N/2, N/2}\right\rangle,
\end{align}
where $\mathbb{S}{=}S_0{=}N/2$. In obtaining the above equation, we have used Eqs.~\eqref{eq: S_square_rho_minus_commutator} and~\eqref{eq: rho_z_S_plus_minus}. At $\nu{=}1$, since $\bar{\rho}_{L, M}^{~z}\left|\Psi_{\nu{=}1}^{N/2, N/2}\right\rangle{=}0$ for $L{\geq}1$, $\bar{\rho}^{~-}_{L, M}$ generates a spin-flip density wave with definite spin $S_0{-}1$. On the contrary, for FQH states, $\bar{\rho}_{L, M}^{~z}\left|\Psi_{\nu}^{N/2, N/2}\right\rangle{=}0$ only at $L{=}1$, and therefore, the corresponding spin-flip density wave excitation does not have a definite spin except at $L{=}1$. 

Next, we consider other states in the maximal spin multiplet with $\mathbb{S}^{z}{<}N/2$, i.e., states like $\left|\Psi_{\nu}^{N/2, N/2-\mathfrak{n}}\right\rangle{=}\left(\mathbb{S}^{-}\right)^{\mathfrak{n}}\left|\Psi_{\nu}^{N/2, N/2}\right\rangle$, where $\mathfrak{n}{=}1,2,{\cdots}, N$. To infer the spin quantum number of these spin-flip density wave states, we note the following identities, which can be derived from Eqs.~\eqref{eq: S_plus_minus_rho_minus_plus} and~\eqref{eq: rho_z_S_plus_minus}:
\begin{align}
 \label{eq: rho_z_S_minus_power_n}
    \bar{\rho}^{~z}_{L, M}\left(\mathbb{S}^{-}\right)^{\mathfrak{n}}&=\left(\mathbb{S}^{-}\right)^{\mathfrak{n}}\bar{\rho}^{~z}_{L, M}-\mathfrak{n}\left(\mathbb{S}^{-}\right)^{\mathfrak{n}-1}\bar{\rho}^{~-}_{L, M},\\[6pt]
\label{eq: rho_plus_S_minus_power_n}
    \bar{\rho}^{~+}_{L, M}\left(\mathbb{S}^{-}\right)^{\mathfrak{n}}&=\left(\mathbb{S}^{-}\right)^{\mathfrak{n}}\bar{\rho}^{~+}_{L, M} + 2\mathfrak{n}\left(\mathbb{S}^{-}\right)^{\mathfrak{n}-1}\bar{\rho}^{~z}_{L, M}\nonumber\\
    &+\mathfrak{n}\left(\mathfrak{n}{-}1\right)\left(\mathbb{S}^{-}\right)^{\mathfrak{n}-2}\bar{\rho}^{~-}_{L, M},\\[6pt]
    \bar{\rho}^{~-}_{L, M}\left(\mathbb{S}^{-}\right)^{\mathfrak{n}}&=\left(\mathbb{S}^{-}\right)^{\mathfrak{n}}\bar{\rho}^{~-}_{L, M}.
\end{align}
From the above equations, it follows that, at $\nu{=}1$, $\bar{\rho}_{L{\geq}1, M}^{\alpha}\left|\Psi_{\nu{=}1}^{N/2, N/2-\mathfrak{n}}\right\rangle {\propto}\bar{\rho}_{L{\geq}1, M}^{-}\left|\Psi_{\nu{=}1}^{N/2, N/2}\right\rangle$, and therefore $\bar{\rho}_{L{\geq}1, M}^{~\alpha}\left|\Psi_{\nu{=}1}^{N/2, N/2-\mathfrak{n}}\right\rangle$ has a definite spin $S_0{-}1$. Similarly, for FQH states, only $\bar{\rho}_{L{=}1,M}^{~\alpha}\left|\Psi_{\nu{=}1}^{N/2, N/2-\mathfrak{n}}\right\rangle$ has definite spin $S_{0}{-}1$, whereas $\bar{\rho}_{L{>}1,M}^{~\alpha}$ do not create a definite spin state. Note that for $\alpha{=}+$, and $\mathfrak{n}{=}0,1$, $\bar{\rho}_{L{\geq}1, M}^{~\alpha}\left|\Psi_{\nu{=}1}^{N/2, N/2-\mathfrak{n}}\right\rangle{=}0$ and $\bar{\rho}_{L{=}1,M}^{~\alpha}\left|\Psi_{\nu{=}1}^{N/2, N/2-\mathfrak{n}}\right\rangle{=}0$, as is evident from Eq.~\eqref{eq: rho_plus_S_minus_power_n}. 

\subsection{Spin-singlet states}
\label{ssec: spin_of_density_wave_singlet}
For a spin-singlet state $\left|\Psi_{0}^{\mathbb{S}{=}0}\right\rangle$, it follows straightforwardly from Eqs.~\eqref{eq: S_square_rho_z_commutator}-\eqref{eq: S_square_rho_minus_commutator} that its spin-flip density-wave states have definite spin $\mathbb{S}{=}1$, i.e.,
\begin{align}
   \vec{\mathbb{S}}^{2} \bar{\rho}_{L, M}^{~\alpha}\left|\Psi_{0}^{\mathbb{S}{=}0}\right\rangle&=2\bar{\rho}_{L, M}^{~\alpha}\left|\Psi_{0}^{\mathbb{S}{=}0}\right\rangle,
\end{align}
where we have used the fact $\mathbb{S}^{\alpha}\left|\Psi_{0}^{\mathbb{S}{=}0}\right\rangle{=}0$.

\subsection{Partially polarized states}
\label{ssec: spin_density_wave_PP}
Similar to the fully polarized and singlet states, to infer the spin of the density waves in a PP state $\left|\Psi_{0}^{\rm PP}\right\rangle$ with $\mathbb{S}{=}S_0{<}N/2$ and $\mathbb{S}^{z}{=}S_0$, we act the $\vec{\mathbb{S}}^{2}$ operator on the resulting density-wave states. Consequently, one obtains
\begin{align}
\label{eq: spin_PP_rho_z_S_0_S_0}
  \vec{\mathbb{S}}^{2} \bar{\rho}^{~z}_{L, M} \left|\Psi_{0}^{\rm PP}\right\rangle&= S_{0}(S_0+1)\bar{\rho}^{~z}_{L, M}\left|\Psi_{0}^{\rm PP}\right\rangle\nonumber\\
 & -\mathbb{S}^{-}\bar{\rho}_{L, M}^{~+}\left|\Psi_{0}^{\rm PP}\right\rangle,\\[6pt]
 \label{eq: spin_PP_rho_minus_S_0_S_0}
  \vec{\mathbb{S}}^{2} \bar{\rho}^{~-}_{L, M} \left|\Psi_{0}^{\rm PP}\right\rangle&= (S_{0}-1)S_0\bar{\rho}^{~-}_{L, M}\left|\Psi_{0}^{\rm PP}\right\rangle\nonumber\\
 & +2\mathbb{S}^{-}\bar{\rho}_{L, M}^{~z}\left|\Psi_{0}^{\rm PP}\right\rangle,\\[6pt]
 \label{eq: spin_PP_rho_plus_S_0_S_0}
  \vec{\mathbb{S}}^{2} \bar{\rho}^{~+}_{L, M} \left|\Psi_{0}^{\rm PP}\right\rangle&= (S_{0}+1)(S_0+2)\bar{\rho}^{~+}_{L, M}\left|\Psi_{0}^{\rm PP}\right\rangle.
\end{align}
Since $\bar{\rho}_{L, M}^{~+}\left|\Psi_{0}^{\rm PP}\right\rangle$ and $\bar{\rho}_{L, M}^{~z}\left|\Psi_{0}^{\rm PP}\right\rangle$ with $L{\geq}1$ do not vanish for a PP ground state, it follows that $\bar{\rho}^{~z}_{L, M}$ and $ \bar{\rho}^{~-}_{L, M}$ do not produce states with a definite spin.  Interestingly, $\bar{\rho}^{+}_{L{\geq}1, M}$ does yield a definite spin–density-wave state with spin $S_0{+}1$ when acting on a PP ground state $\left|\Psi_{0}^{\rm PP}\right\rangle$ with $\mathbb{S}^{z}{=}S_0$. However, for $\mathbb{S}^{z}{<}S_0$, the action of $\bar{\rho}^{~+}_{L{\geq}1, M}$ does not produce a definite spin state. This follows analogously from the previous discussion in App.~\ref{ssec: spin_states_in_the_maximal_spin_multiplet}, particularly from Eq.~\eqref{eq: rho_plus_S_minus_power_n}.

\section{Mapping the spherical total orbital angular momentum $L$ to the planar momentum $q$}
\label{app: L_to_q}
To map the the spherical orbital angular momentum $L$ to the planar momentum $q$, we consider the total unprojected structure factor $S^{I}\left(L\right)$ on the sphere for the $\nu{=}1$ state and consider its  $Q{\rightarrow\infty}$ limit, and compare it with its planar counterpart $S^{I}\left(q\right){=}1{-}e^{{-}q^2/2}$. The unprojected structure factor $S^{I}\left(L\right)$ is given by~\cite{He94, Dora24} 
\begin{align}
\label{eq: limit_S(L)_derivation_0}
S^{I}\left(L\right)&=1-(2Q+1)~\left(\begin{array}{ccc}
Q & Q & L \\
-Q & Q & 0
\end{array}\right)^2 .
\end{align}
The above equation is obtained from Eq.~\eqref{eq: relaton_unprojected_projected_structure_factor_fully_polarized_states} by noting that $\bar{S}^{I}\left(L{>0}\right){=}0$ for the $\nu{=}1$ state. For convenience, we have redefined $S^{I}\left(0\right){=}N$, which is constant for a given system, to be $S^{I}\left(0\right){=}0$. Next, we rewrite $S^{I}\left(L\right)$ as follows
\begin{align}
   \label{eq: limit_S(L)_derivation_1}
S^{I}\left(L\right)&=1-(2Q+1)\frac{((2Q)!)^{2}}{(2Q-L)! (2Q+1+L)!}\nonumber\\
&=1-(2Q+1)e^{\ln W\left(Q,L\right)},
\end{align}
where
\begin{align}
\label{eq: limit_S(L)_derivation_2}
   W\left(Q,L\right)&=\frac{((2Q)!)^{2}}{(2Q-L)! (2Q+1+L)!}.
\end{align}
To obtain the $Q{\rightarrow}\infty$ planar limit of $\ln W\left(Q,L\right)$ and hence of $S^{I}\left(L\right)$, we use Stirling's approximation, i.e., $\lim_{n{\to}\infty}\ln\left(n!\right){\approx}n\ln n{-}n$. In other words,
\begin{align}
 \label{eq: limit_S(L)_derivation_3}   
 \lim_{Q\rightarrow\infty}\ln W\left(Q,L\right)&\approx 4Q\ln(2Q)-4Q  \\ \nonumber 
 &-(2Q-L)\ln(2Q-L)+(2Q-L) \\ \nonumber 
 &-(2Q+1+L)\ln(2Q+1+L)  \\ \nonumber 
 &+(2Q+1+L).
\end{align}
Next, we rearrange the terms and express the above equation as a function of $L/(2Q)$ to obtain
\begin{align}
    \label{eq: limit_S(L)_derivation_4} 
   \lim_{Q\rightarrow\infty}\ln W\left(Q,L\right)&\approx -2Q \ln\left(1-\frac{L}{2Q}\right)-2Q\ln\left(1+\frac{L+1}{2Q}\right) \nonumber\\ 
   &+ L\ln\left(1-\frac{L}{2Q}\right)-L\ln\left(1+\frac{L+1}{2Q}\right)   \nonumber\\
   &-\ln\left(2Q+1\right)-\ln\left(1+\frac{L}{2Q+1}\right)+1.  \nonumber
\end{align}
In the the above equation, we further approximate $\ln\left(1{+}L/\left(2Q{+}1\right)\right){\approx}\ln\left(1{+}L/\left(2Q\right)\right)$. As $L/\left(2Q\right){\ll}1$ [since for a fixed $L$ we take the $Q{\rightarrow}\infty$ limit], we can use the Taylor expansion of $\ln\left(1{+}x\right)$, i.e.,
\begin{align}
    \ln\left(1+x\right)&=x-\frac{x^2}{2}+\frac{x^{3}}{3}+\cdots, ~~\text{where}~~x\ll 1.
\end{align}
Consequently, the first leading order term in Eq.~\eqref{eq: limit_S(L)_derivation_4} is given by
\begin{align}
\label{eq: limit_S(L)_derivation_5} 
 \lim_{Q\rightarrow\infty}\ln W\left(Q,L\right)&\approx -2Q\left(-\frac{L}{2Q}-\frac{1}{2}\frac{L^2}{(2Q)^2}\right)  \\
 &-2Q \left(\frac{L+1}{2Q}-\frac{1}{2}\frac{(L+1)^2}{(2Q)^2}\right)+L\left(-\frac{L}{2Q}\right) \nonumber\\
 &-L\left(\frac{L+1}{2Q}\right)-\frac{L}{2Q}+1-\ln\left(2Q+1\right)  \nonumber\\
 &+\mathcal{O}\left((L/2Q)^2\right). \nonumber
\end{align}
The above equation simplifies to
\begin{align}
    \label{eq: limit_S(L)_derivation_6} 
    \lim_{Q\rightarrow\infty}\ln W\left(Q,L\right)&\approx-\frac{L(L+1)}{2Q}-\ln\left(2Q+1\right) \nonumber \\
    &+\mathcal{O}\left((L/2Q)^2\right) \nonumber \\
    \implies  \lim_{Q\rightarrow\infty}W\left(Q,L\right)&\approx\frac{e^{-\frac{L(L+1)}{2Q}+\mathcal{O}\left((L/2Q)^2\right)}}{2Q+1} \nonumber \\
    &\approx\frac{e^{-\frac{L(L+1)}{2Q}}}{2Q+1}.
\end{align}
Substituting Eq.~\eqref{eq: limit_S(L)_derivation_6} into Eq.~\eqref{eq: limit_S(L)_derivation_1}, one obtains
\begin{equation}
\lim_{Q\rightarrow\infty}S^{I}\left(L\right)=1-e^{-\frac{L(L+1)}{2Q}}.    
\end{equation}
After restoring the magnetic length $\ell$, this suggests that, unlike the usual definition of the planar momentum, $q{=}L/\sqrt{Q}\ell$~\cite{Haldane85a} (which we had also used in a previous work of ours~\cite{Dora24}), one should identify the linear momentum $q$ with $q\ell{=}\sqrt{L(L{+}1)/Q}$ (see also Refs.~\cite{He94, Anakru25}), for any finite wave number since $L{\ll}2Q$ for fixed $L$ and $Q{\to}\infty$, and in particular, in the long wavelength limit, since with that, the above equation correctly reproduces the planar result~\cite{Girvin86}
\begin{align}
   S^{I}\left(q\right)&=1-e^{-(q\ell)^2/2}. 
\end{align}

\begin{figure}[tbh!]
        \includegraphics[width=0.99\columnwidth]{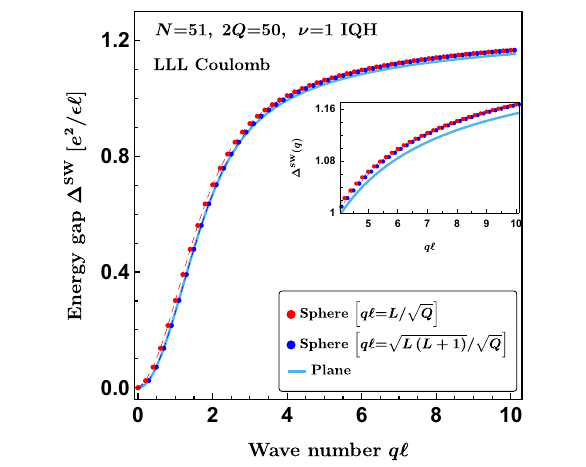}
          \caption{Comparison of the $N{=}51$-electron spherical and thermodynamic planar spin-wave gaps in the $\nu{=}1$ integer quantum Hall state. The spherical angular momentum $L$ is mapped to the planar momentum $q$ in two different ways: $(i)~q\ell{=}L/\sqrt{Q}$ [red dots], and $(ii)~q\ell{=}\sqrt{L(L{+}1)}/\sqrt{Q}$ [blue dots].}
          \label{fig: SF_dispersion_IQH_state_N_51}
\end{figure}
In Fig.~\ref{fig: SF_dispersion_IQH_state_N_51}, we show the SW dispersion for the $\nu{=}1$ IQH state for a system of $N{=}51$ electrons [in the main text, an analogous plot for a larger system of $N{=}300$ electrons is shown in Fig.~\ref{fig: SF_dispersion_IQH_state}] with both the conventional $q\ell{=}L/\sqrt{Q}$~\cite{Haldane85a} and the refined $q\ell{=}\sqrt{L(L{+}1)}/\sqrt{Q}$ mappings. For comparison, we have also shown the planar SW gap. We find that the refined mapping gives better agreement with the planar gap compared to the conventional one at small wave numbers. Since we are primarily interested in density modes, which, if accurate, are only so in the long-wavelength limit, we will use the refined mapping throughout this work across all fillings, although, strictly speaking, the refined mapping was derived specifically in the context of $\nu{=}1$. The spherical gaps deviate from the planar results at intermediate to large $q\ell$ since here the finite-size curvature effects of the sphere become important and the mapping $q\ell{=}\sqrt{L(L{+}1)}/\sqrt{Q}$ becomes less accurate.  

\section{Spin-wave dispersion for model interactions in planar and sphere geometry at $\nu{=}1$}
\label{app: SF_mode_model_interactions_IQH_state}
In this section, we further validate the identification $q\ell{=}\sqrt{L(L{+}1)/Q}$ that maps the spherical angular momentum $L$ to the planar momentum $q$. Specifically, we demonstrate that the SW gap for the $\mathcal{R}{-}$ranged TK interaction on the sphere, given by $v^{(\mathcal{R}-TK)}(|\boldsymbol{\Omega} {-}\boldsymbol{\Omega^{'}}|){=}4\pi(\nabla^{2}_{\boldsymbol{\Omega}})^{\mathcal{R}}~\delta(\boldsymbol{\Omega} {-}\boldsymbol{\Omega^{'}})$, approaches the SW gap of the planar TK interaction $v^{(\mathcal{R}-TK)}(|\boldsymbol{r} {-}\boldsymbol{r^{'}}|){=}4\pi(\nabla^{2}_{\boldsymbol{r}})^{\mathcal{R}}~\delta(\boldsymbol{r} {-}\boldsymbol{r^{'}})$ for the $\nu{=}1$ state as $Q{\to}\infty$, under the identification that $q\ell{=}\sqrt{L(L{+}1)/Q}$ for $L{\ll}2Q$. For convenience, we set $\ell{=}1$ in the following discussion.

The planar SW gap for the TK-interaction obtained from Eq.~\eqref{eq: planar_general_SF_gap_equation} reads as
\begin{align}
\label{planar_SF_gap_model_interaction}
\Delta^{\rm SW,\left(\mathcal{R}-\rm TK\right)}_{\nu{=}1}\left(q\right)&=\frac{1}{2\pi}\int_{0}^{\infty} dk ~k ~e^{-k^2/2}\left[4\pi \left(-k^2\right)^{\mathcal{R}}\right]\nonumber\\
&~~~~~~~~~~~~~~~~~~~~~\times\left[1-J_{0}\left(qk\right)\right] \nonumber\\ 
&=-2^{1+\mathcal{R}} ~\Gamma\left(1+\mathcal{R}\right)~L_{-1-\mathcal{R}}\left(-\frac{q^2}{2}\right).
\end{align}
Here, $J_{0}(x)$ is the zeroth-order Bessel function and $L_{n}(x)$ is the $n^{\rm th}$ order Laguerre polynomial. In Eq.~\eqref{planar_SF_gap_model_interaction}, we have used the Fourier component of the TK-interaction, $v^{\mathcal{R}-\rm TK}\left(k\right){=}4\pi \left(-k^2\right)^{\mathcal{R}}$. The following simplified expressions of the SW gap for the cases when $\mathcal{R}{=}0,1,2$, which correspond to $V_0^{0-\rm TK}, V_1^{1-\rm TK}, V_2^{2-\rm TK}$ planar Haldane pseudopotentials, respectively, are noteworthy:  
\begin{align}
\label{eq: simplified_expression_SF_gap_V0_V1_V2_pps}
\Delta^{\rm SW,\left(0-\rm TK\right)}_{\nu{=}1}\left(q\right)&=2\left(1-e^{-q^2/2}\right),\nonumber\\ 
\Delta^{\rm SW,\left(1-\rm TK\right)}_{\nu{=}1}\left(q\right)&=-4+\left(4-2q^2\right)e^{-q^2/2},\nonumber\\
\Delta^{\rm SW,\left(2-\rm TK\right)}_{\nu{=}1}\left(q\right)&=16-\left(16-16q^2+2q^4\right)e^{-q^2/2}.
\end{align}
Next, we compute the SW gap on the sphere for range-$\mathcal{R}{\ll}2Q$, and demonstrate that, in the thermodynamic limit, these match the corresponding planar gaps at small wave numbers. We illustrate this for the  $v^{(2-TK)}(|\boldsymbol{\Omega} {-}\boldsymbol{\Omega^{'}}|)$ interaction; results for other values of $\mathcal{R}$ follow in a similar manner. For simplicity, we follow Eq.~\eqref{eq: SF_gap_Nakajima_Aoki_IQH_state} to determine the SW gap for the $v^{(2-TK)}(|\boldsymbol{\Omega} {-}\boldsymbol{\Omega^{'}}|)$ interaction, which requires the knowledge of its pseudopotentials. These pseudopotentials are as follows
\begin{align}
\label{eq: Haldane_pps_TK_2}
    V_{0}^{2-\rm TK}&=\frac{-2(2Q+1)^2}{\pi(1-16Q^2)}, \nonumber\\ 
    V_{1}^{2-\rm TK}&=\frac{(2Q+1)^2}{\pi Q(1-4Q)},    \nonumber\\ 
    V_{2}^{2-\rm TK}&=\frac{(2Q-1)(2Q+1)^2}{\pi Q(-3+4Q)(-1+4Q)}, \nonumber\\ 
    V_{\mathfrak{m}}^{2-\rm TK}&=0 ~\forall~\mathfrak{m}>2.
\end{align}
From Eq.~\eqref{eq: SF_gap_Nakajima_Aoki_IQH_state}, one computes the SW gap of the range-$2$ TK interaction as
\begin{align}
\label{eq: TK_2_SF_gap_NAkajima_Aoki_IQH_state}
    \Delta^{\rm SW, (\rm 2-\rm TK)}_{\nu{=}1}\left(L\right)&=\sum_{\mathfrak{m}{=}0}^{2}\left(2\left(2Q-\mathfrak{m}\right)+1\right)(-1)^{\mathfrak{m}}  \\ 
    &\times V_{\mathfrak{m}}
    \left[\frac{1}{2Q+1}-(-1)^{\mathfrak{m}}\left\{\begin{array}{lll}
Q & Q & ~~~~L \\
Q & Q & 2Q-\mathfrak{m}
\end{array}\right\}\right].\nonumber
\end{align}
To obtain the SW gap in the thermodynamic limit, it is helpful to note the following explicit expressions of the Wigner $6j$ symbols appearing in the above Eq.~\eqref{eq: TK_2_SF_gap_NAkajima_Aoki_IQH_state} for $\mathfrak{m}{=}0,1,2$:
\begin{align}
\label{eq: Wigner_6_j_symbol_m_0}
    \left\{\begin{array}{lll}
Q & Q & L \\
Q & Q & 2Q
\end{array}\right\}&=W\left(Q,L\right),\\
\label{eq: Wigner_6_j_symbol_m_1}
\left\{\begin{array}{lll}
Q & Q & ~~~L \\
Q & Q & 2Q-1
\end{array}\right\}&=W\left(Q,L\right)\left[\frac{L(L+1)}{Q}-1\right],\\
\label{eq: Wigner_6_j_symbol_m_2}
\left\{\begin{array}{lll}
Q & Q & ~~~L \\
Q & Q & 2Q-2
\end{array}\right\}&=W\left(Q,L\right)\Bigg[\frac{(L(L+1))^2}{4Q^2}-\frac{L(L+1)}{Q} \nonumber \\
&{+} \frac{(L(L{+}1))^2}{2Q(2Q{-}1)}-2\frac{L(L{+}1)}{2Q{-}1}+1\Bigg].  
\end{align}
For a definition of $W\left(Q,L\right)$, see Eq.~\eqref{eq: limit_S(L)_derivation_2}. Next, one can substitute Eq.~\eqref{eq: Haldane_pps_TK_2}, along with Eqs.~\eqref{eq: Wigner_6_j_symbol_m_0}-\eqref{eq: Wigner_6_j_symbol_m_2} in Eq.~\eqref{eq: TK_2_SF_gap_NAkajima_Aoki_IQH_state}, and identify $q{=}\sqrt{L(L{+}1)/Q}$, and then take the limit $Q{\to}\infty$. In doing so, one can make use of the $Q{\to}\infty$ limit of $W\left(Q,L\right)$ as given in Eq.~\eqref{eq: limit_S(L)_derivation_6}. Consequently one obtains 
\begin{align}
    \lim_{Q\to\infty} \Delta^{\rm SW, (\rm 2-\rm TK)}_{\nu{=}1}\left(L\right)&=16-\left(16-16q^2+2q^4\right)e^{-q^2/2},
\end{align}
which is identical to the planar result presented in Eq.~\eqref{eq: simplified_expression_SF_gap_V0_V1_V2_pps}. Interestingly, although the pseudopotentials of the spherical $v^{(2-TK)}(|\boldsymbol{\Omega} {-}\boldsymbol{\Omega^{'}}|)$ interaction do not map onto those of the planar interaction as $Q{\to}\infty$~\cite{Dora24}, the $\nu{=}1$ SW gap in the small wave number limit nevertheless agrees on the two geometries.

\section{Projected structure factor from pair-correlation function}
\label{app: gr_SL}
To obtain the planar GMP gaps, we first compute the pair-correlation function for the states of our interest using their trial wave function for large systems on the sphere. Assuming that these systems are large enough that they are representative of the thermodynamic limit, we switch to the planar geometry and fit this pair-correlation function to a particular form that is constrained by the topological quantum numbers of the state, as explained in Refs.~\cite{Girvin86, Fulsebakke23, Dora24}, and then Fourier transform it to obtain the unprojected static structure factor on the plane from it. The planar projected structure factor $\bar{S}^{(\rm p)}(q)$ is related it is unprojected version, $S^{(\rm p)}(q)$, as $\bar{S}^{(\rm p)}(q){=}S^{(\rm p)}(q){-}\left(1-e^{-(q\ell)^2/2}\right)$~\cite{Girvin86}.

The constraints on the pair-correlation function are imposed by the leading coefficients in the long-wavelength expansion of the static structure factor, which for many FQH states can be related to its topological quantum numbers~\cite{Kalinay00, Can14, Gromov15, Nguyen17, Dwivedi19}. The topological properties of Abelian FQH states, which are what we will consider here, can be encoded entirely in their $K$-matrix, charge vector $\vec{t}$, and spin-vector $\vec{\mathfrak{s}}$~\cite{Wen95}. These determine the topological numbers as follows: the filling factor $\nu{=}\vec{t}^{\rm T}{\cdot}K^{{-}1}{\cdot}\vec{t}$, Wen-Zee shift~\cite{Wen92} $\mathcal{S}{=}(2/\nu)\vec{t}^{\rm T}{\cdot}K^{{-}1}{\cdot}\vec{\mathfrak{s}}$, and the chiral central charge $c_{-}$ is the number of positive minus the number of negative eigenvalues of the $K$ matrix. The orbital spin variance $\beta{=}\nu^{s}{-}\nu(\mathcal{S}/2)^{2}$, where $\nu^{s}{=}\vec{\mathfrak{s}}^{\rm T}{\cdot}K^{{-}1}{\cdot}\vec{\mathfrak{s}}$ is the ``spin filling fraction"~\cite{Wen92}. In this appendix, we limit our discussion to the fully polarized FQH states. Topological quantum field theories predict that the leading coefficients in the expansion of the structure factor $S(q){=}S_{2}q^{2}{+}S_{4}q^{4}{+}S_{6}q^{6}{+}{\cdots}$ for chiral FQH states and their hole-conjugate states are related to these quantities as~\cite{Gromov15, Nguyen17} as $S_{2}{=}1/2$, $S_{4}{=}(\mathcal{S}{-}2)/8$ and $S_{6}{=}{-}\left[b/(8\nu){-}(\mathcal{S}{-}2)/16\right]$, where $b{=}\nu \mathcal{S}/2 (\mathcal{S}/2{-}1){+}\tilde{c}/12$, where $\tilde{c}{=}c_{-}{-}12\beta$. For Laughlin states, these relations were derived in Ref.~\cite{Can14}, and these were conjectured to hold for Jain states, and more generally for chiral states in Refs.~\cite{Gromov15, Nguyen17}. In Sec.~\ref{ssec: eft_Jain}, we present these topological quantum numbers for the Jain states. Then, in Sec.~\ref{app: 4_11_and_4_13_and_6_17}, we present the GMP gaps for certain non-Jain, but Abelian FQH states, updating and improving upon some previous results. We wrap this appendix in Sec.~\ref{app: gr_SL_sphere} by providing a relation between the pair correlation and structure factor in the spherical geometry. 

\subsection{Effective field theory of Jain states}
\label{ssec: eft_Jain}
For the fully polarized Jain states at $\nu{=}n/(2pn{+}\eta)$, where $\eta{=}{\pm}1$ (with $\eta{=}{+}1$ for parallel-vortex attachment and $\eta{=}{-}1$ for reverse-vortex attachment), the $K$-matrix, charge vector $\vec{t}$ and spin-vector $\vec{\mathfrak{s}}$ are given as
\begin{eqnarray}
    \label{eq: eft_Jain}
    K&=&2p \mathbb{C}_{n} + \eta \mathbb{I}_{n}, \\
    \vec{t}[i]&=&1,~i=1,2,\cdots,n,  \\
    \vec{\mathfrak{s}}[i]&=&(2p-\eta)/2+\eta i,~i=1,2,\cdots,n,
\end{eqnarray}
where $\mathbb{C}_{n}$ is the $n{\times}n$ matrix of all ones that represents vortex-attachment and $\mathbb{I}_{n}$ is the $n{\times}n$ identity matrix that represents the IQH state. From these, we get the Wen-Zee shift, $\mathcal{S}{=}ln{+}2p$, chiral central charge, $c_{-}{=}1{+}\eta(n{-}1)$ and orbital spin variance $\beta{=}\eta n(n^2{-}1)/12$~\cite{Gromov15, Nguyen17}. 

We note here that using the $SL(n,\mathbb{Z})$ transformation $W{=}\mathbb{I}_{n}{-}^{({-}1)}\mathbb{D}_{n}$ (or its transpose $W^{T}$), where $^{({-}1)}\mathbb{D}_{n}$ is the $n{\times}n$ matrix with all entries in the sub-diagonal (diagonal just below the main diagonal) equal to $1$ and the rest $0$, we can transform the above $K$-matrix, charge vector and spin-vector as $\tilde{K}{=}W{\cdot}K{\cdot}W^{T}$, $\tilde{\vec{t}}{=}W{\cdot}\vec{t}$, and $\tilde{\vec{\mathfrak{s}}}{=}W{\cdot}\vec{\mathfrak{s}}$ to obtain a new set of $K$-matrix $\tilde{K}$, charge vector $\tilde{\vec{t}}$ and spin-vector $\tilde{\vec{\mathfrak{s}}}$ which are equivalent to those given in Eq.~\eqref{eq: eft_Jain}, and have also been used extensively in the literature~\cite{Wen95}.

\subsection{Anomalously low magnetoroton modes at $4/11$ and $4/13$}
\label{app: 4_11_and_4_13_and_6_17}

\begin{figure*}[tbh]
        \includegraphics[width=1\columnwidth]{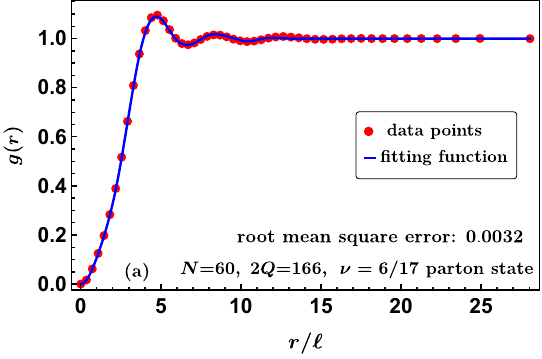}
        \includegraphics[width=1\columnwidth]{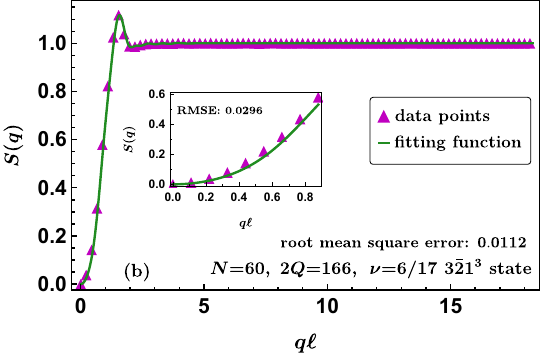}\\
        \includegraphics[width=1\columnwidth]{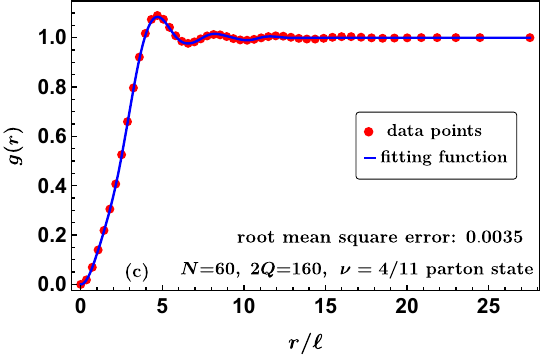}
        \includegraphics[width=1\columnwidth]{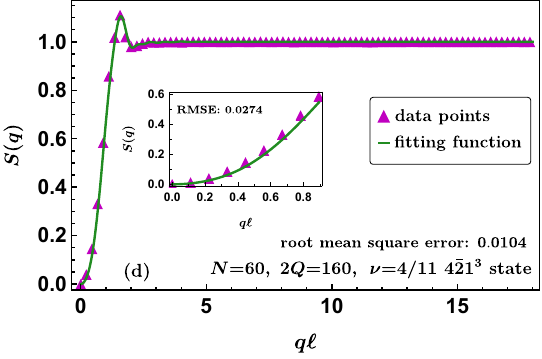} \\
        \includegraphics[width=1\columnwidth]{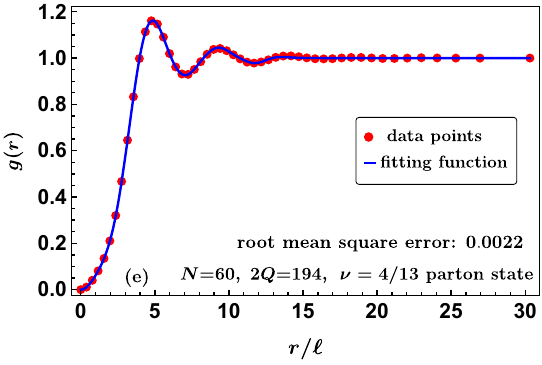}
        \includegraphics[width=1\columnwidth]{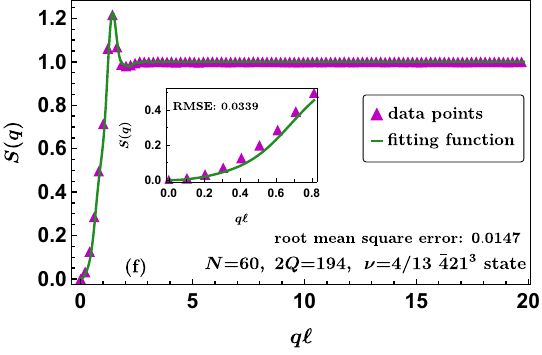} \\
          \caption{Comparison of the fitted and computed pair-correlation function $g(r)$ and the unprojected static structure factor $S(q)$ for the $3\bar{2}1^{3}$ [top panels], $4\bar{2}1^{3}$ [middle panels], and $\bar{4}21^{3}$ [bottom panels] parton fractional quantum Hall states. The fitting is done for the planar geometry, while the actual computations are done on the spherical geometry. }
          \label{fig: comparision_fitted_computed_gr_Sq}
\end{figure*}

    \begin{figure}[tbh]
        \includegraphics[width=0.48\textwidth]{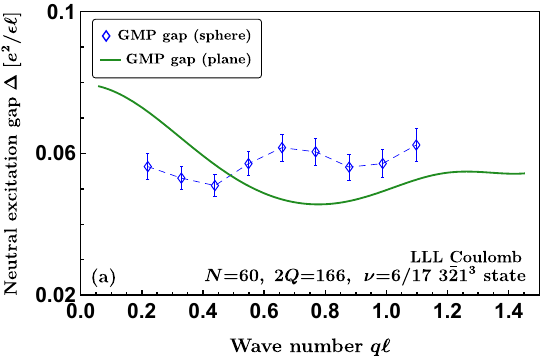}\\
        \includegraphics[width=0.48\textwidth]{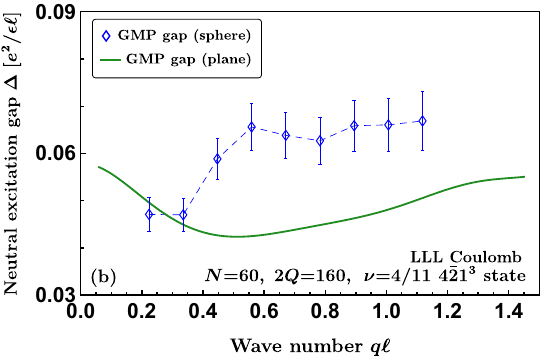}\\
        \includegraphics[width=0.48\textwidth]{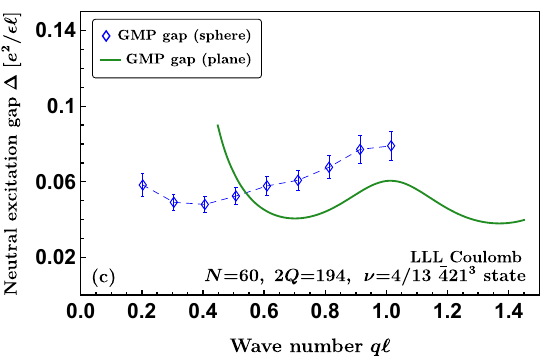}\\
          \caption{LLL Coulomb GMP gap computed on planar and spherical geometries for the $3\bar{2}1^{3}$ [top panel], $4\bar{2}1^{3}$ [middle panel], and $\bar{4}21^{3}$ [bottom panel] parton fractional quantum Hall states. 
          }
          \label{fig: planar_sphere_GMP_gap_parton_states}
\end{figure}

Refs.~\cite{Mukherjee15, Das20} considered the unconventional FQH states at $\nu{=}4/11$ and $\nu{=}4/13$, and found that the GMP mode for them has an anomalously low roton. The results of Refs.~\cite{Mukherjee15, Das20} were obtained by assuming that as $q{\to}0$, the projected static structure factor $\bar{S}(q)$ goes as $(q\ell)^{4}(1{-}\nu)/(8\nu)$. However, as we recently found~\cite {Dora24}, aside from Laughlin states, the aforementioned dependence of $\bar{S}(q)$ does not capture the correct behavior in other FQH states. Instead, this $\bar{S}_{4}$ coefficient is related not just to the filling factor $\nu$ but also to the Wen-Zee shift $\mathcal{S}$~\cite{Wen92, Nguyen17}. Expanding up to $(q\ell)^{6}$, and using the $4\bar{2}1^{3}$ parton state at $4/11$~\cite{Balram21c} and $\bar{4}21^{3}$ parton state at $4/13$~\cite{Dora22}, which likely lie in the same universality class as the $4/11$ and $4/13$ states considered in Ref.~\cite{Mukherjee15}, we evaluate the GMP gaps. The topological quantum numbers of the various parton states considered in this work, which are obtained from their $K$-matrix, charge, and spin vectors~\cite{Balram21a, Balram21c, Dora22}, are tabulated in Table~\ref{tab: parton_state_properties}. In particular, the quantum number $\beta$, which is also equal to the variance in orbital spin, is $\beta{=}\vec{\mathfrak{s}}^{T}{\cdot} K^{-1} {\cdot} \vec{\mathfrak{s}}{-}(\nu/4)\mathcal{S}^{2}$~\cite{Gromov15} (We note that Ref.~\cite{Nguyen17} defines the orbital spin variance as $\beta/\nu$.). We note here that the parton states we considered are not chiral; however, we will assume the relations connecting $S_{4}$ and $S_{6}$ to topological quantum numbers hold and see what we obtain. Additionally, we also consider the $3\bar{2}1^{3}$ parton state at $\nu{=}6/17$~\cite{Balram21a}. 

A comparison between the unprojected structure factor on the plane, obtained as explained above from its pair-correlation function by imposing the constraints arising from the topological quantum numbers, and on the sphere obtained via a direct computation in the spherical geometry is presented in Fig.~\ref{fig: comparision_fitted_computed_gr_Sq}. The unprojected structure factors on the plane and sphere are in good agreement with each other. We use the planar unprojected structure factor to compute the planar GMP gaps, while the spherical unprojected structure factor is used to evaluate the spherical GMP gaps.

\begin{table}[htbp]
	\centering
	\begin{tabular}{|c|c|c|c|c|}
		\hline 
		$\nu$ 	&  state 						&       $\mathcal{S}$~\cite{Wen92} 		   & $c_{-}$~\cite{Kane97}      & $\beta$~\cite{Gromov15} \\ \hline
		$4/11$	& $4\bar{2}1^{3}$~\cite{Balram21c}  &   $5$					&  $3$  	   &	$9/2$		\\ \hline
		$4/13$	& $\bar{4}21^{3}$~\cite{Dora22} 	&   $1$					&  ${-}1$  	   &	${-}9/2$		\\ \hline
        $6/17$	& $3\bar{2}1^{3}$~\cite{Balram21a}  &   $4$					&  $2$  	   &	$3/2$		\\ \hline
	\end{tabular} 
	\caption{\label{tab: parton_state_properties} This table gives the Wen-Zee shift $\mathcal{S}$, the chiral central charge $c_{-}$, and the orbital spin variance $\beta$ of the various parton states considered here.}
\end{table}

The GMP gap in the limit $q{\to}0$ saturates and does not diverge for $4/11$ and $6/17$ but appears to diverge for $4/13$ [see Fig.~\ref{fig: planar_sphere_GMP_gap_parton_states}]. Furthermore, for $4/11$ and $6/17$, the planar and spherical gaps are in agreement with each other, particularly in the long-wavelength limit, while that is not the case for $4/13$. This suggests that fully chiral states or their hole conjugates exist, which are equivalent to the $4/11$ and $6/17$ parton states considered, but not for the $4/13$ parton state. For example, the $3\bar{2}1^{3}$ parton state at $\nu{=}6/17$ is topologically equivalent to the fully chiral $\Phi_{1{+}1/5}\Phi_{1}^{4}$ state, where $\Phi_{1{+}1/5}$ is the state where the lowest $\Lambda$L is filled and the 1/5 Laughlin state is constructed in the second $\Lambda$L~\cite{Balram21a}.

Even if a FQH state is not chiral, the presence of disorder can gap out pairs of counter-propagating modes, thereby rendering the state chiral~\cite{Haldane95}. The condition for gapping out such pairs is captured by the null-vector criterion: a vector $\boldsymbol{\Lambda}$ satisfying $\boldsymbol{\Lambda}^{T}\cdot K\cdot \boldsymbol{\Lambda}{=}0$~\cite{Haldane95, Levin13, Balram20}. The 4/11 and 6/17 $K$-matrices have only one negative eigenvalue along with four and three positive eigenvalues~\cite{Balram21c, Balram21a}. The existence of a null-vector in these states ($\boldsymbol{\Lambda}^{4\bar{2}1^{3}}{=}\{0, 0, 1, 1, 0\}^{T}$ and $\boldsymbol{\Lambda}^{3\bar{2}1^{3}}{=}\{0, 1, 1, 0\}^{T}$) gaps out a pair of counter-propagating modes, leaving the gapless edge modes co-propagating, rendering the edge theory fully chiral. In contrast, the $4/13$ $K$-matrix has three negative and two positive eigenvalues~\cite{Dora22}. It does have a null-vector, $\boldsymbol{\Lambda}^{\bar{4}21^{3}}{=}\{1, 0, 0, 1, 0\}^{T}$, but that only gaps out a pair of counter-propagating edge modes, leaving behind two upstream and one downstream mode, which still keeps the edge theory non-chiral.

In all these parton states, we expect that the GMP mode in the long-wavelength limit, i.e., the GMP graviton, will split into multiple parton gravitons as is the case for the secondary Jain states at $n/(4n{\pm}1)$ for $n{\geq}2$~\cite{Balram21d, Nguyen22, Balram24, Bose25}. These parton gravitons can be described as the long-wavelength limit of the parton exciton in the IQH wave function $\Phi_{n}$ with $|n|{\geq}2$.

\subsection{Relation between pair-correlation function and structure factor on the sphere}
\label{app: gr_SL_sphere}
This appendix provides a derivation connecting the total pair-correlation function $g^{I}$ and the structure factor $S^{I}$ on the sphere for a uniform state. On the sphere, the pair-correlation function $g^{I}\left(\Omega\right)$ in a uniform state is defined as
\begin{align}
\label{eq: g(r)_definition}
   g^{I}\left(\Omega\right)&= \frac{\left\langle\rho^{I}\left(\boldsymbol{\Omega}_{1}\right)\rho^{I}\left(\boldsymbol{\Omega}_{2}\right)\right\rangle - \delta^{(2)}\left(\boldsymbol{\Omega}_{1}-\boldsymbol{\Omega}_{2}\right)\left\langle\rho^{I}\left(\boldsymbol{\Omega}_1\right)\right\rangle}{(\rho_{0})^2},
\end{align}
where $\Omega{=}\left|\boldsymbol{\Omega}_{1}{-}\boldsymbol{\Omega}_{2}\right|$ is the chord distance between two particles at $\boldsymbol{\Omega}_1{=}(\theta_1,\phi_1)$ and $\boldsymbol{\Omega}_2{=}(\theta_2,\phi_2)$, and the uniform density $\rho_{0}{=}N/(4\pi Q)$ [here, we have set $\ell{=}1$]. Using Eq.~\eqref{eq: angular_momentum_decomposition}, $g^{I}\left(\Omega\right)$ can be written as
\begin{align}
\label{eq: g(r)_S_L_derivation_step_1}
     g^{I}\left(\Omega\right)&=\frac{N}{4\pi Q^2 (\rho_{0})^2}\sum_{L{=}0}^{\infty}\bigg[S^{I}\left(L\right)\sum_{M{=}-L}^{L}Y_{L, M}\left(\boldsymbol{\Omega}_1\right)\times\nonumber\\
     &\left[Y_{L, M}\left(\boldsymbol{\Omega}_2\right)\right]^{*}\bigg]- \frac{1}{\rho_0}\delta^{(2)}\left(\boldsymbol{\Omega}_{1}-\boldsymbol{\Omega}_{2}\right).
\end{align}
Here, we have used the definition of the total structure factor $S^{I}(L){=}(N/4\pi)\left\langle\left(\rho^{I}_{L, M}\right)^{\dagger}\rho^{I}_{L, M}\right\rangle$, and substituted $\left\langle\rho^{I}\left(\boldsymbol{\Omega}_1\right)\right\rangle{=}\rho_0$, since the state is uniform. Next, using the addition theorem of spherical harmonics,
\begin{align}
    P_{L}\left(\cos\left(\theta_{12}\right)\right)&=\frac{4\pi}{2L+1}\sum_{M{=}-L}^{L}Y_{L, M}\left(\boldsymbol{\Omega}_1\right)\left[Y_{L, M}\left(\boldsymbol{\Omega}_2\right)\right]^{*},
\end{align}
one obtains
\begin{align}
    g^{I}\left(\Omega\right)&=\frac{1}{(\rho_{0})^2}\frac{N}{4\pi Q^2}\sum_{L{=}0}^{\infty}\left(2L+1\right)P_{L}\left(\cos\left(\theta_{12}\right)\right)S^{I}\left(L\right)\nonumber\\
    &- \frac{1}{\rho_0}\delta^{(2)}\left(\boldsymbol{\Omega}_{1}-\boldsymbol{\Omega}_{2}\right).
\end{align}
Here, $\theta_{12}$ is the polar angle between two unit vectors $\boldsymbol{\Omega}_1$ and $\boldsymbol{\Omega}_2$, and $P_{L}\left(x\right)$ denotes the $L^{\rm th}$ order Legendre polynomial. Since $g^{I}\left(\Omega\right)$ depends only on the distance between two particles, we fix one particle at the north pole such that $\boldsymbol{\Omega}_1{=}(0,0)$, and the other particle at $\boldsymbol{\Omega}_2{=}(\theta,\phi)$, so that, $\theta_{12}{=}\theta$. Using the fact that $\delta^{(2)}\left(\boldsymbol{\Omega}_{1}{-}\boldsymbol{\Omega}_{2}\right){=}\delta(\theta)\delta(\phi)/(Q \sin(\theta))$, which for an azimuthally symmetric case of our interest, can be further reduced to $\delta(\theta)/(2 \pi Q \sin(\theta))$, the above equation then becomes
\begin{align}
\label{eq: pair_correlation_unprojected_structure_factor_relation}
    g^{I}\left(\theta\right)&=\frac{1}{N}\sum_{L{=}0}^{\infty}\left(2L+1\right)P_{L}\left(\cos\left(\theta\right)\right)S^{I}\left(L\right)\nonumber\\
    &- \frac{2}{N} \frac{\delta\left(\theta\right)}{\sin(\theta)},
\end{align}
where we have substituted the value for $\rho_{0}$. This is the desired relation between the pair-correlation function and the unprojected structure factor.

Furthermore, one can also relate $ g^{I}\left(\theta\right)$ to the projected structure factor $\bar{S}^{I}\left(L\right)$ by using Eq.~\eqref{eq: relaton_unprojected_projected_structure_factor_fully_polarized_states}, which results in
\begin{align}
\label{eq: pair_correlation_projected_structure_factor_relation}
    g^{I}\left(\theta\right)&=\frac{1}{N}\sum_{L{=}0}^{2Q}(2L+1) P_{L}\left(\cos\left(\theta\right)\right) \\ 
    &\times \left[\bar{S}^{I}\left(L\right)-(2Q+1)\left(\begin{array}{ccc}
Q & Q & L \\
-Q & Q & 0
\end{array}\right)^{2}\right]. \nonumber
\end{align}

The Eqs.~\eqref{eq: pair_correlation_unprojected_structure_factor_relation} and~\eqref{eq: pair_correlation_projected_structure_factor_relation} can equivalently be expressed in terms of the arc distance, $R_{a}$, or the chord distance, $R_{c}$, by noting their relations with $\theta$: $R_{a}{=}\sqrt{Q}\theta$ and $R_{c}{=}2\sqrt{Q}\sin\left(\theta/2\right)$. For fermionic systems, as $g^{I}\left(\theta{=}0\right){=}0$, one obtains the following sum rule satisfied by $\bar{S}^{\rm fermionic}\left(L\right)$:
\begin{align}
    \sum_{L{=}0}^{2Q}(2L+1)\left[\bar{S}^{\rm fermionic}\left(L\right)\right]&=2Q+1.
\end{align}
The pair-correlation function of the $\nu{=}1$ IQH state, which has $\bar{S}(L){=}N\delta_{L,0}$, is given by
    \begin{align}
   g_{\nu{=}1}^{I}\left(\theta\right)&= 1-\frac{1}{N}\sum_{L{=}0}^{2Q}(2L+1)P_{L}\left(\cos\left(\theta\right)\right) \\
   &\times \left[(2Q+1)\left(\begin{array}{ccc}
Q & Q & L \\
-Q & Q & 0
\end{array}\right)^{2}\right]. \nonumber
\end{align}
Interestingly, the right-hand side of the above equation satisfies the identity
\begin{align}
&1-\frac{1}{N}\sum_{L{=}0}^{2Q}(2L+1)P_{L}\left(\cos\left(\theta\right)\right)\left[(2Q+1)\left(\begin{array}{ccc}
Q & Q & L \\
-Q & Q & 0
\end{array}\right)^{2}\right] \nonumber \\
&=1-\left(1-\frac{R_{c}^{2}/2}{2Q}\right)^{2Q},~~~(N{=}2Q{+}1 \text{ for }\nu{=}1)
\end{align}
which for fixed chord distance $r{\equiv}R_{c}$ in the limit $2Q{\to}\infty$ results in $g^{\nu{=}1}(r){=}1{-}e^{-\frac{r^{2}}{2\ell^2}}$ [we have restored the magnetic length $\ell$ here], which matches the well-known planar result~\cite{Giuliani08}.

One can also invert Eq.~\eqref{eq: pair_correlation_unprojected_structure_factor_relation} to express $S^{I}\left(L\right)$ in terms of $g^{I}\left(\theta\right)$ by using the orthogonality of Legendre polynomials, which results in
\begin{align}
    S^{I}\left(L\right)&=1+\frac{N}{2}\int_{-1}^{1} d\left(\cos\left(\theta\right)\right)g^{I}(\theta) P_{L}\left(\cos\left(\theta\right)\right).
\end{align}

We conclude this appendix by noting the normalization of the pair-correlation function in the thermodynamic limit, for which we switch to the planar geometry. First, we note that $g^{I}\left(r\right){=}g_{\uparrow,\uparrow}\left(r\right){+}g_{\uparrow,\downarrow}\left(r\right){+}g_{\downarrow,\uparrow}\left(r\right){+}g_{\downarrow,\downarrow}\left(r\right)$. By definition $g_{\alpha,\beta}(r)$ represents the probability of finding a spin-$\beta$ particle at a distance $r$ from a spin-$\alpha$ particle, where $\alpha,\beta{=}\uparrow,\downarrow$. We normalize the pair-correlation function such that as $r{\to}\infty$, $g_{\alpha,\beta}\left(r\right)$ reduces to $N_{\alpha}N_{\beta}/N^2$, which is the probability of finding two far-separated, uncorrelated particles with spin-$\alpha$ and spin-$\beta$. Consequently, $g^{I}\left(r{\to}\infty\right){\to}1$. 

\section{Spin-flip and spin-conserving gaps in the Haldane-Rezayi singlet state.}
\label{app: gaps_Haldane_Rezayi}
\begin{figure*}[tbh!]
         \includegraphics[width=0.66\columnwidth]{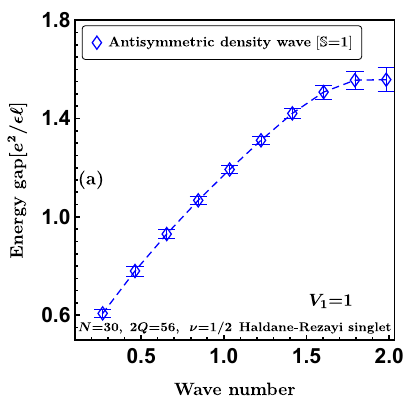}
        \includegraphics[width=0.66\columnwidth]{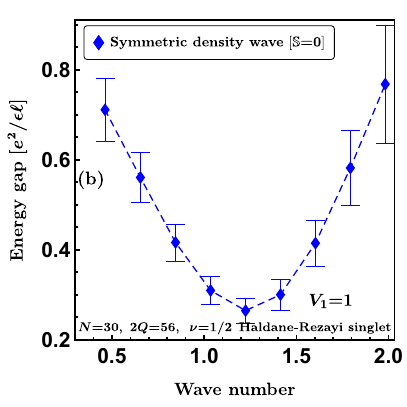}
      \includegraphics[width=0.66\columnwidth]{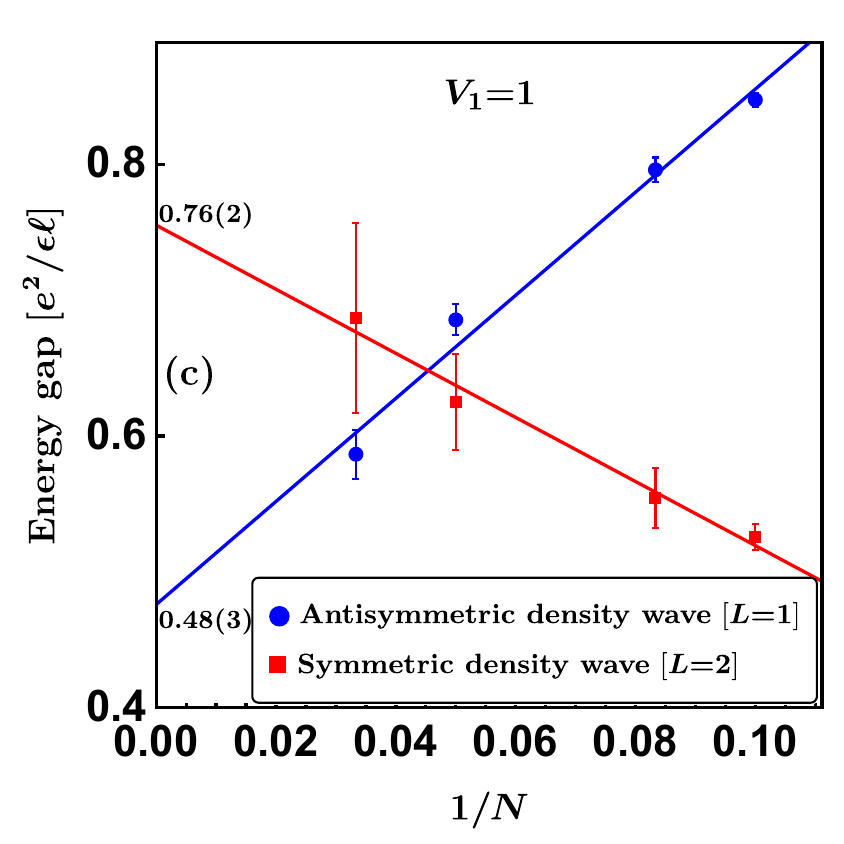}
        \caption{Spin-flip and spin-conserving density-wave gaps in the $\nu{=}1/2$ Haldane-Rezayi singlet state. The rightmost panel shows the thermodynamic extrapolation of the long-wavelength density-corrected~\cite{Morf86b} gaps of the antisymmetric and symmetric density-wave modes.} 
          \label{fig: gaps_Haldane_Rezayi_singlet}
        \end{figure*}
The spin-flip and spin-conserving density-wave gaps for the $\nu{=}1/2$ Haldane-Rezayi singlet state [see Eq.~\eqref{eq: Haldane_Rezayi_WF}]~\cite{Haldane88a, Haldane88b} are shown in Figs.~\ref{fig: gaps_Haldane_Rezayi_singlet}$(a)$ and~~\ref{fig: gaps_Haldane_Rezayi_singlet}$(b)$, respectively, for the $V_1{=}1$ interaction, which is its model Hamiltonian. Fig.~\ref{fig: gaps_Haldane_Rezayi_singlet}$(c)$ depicts the thermodynamic extrapolation of the long-wavelength density-corrected~\cite{Morf86b} gaps: $L{=}1$ gap for ADW and $L{=}2$ for SDW modes, as a linear function of $1/N$. Moreover, the gaps are multiplied by a density-correction factor of $\sqrt{2Q\nu/N}$ before the extrapolation to minimize the finite size effects on the sphere~\cite{Morf86}. 

The extrapolated gap at $L{=}2$ for the SDW state, equivalently the GMP graviton, $\Delta^{\rm GMP-graviton}_{\rm HR}{=}0.76(2)e^2/(\epsilon\ell)$, differs significantly from that computed from ED, $\Delta^{\rm ED-graviton}_{\rm HR}{=}0.21(6)e^2/(\epsilon\ell)$~\cite{Nguyen23}. This suggests that, in the Haldane-Rezayi singlet state, the GMP graviton splits into multiple gravitons. 

\bibliography{biblio_fqhe}

\end{document}